\documentclass[varenna]{cimento}
\usepackage[pdftex]{graphicx}
\usepackage{amsmath}
\usepackage{amsfonts}
\usepackage[latin1]{inputenc}

 \newcommand{\bs}{\boldsymbol}
 \newcommand{\D}{{\rm d}}
 \newcommand{\I}{{\rm i}}
 \newcommand{\E}{{\rm e}}
 \newcommand{\NN}{\mathbb{N}}
 \newcommand{\ZZ}{\mathbb{Z}}
 \newcommand{\oc}{\omega_{\rm c}} 
 \newcommand{\kB}{k_{\rm B}} 
  \newcommand{\hr}{\hat{\bs r}}
  \newcommand{\hp}{\hat{\bs p}}{}
  \newcommand{\hv}{\hat{\bs v}}
  
  \newcommand{\lm}{\ell_{\rm mag}} 
  \newcommand{\ie}{\emph{i.e.},}
  \newcommand{\intint}{\int \hskip-3mm\int}   
  \newcommand{\beq}{\begin{equation}}
\newcommand{\eeq}{\end{equation}}
\newcommand{\bpm}{\begin{pmatrix}}
\newcommand{\epm}{\end{pmatrix}}
 \newcommand{\Er}{E_{\rm r}}
   \newcommand{\ketg}{|g\rangle}
  \newcommand{\kete}{|e\rangle}

  \newcommand{\unitx}{{\bs u_x}}
  \newcommand{\unity}{{\bs u_y}}
  \newcommand{\unitz}{{\bs u_z}}

\title{Introduction to the physics of artificial gauge fields}
\author{J.~Dalibard}
\institute{Collège de France and Laboratoire Kastler Brossel, CNRS, ENS-PSL Research University, UPMC-Sorbonne Universités, 
11 place Marcelin Berthelot, 75005 Paris, France}

 \PACSes{  
 \PACSit{03.65.Vf}{}
 \PACSit{03.75.-b}{} 
 \PACSit{67.85.-d}{}}

\begin{document}

\maketitle

\begin{abstract}
Simulating magnetic effects with cold gases of neutral atoms is a challenge. Since these atoms have no charge, one needs to create artificial gauge fields by taking advantage of the geometric phases that can result for instance from atom-light interaction. We review here some schemes that lead to the desired Hamiltonians, either in a bulk geometry or in a lattice configuration. We also detail the relations between some general concepts of magnetism, such as gauge invariance, Landau levels, topological bands, and the features that can be generated in cold atoms setups. 
\end{abstract}

%\tableofcontents

Magnetic effects play an essential role in quantum physics. Notions or phenomena as diverse as gauge invariance, quantum Hall and Aharonov--Bohm effects, topological insulators, find their origin in the coupling between electromagnetic fields and particles carrying an electric charge. For a particle with mass $M$, charge $q$ and velocity $\bs v$, this interaction is described in terms of the Lorentz force $\bs F=q\, \bs v \times \bs B$ or equivalently the Hamiltonian
\begin{equation}
\hat H =\frac{\left( \hp -q \bs A (\hr)\right)^2}{2M},
\label{eq:mag_Hamiltonian}
\end{equation}
where $\bs A$ is the vector potential associated to the field $\bs B$ and $\hp$ the momentum operator of the particle. The study of magnetism constitutes an important aspect of the general \emph{quantum simulation} program based on cold atomic gases, in which one hopes to emulate with these gases a large variety of quantum phenomena that one encounters in other fields of many-body physics, for example condensed matter physics \cite{Bloch:2012}. However the fact that the atoms are neutral ($q=0$) forces one to look for ``substitutes" to real magnetism, for example using light beams with well chosen frequencies and directions. 

This course will constitute an introduction to the general subject of artificial (or synthetic) magnetic fields. Let us emphasize in advance that this subject is developing extremely fast and it is not possible in the limited space of these lecture notes to give an exhaustive presentation. Our  goal is to present an overview of the domain, and orient the reader to more complete reviews such as  \cite{Goldman:2013b,Galitski:2013,Dalibard:2011} in order to deepen the subject.

The outline of this lecture is the following. In the first section, we present some important features of magnetism such as gauge invariance and Landau spectrum. We discuss the Aharanov--Bohm effect, which will constitute later a guideline for identifying magnetic-like phenomena. We also present a first option to simulate magnetism, which consists in setting a gas of neutral particles in rotation. Sect.~\ref{sec:free_atom} is devoted to the simulation of magnetism using geometrical phases, such as Berry's phase. We  show how to transpose this notion to particles moving in a light field and following adiabatically a dressed state, \ie\ an eigenstate of the atom-light coupling. In sect.~\ref{sec:non_abelian} we generalize the notion of geometric phase to the case of non-Abelian potentials, and discuss the particular case of spin-orbit coupling. Then sect.~\ref{sec:lattice} is devoted to the important case of a particle moving in a lattice in the presence of a magnetic flux. The discretization of space introduced by the lattice considerably enriches the problem, with a single-particle energy spectrum that acquires a fractal nature. We then discuss various approaches to simulate this lattice problem with neutral atoms, based either on optical lattice shaking or modulation (sect.~\ref{sec:lattice_shaking}) or on the use of a resonant coupling between atomic internal states (sect.~\ref{sec:lattice_internal}). Finally we conclude by outlining some perspectives of this field of research, when one tries to go beyond the single-particle problem considered in these notes. The first appendix attached to this chapter discusses the eigenvalue problem for a particle in a uniform magnetic field (Landau levels). The second appendix 
details the notion of topology for a particle in a lattice, and outlines the characterization of these topological properties in terms of an integer number, the Chern index. 

\section{Magnetism and quantum physics}
\label{sec:introduction}

\subsection{Gauge invariance}
\label{subsec:gauge_invariance}
We will essentially consider in these notes magnetic effects induced by a static magnetic field $\bs B(\bs r)$. This magnetic field is divergence-free, $\bs \nabla \bs B=0$, which expresses the fact that the flux of $\bs B$ across any closed surface is zero or, in other words, that there does not exist free magnetic monopoles. From this constraint, one deduces that  $\bs B$ can be written as $\bs B=\bs \nabla \times \bs A$, where $\bs A$ is a vector field. It is clear that this writing is not unique since two vector potentials $\bs A$ and $\bs A'$ such that $\bs \nabla \times(\bs A-\bs A')=0$ correspond to the same magnetic field and have the same physical effects. More precisely, since a vector field with zero curl can be written as the gradient of a scalar function $\chi(\bs r)$, the set of vector potentials that correspond to a given magnetic field $\bs B(\bs r)$ are related to each other by
\begin{equation}
\bs A'(\bs r)=\bs A(\bs r)+\bs \nabla \chi(\bs r),
\label{eq:gauge_change}
\end{equation} 
where $\chi(\bs r)$ is an arbitrary (sufficiently regular) function.

Consider for example the case of a uniform field $\bs B=B\, \unitz$, where $\bs u_j$ denotes the unit vector along the $j$ axis. Among an infinite number of options, three convenient choices  for $\bs A$ are the following vector fields in the $xy$ plane: 
\begin{equation}
\bs A(\bs r)=Bx\,\unity, \qquad \bs A(\bs r)=-By\,\unitx,\qquad \bs A(\bs r)=\frac{B}{2}\left(
x\,\unity -y\,\unitx
\right),
%\abel{}
\end{equation}
where the two first expressions correspond to the \emph{Landau gauge} and the third one to the \emph{symmetric gauge}.  The function $\chi$ allowing one to change from one gauge to the other is in this case proportional to $Bxy$. We note that these gauge choices break the in-plane translational and rotational invariances of the initial problem.

We now turn to the description of the motion of a charged particle in the magnetic field $\bs B$ within Lagrangian and Hamiltonian formalisms. Imposing the equation of motion $m\ddot {\bs r}= q\dot {\bs r} \times \bs B$, one finds that the class of suitable Lagrange functions is 
\begin{equation}
L(\bs r, \dot {\bs r})=\frac{1}{2}M\dot{\bs r}^2 + q \;\dot{\bs r}\cdot \bs A(\bs r).
\label{eq:lagrangien_B}
\end{equation}
The Euler--Lagrange equations
\begin{equation}
\frac{\partial L}{\partial r_i}=\frac{\D}{\D t} \left( \frac{\partial L}{\partial \dot r_i} \right), \qquad r_i=x,y,z,
\label{eq:Euler_Lagrange}
\end{equation}
then provide the desired result. The Lagrange function (\ref{eq:lagrangien_B}) clearly depends on the gauge choice. However one can check from eq.\;(\ref{eq:gauge_change}) that Lagrange functions corresponding to gauge choices for the same $\bs B$ field differ only by a total derivative with respect to time, hence do correspond to the same physical problem.

The Hamilton function $H(\bs r,\bs p)$ associated to the Lagrange function $L(\bs r, \dot {\bs r})$ is obtained by calculating first the conjugate momentum
\begin{equation}
\bs p=\bs \nabla_{\dot{\bs r}}L(\bs r,\dot{\bs r})=M\dot{\bs r}+ q \bs A(\bs r),
\label{eq:moment_Hamilton}
\end{equation}
and then making the Legendre transform $H(\bs r,\bs p)=\bs p\cdot \dot{\bs r} - L(\bs r, \dot {\bs r})$, which leads to the result (\ref{eq:mag_Hamiltonian}). When turning to quantum mechanics, canonical quantization amounts to associate operators $\hat {\bs r}$ and $\hat {\bs p}$ to the position and canonical momentum of the particle, with the quantization rule
$[\hat r_j,\hat p_k]=\I\hbar\,\delta_{j,k}$ \cite{Dirac:1931,Dirac:livre}. We will always take in the following the standard choice $\hp =-\I \hbar \bs \nabla_{\bs r}$, which ensures that these commutation relations are fulfilled.

It is clear that if a given wavefunction $\psi(\bs r,t)$ is a solution of the Schrödinger equation
\begin{equation}
\I \hbar \frac{\partial \psi(\bs r,t)}{\partial t}=\frac{(-\I \hbar \bs \nabla -q \bs A(\bs r))^2 }{2M} \psi(\bs r,t),
\label{eq:Schro}
\end{equation}
it will generally not be a solution of the Schrödinger equation for another gauge choice $\bs A'(\bs r)$ that is deduced from $\bs A(\bs r)$ by the gauge change of eq.\;(\ref{eq:gauge_change}). The remedy to this problem is simple: one has to impose that the wavefunction is also modified in a gauge change. More precisely a gauge transformation in quantum mechanics is defined as the simultaneous substitutions 
\begin{eqnarray}
\bs A(\bs r) &\longrightarrow& \bs A'(\bs r)=\bs A(\bs r)+\bs \nabla \chi(\bs r), \\
\psi(\bs r,t) &\longrightarrow& \psi'(\bs r,t) =\hat T \psi(\bs r,t), \qquad \mbox{with} \quad \hat T=\exp[\I q\chi(\hr)/\hbar].
\label{eq:modif_psi_gauge}
\end{eqnarray}
One can then check that if $\psi$ is a solution of the Schrödinger equation (\ref{eq:Schro}) for the vector potential $\bs A$, then $\psi'$ is a solution of the Schrödinger equation for  $\bs A'$.

\subsection{Cyclotron motion and Landau levels}
\label{subsec:cycloton_LL}

Consider a charged particle placed in a uniform magnetic field $\bs B$ parallel to the $z$ axis. We will restrict in the following to the dynamics of the particle in the $xy$ plane. If the motion of the particle is described by classical mechanics, the particle undergoes a uniform, circular motion
with angular frequency
\begin{equation}
\oc =\frac{|q|B}{M},
%\abel{}
\end{equation}
called the \emph{cyclotron frequency}. In quantum mechanics, dimensional analysis indicates that  a natural energy scale appears, $\hbar \oc$, as well as the length scale
\begin{equation}
\lm=\sqrt {\frac{\hbar}{M\oc}}
%\abel{}
\end{equation}
called the \emph{magnetic length}. For an electron in a field of 1\,T, the cyclotron frequency and the magnetic length take the value $\oc/2\pi=28$\,GHz and $\lm=26$\,nm.
 
The reason for which the magnetic length scale appears physically in the quantum problem can be understood as a consequence of Heisenberg inequality. Classically, the size $r_0$ and the velocity $v_0$ of a cyclotron orbit are linked by the linear relation $v_0=\oc r_0$: the smaller the orbit $r_0$, the smaller the corresponding velocity $v_0$. In quantum terms one cannot prepare the particle in a state where both the position and the velocity are arbitrarily well known and $\Delta r\, \Delta v \geq \hbar/(2M)$. The magnetic length can be understood as the minimal cyclotron orbit size compatible with this inequality. 

The energy spectrum of a charged particle in a uniform $B$ field is remarkably simple and consists of equidistant \emph{Landau levels}:
\begin{equation}
E_n=\left( n + \frac{1}{2} \right) \hbar \oc, \qquad n \in \NN.
\label{eq:Landau_level_spectrum}
\end{equation} 
To prove this result we introduce the \emph{kinetic momentum}
\begin{equation}
\hat {\bs \Pi}=\hat {\bs p} -q\bs A(\hat {\bs r})
\label{eq:kinetic_momentum}
\end{equation}
 and rewrite the Hamiltonian (\ref{eq:mag_Hamiltonian}) as
\begin{equation}
\hat H =\frac{1}{2M} \left( \hat \Pi_x^2 +\hat \Pi_y^2
\right).
\end{equation}
Contrarily to $\hat {\bs r}$ and $\hat {\bs p}$, two components of $\hat {\bs \Pi}$ do not commute but their commutator is a constant for a uniform field:
\begin{equation}
[\hat \Pi_x,\hat \Pi_y]=\I \,\hbar q B.
\end{equation}
We are then facing a problem that is formally equivalent to the search of the eigenvalues of a harmonic oscillator Hamiltonian $(\hat P^2+\hat X^2)/2$ with $[\hat X,\hat P]=\I$, hence the structure in equidistant energy levels. 

The reasoning above is gauge-independent; to go further and determine a basis of eigenstates of the Hamiltonian, one needs to specify a gauge choice. The procedure is outlined in Appendix 1 both for the Landau gauge and the symmetric gauge. One finds that each Landau level has a macroscopic degeneracy ${\cal N}$, proportional to the area ${\cal A}$ accessible to the particle in the $xy$ plane, 
$ {\cal N}={\cal A}/{2\pi\lm^2}$.
This result can be interpreted as the fact that each independent state in a given Landau level occupies the area $2\pi \lm^2$. Since $\lm^2 \propto 1/B$, this degeneracy can also be written ${\cal N}={\Phi}/{\Phi_0}$, where we have introduced the flux $\Phi={\cal A}B$ of the magnetic field through the accessible area and the flux quantum $\Phi_0= h/q$.

\subsection{The Aharonov--Bohm effect}
\label{subsec:Aharonov_Bohm}
In their famous 1959 paper \cite{Aharonov-Bohm59PR}, Aharonov and Bohm\footnote{We use here the standard terminology for this effect, although a very similar discussion was made 10 years before by Ehrenberg and Siday \cite{Ehrenberg:1949}.} proposed a \emph{gedanken experiment} that illustrates a remarkable feature of quantum mechanics: One can detect the presence of a magnetic field using measurements made on particles that have never penetrated the regions of non-zero field.

\begin{figure}[t]
\begin{center}
\includegraphics[height=5cm]{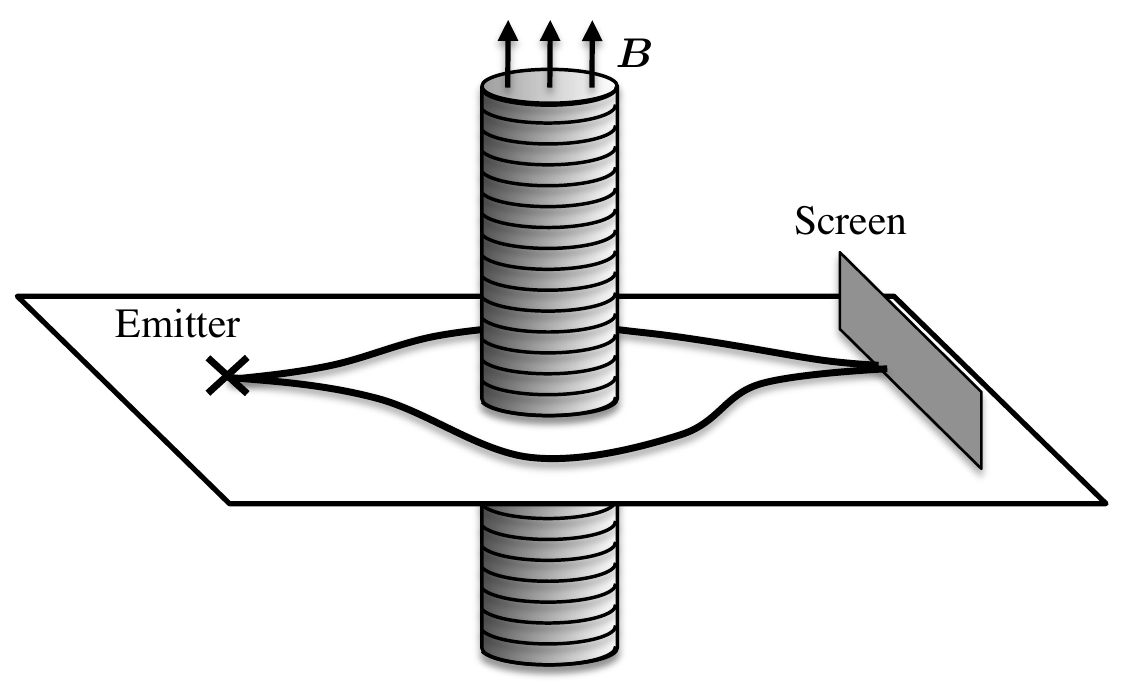}
\end{center}
\caption{Gedanken experiment of Aharonov and Bohm \cite{Aharonov-Bohm59PR}. A non-zero magnetic field inside the solenoid causes a displacement of the interference pattern detected on the screen.}
\label{fig:AB}
\end{figure}

The geometry proposed by Aharonov and Bohm is illustrated in fig.\;\ref{fig:AB}. A two-path interferometer is pierced by an infinite solenoid, which is surrounded by a potential barrier so that the particles cannot enter inside. When a current runs in the solenoid, a non-zero magnetic field appears inside the solenoid, but the field is always zero outside. However the fact of running a current through the solenoid causes a displacement of the interference pattern detected on the screen. This displacement corresponds to a phase shift of one arm of the interferometer with respect to the other one 
\begin{equation}
 \Delta \varphi =2\pi\, \frac{\Phi}{\Phi_0},
\label{eq:AB}
\end{equation}
where $\Phi$ is the magnetic flux through the solenoid and $\Phi_0=h/q$ the flux quantum. 

The Aharonov--Bohm phase is a geometric phase, in the sense that it does not depend on the velocity of the particle, nor on the time it takes to go from the emitter to the detection screen. It can even be called a topological phase in this particular geometry, since one can deform at will the trajectories without changing $\Delta \varphi$, as long as one keeps one path on each side of the solenoid.

The phase factor $\E^{\I \Delta \varphi}$ is gauge invariant since it is expressed in terms of the magnetic field itself \cite{Wu:1975}. If one insists in expressing $ \Delta \varphi$ in terms of `local' quantities, \emph{i.e.}, quantities defined in the region accessible to the particle, then it has to be written in terms of the vector potential:
\begin{equation}
\Delta \varphi=\frac{2\pi}{\Phi_0}\intint  B_z(x,y)\;\D x\,\D y=\frac{1}{\hbar} \oint_{\cal C} q\bs A(\bs r)\cdot \D \bs r,
\label{eq:AB_alter}
\end{equation} 
where the line integral is taken along the oriented contour ${\cal C}$ consisting of the two paths of the interferometer.

\begin{figure}[t]
\begin{center}
\includegraphics{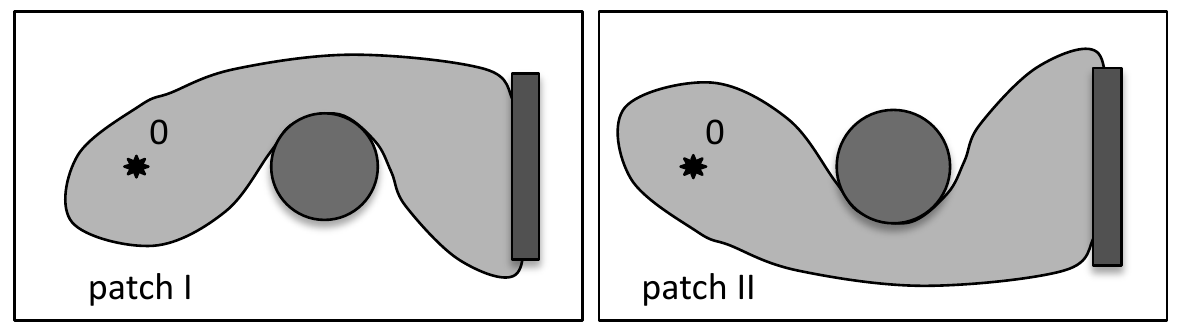}
\end{center}
\caption{Aharonov--Bohm geometry seen from above. The light grey zones are two patches over which the vector potential in the presence of current can be written as $\bs A(\bs r) =\bs \nabla \chi(\bs r)$. The function $\chi$ is not the same for the two patches.}
\label{fig:zones_AB}
\end{figure}

We briefly outline below one possible reasoning leading to eqs.\;(\ref{eq:AB}) and (\ref{eq:AB_alter}). The idea is to choose different gauges on different patches of space, and then connect these patches together \cite{Aharonov-Bohm59PR}. In the absence of current, we can make the gauge choice $\bs A=0$ in the whole space. When a current circulates in the solenoid,  $\bs A(\bs r)$ is non-zero outside the solenoid but $\bs B$ remains null in this outer region, hence we have
\begin{equation}
\bs \nabla \times \bs A(\bs r) = 0\quad \mbox{out of the solenoid.}
\label{eq:rot_A_egal_zero}
\end{equation}
Once a choice for $\bs A(\bs r)$ has been made over the whole space, consider the two simply connected regions (\emph{i.e.}, regions without a hole) sketched in fig.\;\ref{fig:zones_AB}, which we label I and II. Over each region we can unambiguously define the two scalar functions $\chi_{\rm I}$ and $\chi_{\rm II}$ by
\begin{equation}
\chi_{\rm I,II}(\bs r) = \int_{\bs 0}^{\bs r} \bs A (\bs r')\cdot \D \bs r',
\label{eq:_integrale_de_A}
\end{equation}
where the line integral goes from the emitter to a point in region I or II without leaving this region. Since patches I and II are simply connected, the functions  $\chi_{\rm I,II}(\bs r)$ are indeed single-valued. 

Consider the matter wave  $\psi_{\rm l}$ that propagates in the region on the left of the solenoid, inside patch I. For this wave, switching on the current in the solenoid corresponds to the modification
\begin{equation}
\bs A(\bs r)=0 \longrightarrow \bs A(\bs r) =\bs \nabla \chi_{\rm I}(\bs r),
\end{equation}
which can be viewed as a gauge change. This gauge change must be accompanied by the modification of the wavefunction (\ref{eq:modif_psi_gauge}). The same arguments holds for the matter wave $\psi_{\rm r}$
propagating in the region on the right of the solenoid, inside patch II.  Matter-wave interference at a point $\bs r$ on the detection screen is related to
\begin{equation}
\psi_{\rm l}^*(\bs r)\,\psi_{\rm r}(\bs r)= \exp([\I \,q (\chi_{\rm II}(\bs r)-\chi_{\rm I}(\bs r))/\hbar]  \;
{\psi_{\rm l}^{(0)}}^*(\bs r)\,\psi_{\rm r}^{(0)}(\bs r),
\end{equation}
where $\psi_{\rm l,r}^{(0)}$ are the wavefunctions in the absence of current. 
The phase entering the prefactor can be written
\begin{equation}
\chi_{\rm II}(\bs r)-\chi_{\rm I}(\bs r) = \int_{0,{\cal C}_{\rm II}}^{\bs r} \bs A (\bs r)\cdot \D \bs r
-\int_{0,{\cal C}_{\rm II}}^{\bs r} \bs A (\bs r)\cdot \D \bs r,
\end{equation}
where ${\cal C}_{\rm I}$ and ${\cal C}_{\rm II}$ are two paths going from $0$ to $\bs r$ and located respectively in patches I and II. This corresponds to the announced result (\ref{eq:AB_alter}).

The geometrical aspect of the Aharonov--Bohm phase is a hallmark of magnetism in quantum mechanics \cite{Peshkin:1989}. The general program of magnetism simulation can be viewed as a search for possible ways to induce a geometric phase with similar properties on a neutral particle.

\subsection{Rotating gases}
\label{subsec:rotations}
Among the various methods that have been developed in order to simulate magnetism, rotating the  trap holding the particles is probably the simplest one from a conceptual point of view. The idea is to take advantage of the similarity between the magnetic Lorentz force, $q \bs v\times\bs B$, and the Coriolis force which appears in the frame rotating at angular velocity $\Omega$ around the $z$ axis:
\begin{equation}
\bs F_{\rm Coriolis}=2M\,\bs v \times \bs \Omega, \qquad \bs \Omega=\Omega\,\unitz.
%\abel{}
\end{equation}  
More precisely the Hamiltonian in the rotating frame is
\begin{equation}
\hat H=\frac{\hat p^2}{2M}+V(\hr) -\Omega\hat L_z
\label{eq:H_rot1}
\end{equation}
where $V(\bs r)$ is the trapping potential in the $xy$ plane and $\hat L_z=\hat x \hat p_y-\hat y \hat p_x$ is the component of the angular momentum operator along the $z$ direction. This Hamiltonian can be rewritten in the more suggestive way
\begin{equation}
\hat H=\frac{\left( \hp -q \bs A (\hr)\right)^2}{2M} + V(\hr) + V_{\rm centrif.}(\hat r)
\label{eq:H_rot2}
\end{equation} 
where the effective vector potential and the corresponding magnetic field are
\begin{equation}
q \bs A (\bs r)=M\Omega \left( x\unity -y \unitx  \right), \qquad q\bs B = 2 M \bs \Omega ,
%\abel{}
\end{equation}
justifying the analogy between the Coriolis and the Lorentz forces. In addition to the Coriolis force, it is well known that the centrifugal force $M\Omega^2 \bs r$ also appears in the rotating frame. It is indeed present in the Hamiltonian (\ref{eq:H_rot2}), since the centrifugal potential is given by
\begin{equation}
V_{\rm centrif.}(r)=-\frac{1}{2}M\Omega^2 r^2.
%\abel{}
\end{equation}

To implement this method in practice, one can use the harmonic trapping potential:
\begin{equation}
V(\bs r)= V_0(\bs r) + \delta V(\bs r), \quad V_0(\bs r)=\frac{1}{2}M\omega^2 (x^2+y^2), \quad
\delta V(\bs r)=\frac{\epsilon}{2}M\omega^2 (x^2-y^2).
\end{equation}
Here the coordinates $x,y$ refer to basis vectors $\unitx,\unity$ that rotate around $z$ with frequency $\Omega$ with respect to an inertial frame of reference, and the dimensionless parameter $\epsilon$ characterizes the strength of the anisotropy. The success of this method can be tested by implementing it with a superfluid sample, and looking for the vortex lattice that is expected in steady state (see e.g. \cite{Madison:2000a,AboShaeer:2001,Zwierlein:2005}). 

It is interesting to see how one can implement exactly the magnetic Hamiltonian (\ref{eq:mag_Hamiltonian}) with this rotation method. Comparing eq.\;(1) with eq.\;(\ref{eq:H_rot2}), we see that two conditions need to be simultaneously fulfilled: (i) The centrifugal potential has to balance the trapping potential, so that $\Omega=\omega$, which means that one has to resort to an extra  potential (possibly quartic like in \cite{Bretin:2004})  in order to confine the particles; (ii) The trapping anisotropy $\epsilon$ must tend to zero.  At first sight the latter condition seems impossible to fulfill, since nothing sets the gas in rotation in this case. However this difficulty can be circumvented by using the \emph{evaporative spin-up} technique invented by the Boulder group \cite{Schweikhard:2004a}. It consists in preparing first a cloud rotating at a moderate angular frequency (for example $\Omega\sim 0.7\,\omega$) using a non-zero $\epsilon$, switching off this anisotropy, and then performing a selective evaporation that eliminates atoms with an angular momentum below average. In this way the angular momentum per remaining atom increases in the course of evaporation. At the end of the evaporation, the search for the equilibrium state of the gas must take into account the angular momentum $L_z$ of the trapped particles, which can be done using the Lagrange multiplier method. One then ends up  by minimizing the same mathematical quantity as in eq.\;(\ref{eq:H_rot1}), where $\Omega$ now stands for the Lagrange parameter associated to $L_z$. Experimentally the Boulder group could reach $\Omega=0.993\,\omega$ with this method \cite{Schweikhard:2004a}.

%%%%%%%%%%%%%%%%%%%%%%%%%%%%%%%%%%%%%%%%%%%%%%%%%%%

\section{Geometric phases and gauge fields for free atoms}
\label{sec:free_atom}

The notion of adiabatic evolution is frequently encountered in classical and quantum physics, when one looks for the evolution of a system whose state depends on external parameters. Let us consider this type of evolution and focus on the case where the external parameters travel along a closed trajectory. This means that these parameters, which we represent by a vector $\bs \lambda$, evolve slowly from time 0 to time $T$, and take at time $T$ the same value as at initial time:
\begin{equation}
\bs \lambda(0) \longrightarrow \bs \lambda(t) \longrightarrow \bs \lambda(T)=\bs \lambda(0).
%\abel{}
\end{equation}
It is then natural to assume that the system under study is in the same state at time $T$ as at time $0$. However this is not always true, as exemplified by the well-known Foucault pendulum. Suppose that the pendulum oscillates linearly at a initial time. Twenty-four hours later, the Earth has made a full rotation around its axis; hence the suspension point of the pendulum, playing here the role of the external parameter, is back to the same place. However the oscillation plane of the pendulum has rotated by an angle that depends on the latitude at which the pendulum is located.  

The goal of this section is to take advantage of this anholonomy in the quantum case, in order to generate geometric phases (Berry's phase) that can mimic the Aharonov--Bohm phase even if the considered particle has no electric charge. 

\subsection{Berry's phase} 

Consider a quantum system whose Hamiltonian depends on a continuous parameter $\bs \lambda$. We assume for the moment that this parameter is controlled by an external operator.  The parameter $\bs \lambda$ can stand for a real number or a set of numbers. For each value of $\bs \lambda$ we denote by $|\psi_n(\bs \lambda)\rangle$ the eigenstates of the Hamiltonian and $ E_n(\bs \lambda)$ the corresponding energies:
\begin{equation}
\hat H(\bs \lambda) |\psi_n(\bs \lambda)\rangle= E_n(\bs \lambda)\; |\psi_n(\bs \lambda)\rangle.
\label{eq:eq_val_propres}
\end{equation}
We suppose that the set $\{|\psi_n(\bs \lambda)\rangle\}$ forms an orthonormal basis of the Hilbert space for each value of $\bs \lambda$. We are interested in the evolution of the state vector of the system 
\begin{equation}
|\psi(t)\rangle=\sum_n c_n(t)\; |\psi_n[\bs \lambda(t)]\rangle
%\abel{}
\end{equation}
when $\bs \lambda$ slowly evolves in time. More specifically we suppose that the system is prepared at initial time in one particular eigenstate $|\psi_\ell\rangle$:
\begin{equation}
c_\ell(0)=1, \qquad c_n(0)=0 \quad \mbox{if}\ n\neq\ell,
%\abel{}
\end{equation}
 and we suppose that the evolution of $\bs \lambda$ is slow enough for the adiabatic theorem to hold \cite{Messiah_adiab}. In first approximation the state of the system  is thus at any time $t$  proportional to $|\psi_\ell[\bs \lambda(t)]\rangle$, and the evolution of the coefficient $c_\ell(t)$ is obtained from the Schrödinger equation:
\begin{eqnarray}
\I \hbar\; \dot c_\ell&=& \left[ E_\ell(t) -\I \hbar \dot{\bs \lambda}\cdot \langle \psi_\ell|\bs \nabla \psi_\ell \rangle \right]\,c_\ell, \nonumber \\
&=& \left[ E_\ell(t) - \dot{\bs \lambda}\cdot \bs {\cal A}_\ell (\bs \lambda) \right]\,c_\ell.
\label{eq:evol_cl}
\end{eqnarray}
Here we introduced the real vector called \emph{Berry's connection}:
\begin{equation}
\bs {\cal A}_\ell (\bs \lambda)=\I \hbar \; \langle \psi_\ell|\bs \nabla \psi_\ell \rangle,
\label{eq:def_connexion_Berry}
\end{equation}
which will play the role of the vector potential (more precisely of $q\bs A$) in the following.

Suppose now that the parameter $\bs \lambda$ follows a closed contour ${\cal C}$ in parameter 
space, so that $\bs \lambda(T)=\bs \lambda(0)$. The integration of the evolution equation (\ref{eq:evol_cl}) is straightforward and gives:
\begin{equation}
c_\ell (T)= \E^{\I \Phi^{\rm dyn.}(T)}\, \E^{\I \Phi^{\rm geom.}(T)} \,c_{\ell}(0)
%\abel{}
\end{equation}
where we introduced the dynamical phase 
\begin{equation}
\Phi^{\rm dyn.}(T)=-\frac{1}{\hbar}\int_0^T E_\ell(t)\;\D t
\label{eq:phase_dynamique}
\end{equation}
and the geometrical phase  \cite{Berry:1984}
\begin{equation}
\Phi^{\rm geom.}=\frac{1}{\hbar}\int_0^T \dot {\bs \lambda}\cdot \bs {\cal A}_\ell[\bs \lambda]\; \D t=\frac{1}{\hbar}\oint \bs {\cal A}_\ell(\bs \lambda)\cdot\D{\bs \lambda}.
\label{eq:phase_geometrique}
\end{equation}
The dynamical phase is the usual phase that appears also for a time-independent problem. The geometrical phase (Berry's phase) only depends on the ``trajectory" of the parameter $\bs \lambda$ during the evolution, and it is independent of the duration $T$. Both terms $\exp({\I \Phi^{\rm dyn.}})$ and $\exp({\I \Phi^{\rm geom.}})$ are physical quantities in the sense that they are gauge-invariant: they are unchanged if one modifies the definition of the eigenstates $|\psi_n[\lambda]\rangle$ by multiplying them by an arbitrary phase factor. 

In the following we will restrict our discussion to the case where the parameter $\bs \lambda$ evolves in a two-dimensional or a three-dimensional space. For example this parameter can stand for the position of a particle or for its quasi-momentum when it moves in a periodic potential.
We can then introduce \emph{Berry's curvature}, which plays a role similar to a magnetic field:
 \begin{equation}
\bs{\cal B}_\ell=
\bs \nabla \times \bs {\cal A}_\ell.
%\abel{}
\end{equation}
This quantity is a real, gauge-invariant, vector field. In full analogy with the Aharonov--Bohm phase we can rewrite the geometrical phase accumulated by the particle when the parameter ${\bs \lambda}$ moves along the closed contour ${\cal C}$:
\begin{equation}
\Phi^{\rm geom.}(T)=\frac{1}{\hbar}\intint_{\cal S} \bs{\cal B}_\ell\cdot\;\D^2 S
\label{eq:phase_geometrique2}
\end{equation}
where ${\cal S}$ is a surface delimited by the contour ${\cal C}$. 

\subsection{Adiabatic following of a dressed state}
\label{subsec:adiab_following}

The notion of geometric phase can be directly adapted to the case of an atom slowly moving in a monochromatic laser field \cite{Dum:1996}. Two types of degree-of-freedom come into play. First the center-of-mass motion of the atom can be described in terms of its position operator $\hat {\bs r}$ and momentum operator $\hat {\bs p}=-\I \hbar \bs \nabla_{\bs r}$. Second the internal dynamics of the atom corresponds to transitions between electronic states induced by the laser field. Within the rotating-wave approximation, this internal dynamics is described by a time-independent Hamiltonian  $\hat H_{\rm int}(\bs r)$, so that the total Hamiltonian reads:
\begin{equation}
\hat H_{\rm tot}\ =\ \frac{\hat {\bs p}^2}{2M}\ +\ \hat H_{\rm int}(\hat{\bs r}).
\label{eq:Htot}
\end{equation}

Treating first $\bs r$ as an external parameter, we define the \emph{dressed states} as the eigenstates of the internal Hamiltonian
\begin{equation}
\hat H_{\rm int}(\bs r) |\psi_n(\bs r)\rangle= E_n(\bs r)\; |\psi_n(\bs r)\rangle.
%\abel{}
\end{equation}
At any point $\bs r$, the set $\{ |\psi_n(\bs r)\rangle \}$ forms a basis set for the Hilbert space associated to the internal degrees of freedom of the atom. Then we consider the total quantum state of the atom and write it as:
\begin{equation}
\Psi(\bs r, t)=\sum_n \phi_n(\bs r, t) |\psi_n(\bs r)\rangle.
%\abel{}
\end{equation}
The characterization of the atom dynamics amounts to determining the probability amplitudes $\phi_n(\bs r,t)$ to find the atom at point $\bs r$ in the internal state $|\psi_n(\bs r)\rangle$. We now assume that the particle is prepared at initial time in a given dressed state $|\psi_\ell\rangle$. We also suppose that it moves slowly enough so that it essentially remains in this state at any time and the contribution of the $\psi_n$'s for $n\neq \ell$ can be neglected. The validity of this assumption will be discussed in subsect.~\ref{subsec:validity}. Starting from the Schrödinger equation  
\begin{equation}
\I \hbar \frac{\partial \Psi}{\partial t}= \hat H_{\rm tot}\Psi(\bs r,t)=\left( -\frac{\hbar^2}{2M}\Delta + \hat H_{\rm int}(\bs r) \right) \Psi(\bs r,t),
%\abel{}
\end{equation}
we can obtain an equation for the relevant probability amplitude $\phi_\ell$. After some algebra, the result can be written in the form:
\begin{equation}
\I \hbar \frac{\partial \phi_\ell}{\partial t}=\left[\frac{\left( \hat {\bs p} -\bs {\cal A}_\ell( {\bs r})\right)^2 }{2M}
+E_\ell(\bs r) + {\cal V}_\ell(\bs r)\right]\phi_\ell(\bs r, t).
\label{eq:eq_de_S_effective}
\end{equation}
It has exactly the structure of a scalar Schrödinger equation for a charged particle (with $q=1$ by convention) moving in the magnetic field associated to the  
vector potential
\begin{equation}
{\cal A}_\ell (\bs r)=\I\hbar\; \langle \psi_\ell|\bs \nabla \psi_\ell\rangle,
%\abel{}
\end{equation}
which is nothing but Berry's connection introduced in eq.\;(\ref{eq:def_connexion_Berry}). In addition the particle feels a potential that is the sum of two terms. The first one is simply the energy $E_\ell(\bs r)$ of the occupied dressed state, and the second one is the additional scalar potential
\begin{equation}
{\cal V}_\ell(\bs r)=\frac{\hbar^2}{2M} \sum_{n\neq \ell } \left| \langle \bs \nabla \psi_\ell|\psi_n \rangle\right|^2.
\label{eq:potentiel_scalaire}
\end{equation} 
Physically ${\cal V}_\ell(\bs r)$ represents the kinetic energy associated to the micromotion of the atom, as it makes virtual transitions between the effectively occupied dressed state $\psi_\ell$ and all other dressed states $\psi_n$, $n\neq\ell$ \cite{Aharonov:1992,Cheneau:2008}.

\begin{figure}[t]
\begin{center}
\includegraphics[width=11cm]{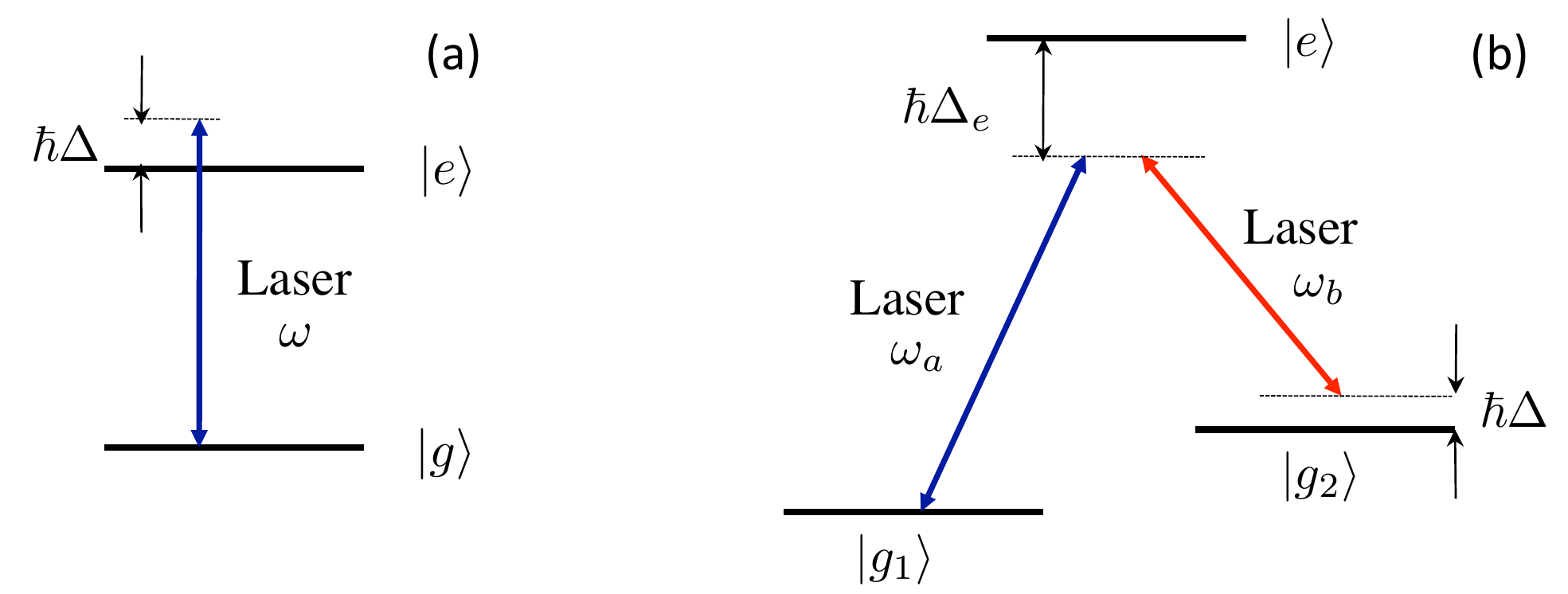}
\end{center}
\caption{(a) Quasi-resonant excitation of a two-level atom. (b) Stimulated Raman transition between two sublevels of the atomic ground state.}
\label{fig:two_level}
\end{figure}

\subsection{The two-level case}
\label{subsec:two_level}
For simplicity we now focus on the case of an atom with only two relevant internal states. We neglect any process related to the spontaneous emission of photons. This situation can occur if we are dealing with the intercombination line of an atom  with two outer electrons, like magnesium, strontium or ytterbium (fig.\;\ref{fig:two_level}a). In this case the radiative lifetime of the excited atomic state is much longer than the relevant experimental time scales, and  the atom-laser coupling is characterized only by the Rabi frequency $\kappa$ and the detuning $\Delta$ between the laser and the atomic frequencies. 

The two-level formalism can also be applied to the case of a $\Lambda$ level scheme 
(fig.\;\ref{fig:two_level}b), with a Raman transition between two sublevels of the  atomic ground state. This scheme is relevant for alkali-metal species as well as erbium or dysprosium. Then if the detuning $\Delta_e$ is large enough,  the excited state $e$ that plays the role of  a relay for the Raman transition can be formally eliminated from the dynamics. One is then left with a two-level problem in the subspace $\{ |g_1\rangle, |g_2\rangle \}$, with again two parameters: the Raman detuning $\Delta$ and the two-photon Rabi frequency $\kappa$:
\begin{equation}
\kappa= \frac{\kappa_a \kappa_b^*}{2\Delta_e},
\label{eq:Raman_coupling}
\end{equation} 
where $\kappa_j$ is the one-photon Rabi frequency for the transition $g_j \leftrightarrow e$.

In both cases the atom-laser coupling, \emph{i.e.,} the Hamiltonian $\hat { H}_{\rm int}$ responsible for the internal dynamics of the atom, can be written as a $2\times  2$ matrix\footnote{For the $\Lambda$ scheme, the levels $g_1$ and $g_2$ are also light-shifted by the lasers $a$ and $b$. For the laser coupling $(\hbar/2) \left(\kappa_a|e\rangle\langle g_1| + \kappa_b|e\rangle\langle g_2|\right) +\mbox{H.c.} $,  these light-shifts are $\hbar |\kappa_j|^2/4\Delta_e$, $j=a,b$. We assume that these light-shifts have been reincorporated into the definition of $\Delta$.} in the basis $\{ |g\rangle, |e\rangle \}$ or $\{ |g_1\rangle, |g_2\rangle \}$:
\begin{equation}
\hat { H}_{\rm int}=\frac{\hbar}{2} \begin{pmatrix}
\Delta & \kappa^* \\
\kappa & -\Delta
\end{pmatrix}.
\label{eq:matrice2par2}
\end{equation}
where both the Rabi frequency $\kappa$ and the detuning $\Delta$ may depend on the atom center-of-mass position. We define the generalized Rabi frequency $\Omega$, the mixing angle $\theta$ and the phase angle $\phi$:
\begin{equation}
 \Omega=\left( \Delta^2+|\kappa|^2\right)^{1/2},\qquad \cos\theta=\frac{\Delta}{\Omega},\qquad \sin \theta=\frac{|\kappa|}{\Omega},\qquad \kappa=|\kappa|\,\E^{\I \phi},
%\abel{}
\end{equation}
so that the atom-laser coupling becomes:
\begin{equation}
\hat { H}_{\rm int}=\frac{\hbar \Omega}{2}
\begin{pmatrix}
\cos \theta & \E^{-\I \phi} \sin \theta\\
\E^{\I \phi}\sin \theta & -\cos \theta
\end{pmatrix}=\frac{\hbar \Omega}{2}\; \bs n\cdot \hat{\bs \sigma},
%\abel{}
\end{equation}
where $\bs n$ is the unit vector characterized by the spherical angles $(\theta,\phi)$ and the $\hat \sigma_j$ ($j=x,y,z$) are the Pauli matrices.
The eigenvalues of this matrix are $\pm \hbar \Omega/2$  and the corresponding eigenstates read:
\begin{equation}
|\psi_+\rangle=\begin{pmatrix}  \cos(\theta/2) \\ \E^{\I \phi} \sin(\theta/2) \end{pmatrix}
\, , \quad
|\psi_-\rangle=\begin{pmatrix} -\E^{-\I\phi}\sin(\theta/2) \\  \cos(\theta/2) \end{pmatrix}.
 \label{eq:eigenstates}
\end{equation}

Suppose now that the state of the atom follows adiabatically one of the dressed states $|\psi_\pm\rangle$. Berry's connection and curvature associated to this state are:
\begin{eqnarray}
\bs {\cal A}_\pm &=&\pm \frac{\hbar}{2}\,  \left(\cos\theta -1 \right) \;\bs \nabla \phi
,\label{eq:connexion_Berry} \\ 
\bs {\cal B}_\pm&=& \pm \frac{\hbar}{2} \bs \nabla (\cos \theta) \times \bs \nabla \phi,
\label{eq:courbure_Berry}
\end{eqnarray}
and the scalar potential reads:
\begin{equation}
 {\cal V}_\pm (\bs r)=\frac{\hbar^2}{8M}\left[
\left(\bs \nabla \theta\right)^2 +\sin^2\theta\,\left( \bs \nabla \phi\right)^2
\right].
%\abel{}
\end{equation}
Since we are interested here in the simulation of magnetism, we look for a non-zero Berry's curvature $\bs {\cal B}_\pm$, which can be achieved only in the presence of non-zero gradients of the phase $\phi$ as well as the mixing angle $\theta$. The latter can be obtained either via a gradient of intensity ($\bs \nabla \kappa$) or a gradient of detuning ($\bs \nabla \Delta$). 

Before going further we note that we can perform a gauge transformation in this problem by multiplying the expression (\ref{eq:eigenstates}) of the dressed states by an arbitrary phase factor. In particular one can choose instead of eq.\;(\ref{eq:eigenstates}) the following dressed-state definition
\begin{equation}
|\psi'_+\rangle=\begin{pmatrix}  \E^{-\I \phi}\cos(\theta/2) \\  \sin(\theta/2) \end{pmatrix}
\, , \quad
|\psi'_-\rangle=\begin{pmatrix} -\sin(\theta/2) \\ \E^{\I\phi} \cos(\theta/2) \end{pmatrix},
 \label{eq:eigenstates2}
\end{equation}
leading to Berry's connection 
\begin{equation}
\bs {\cal A'}_\pm =\pm \frac{\hbar}{2}\,   \left(\cos\theta +1\right) \bs \nabla \phi .
\label{connexion_Berry_2}
\end{equation}
 Berry's curvature $\bs {\cal B}_\pm$ is unchanged in this procedure since it is a gauge-invariant quantity.

\begin{figure}[t]
\begin{center}
\includegraphics[width=6cm]{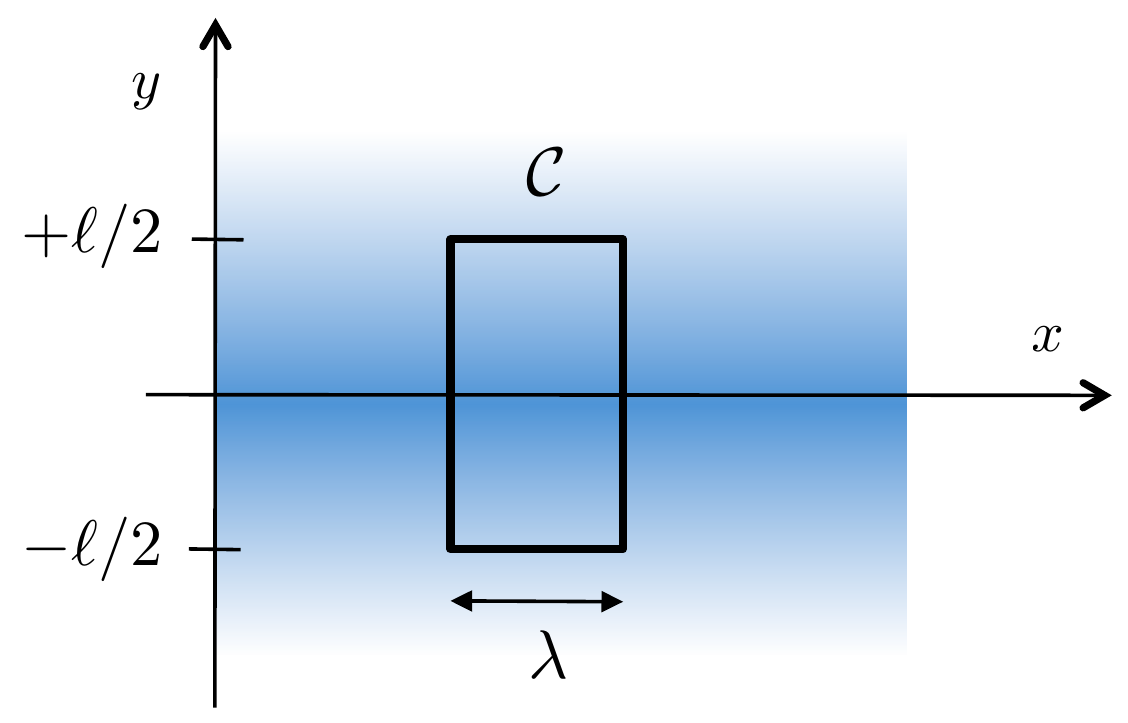}
\end{center}
\caption{Grey scale representation of the magnitude of Berry's connection generated in a $\Lambda$ level scheme (see text). The Aharonov--Bohm--Berry phase for the contour ${\cal C}$ is $\approx 2\pi$.}
\label{fig:contour}
\end{figure}

As an example consider the case of the $\Lambda$ scheme of fig.\;\ref{fig:two_level}b, where the two laser beams are supposed to be plane waves propagating along the $\pm x$ direction, with wave vectors $\bs k_a\approx-\bs k_b =k \unitx$. The two-photon Rabi frequency can thus be written $\kappa(\bs r)=\kappa_0 \,\E^{2\I kx}$, with the constant amplitude $\kappa_0$ and the phase angle $\phi=2kx$. We also suppose that the detuning varies linearly  in space along the $y$ direction: $\Delta(\bs r)=\Delta' y$; this can be achieved for instance using the differential Zeeman shift between the sublevels $g_a$ and $g_b$ in a magnetic field gradient. A natural length scale appears in the problem
\begin{equation}
\ell=\kappa_0/\Delta,
%\abel{}
\end{equation} 
which represents the width of the region centered on the $x$ axis where the Raman excitation is quasi-resonant (Rabi frequency larger than detuning). This geometry provides the necessary ingredients for a non-zero Berry curvature, {\it i.e.}, non-zero and non-collinear gradients for the two angles $\phi$ and $\theta$. It leads to 
\begin{equation}
\bs {\cal B}_\pm(\bs r)=\pm B_0 \;{\cal L}^{3/2}(y)\;\unitz
%\abel{}
\end{equation}
where
\begin{equation}
B_0=\frac{\hbar k}{\ell},\qquad {\cal L}(y)=\frac{1}{1+y^2/\ell^2}.
%\abel{}
\end{equation}
This artificial magnetic field is translationally invariant along the $x$ axis. It is maximal for $y=0$ and decreases as $|y|^{-3}$ when $y\to \infty$. Let us comment briefly on its magnitude $B_0$. Consider the rectangular contour of fig.\;\ref{fig:contour}, of extension $\lambda=2\pi/k$ along $x$ and $\ell$ along $y$,   inside which the field is approximately uniform.  The Aharonov--Bohm--Berry phase associated to this contour is
\begin{equation}
\Phi^{\rm geom.}=\frac{1}{\hbar}\intint_{\cal S} \bs {\cal B}\cdot \bs u\; \D^2 r \approx 2\pi.
%\abel{}
\end{equation}
This means that in this geometry, a superfluid gas should exhibits vortices essentially localized around the line $y=0$, with a typical linear density of one vortex every wavelength $\lambda$ along the $x$ axis. 

A scheme very similar to what we just described was implemented at NIST in 2009 in the group of I. Spielman \cite{Lin:2009b}. The non-zero value of $\Delta'$ was achieved with a magnetic gradient of a few gauss per centimeters, and a value of $\ell$ of a few tens of optical wavelength $\lambda$. When placing a rubidium Bose--Einstein condensate in this configuration, the NIST group observed the nucleation of quantized vortices, proving thus the existence of a non-zero artificial gauge field.

\subsection{Validity of the adiabatic approximation}
\label{subsec:validity}
In what precedes we have assumed that the two-level atom prepared in a given internal dressed state could follow adiabatically the dressed state when moving in the light field. Here we briefly discuss the validity of this approximation.  Generally speaking the validity criterion for the adiabatic approximation is that the angular velocity of the eigenstate $|\psi_\ell\rangle$ should be much smaller than all relevant Bohr frequencies involving this eigenstate (for more details, we refer the reader to \cite{Messiah_adiab}). Here, for an atom moving at velocity $v$, the angular velocity for a dressed state is $\sim kv$, since the expressions (\ref{eq:eigenstates}) of the dressed states  vary typically on the length scale $k^{-1}=\lambda/2\pi$.  For a two-level atom, the relevant Bohr frequency is the generalized Rabi frequency  $\Omega$ so that the validity criterion reads $kv \ll \Omega$.
Since the relevant velocities for an atom absorbing and emitting photons from counter-propagating light waves are at least of the order of the recoil velocity $\hbar k/M$, we find as a necessary condition
\begin{equation}
E_{\rm r} \ll \hbar \Omega, 
\label{eq:valid_adiab}
\end{equation}
where we introduced the recoil energy $E_{\rm r}=\hbar^2k^2/2M$. In practice,  eq.\;(\ref{eq:valid_adiab})  constitutes a relevant criterion to check the applicability of the adiabatic Schrödinger-type equation (\ref{eq:eq_de_S_effective}). 

The recoil energy also enters when one evaluates the maximum Berry's curvature that can be created with laser light. Assuming that both the phase and the mixing angles vary significantly over the  reduced wavelength $k^{-1}=\lambda/2\pi$, we find from eq.\;(\ref{eq:courbure_Berry}) that the corresponding cyclotron frequency $\omega_{\rm c}=B/M$ is such that
\begin{equation}
\hbar \omega_{\rm c} \sim E_{\rm r}.
%\abel{}
\end{equation}
In order to reach such large values, a natural strategy is to switch to optical lattice configurations, in which case atom-laser couplings indeed vary significantly over  $\lambda/2\pi$. This will addressed in sect.~\ref{sec:lattice}.

\subsection{Spontaneous emission and recoil heating}

The use of a Raman transition between sublevels of the electronic ground state of an atom is an efficient way to reduce the heating originating from the random momentum recoils caused by spontaneous emission processes. However the heating rate may not always be lowered down to an acceptable level, especially for species from the alkali-metal series. Another more favorable class of atoms is the lanthanide family, with species like erbium or dysprosium that have  recently been brought to quantum degeneracy (see \emph{e.g.} \cite{Aikawa:2012,Lu:2012}). To facilitate the comparison between species, we define the dimensionless merit factor
\begin{equation}
{\cal M}=\frac{\kappa_{\rm eff}}{\gamma}\ 
\label{eq:merit_factor}
\end{equation}
where $\gamma$ is the spontaneous emission rate, and we analyze how it can be maximized.  Our discussion follows closely that of refs. \cite{Cui:2013,Nascimbene:2013c}. 

\begin{figure}[t]
\begin{center}
\includegraphics{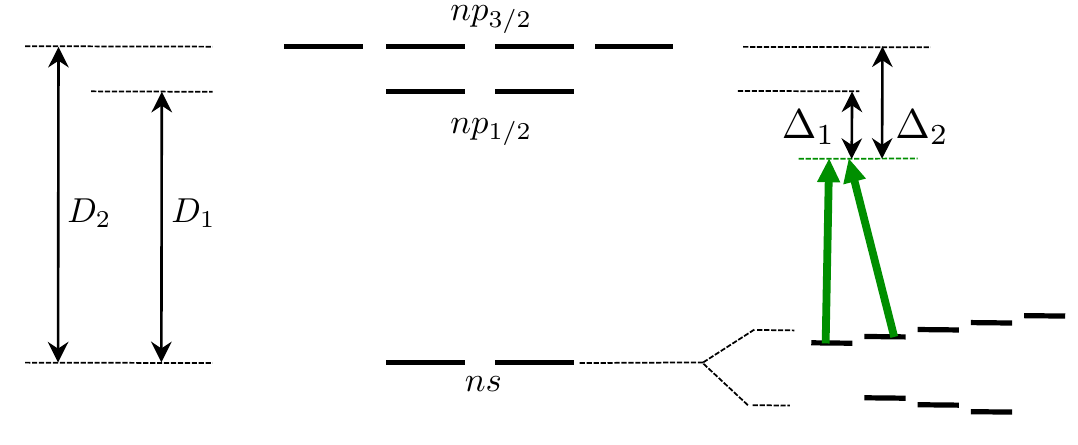}
\end{center}
\caption{Left: Resonance transition $ns\leftrightarrow np$ of an alkali-metal atom. The fine structure of the excited state leads to a splitting of the resonance line into two components $D_1$ and $D_2$. Right: hyperfine structure of the ground state represented for a nuclear spin  $I=3/2$ (case of $^{7}$Li, $^{23}$Na, $^{39}$K,$^{41}$K,$^{87}$Rb). The splitting between the various Zeeman sublevels is produced by an external magnetic field.}
\label{fig:alcalin}
\end{figure}

\paragraph{Alkali-metal species}
Consider an alkali-metal atom with a electronic ground state  $|g\rangle\equiv |ns\rangle$, irradiated by a monochromatic light beam with a frequency close to the resonance transition 
$ns\leftrightarrow np$. Because of fine-structure coupling, the first excited electronic level is split into two sub-levels $np_{1/2}$ and $np_{3/2}$, giving rise to two resonance lines  
 $D_1$ and $D_2$. In addition, because of the non-zero value of the spin $I$ of the nucleus, the ground state $ns$ and  the excited states 
 $np_{1/2}$ and $np_{3/2}$ are also split by hyperfine interaction. In particular the ground state is split into two sublevels with angular momentum $F=I\pm 1/2$ (fig. \ref{fig:alcalin}). The degeneracy between the various Zeeman states can then be lifted by an external magnetic field and resonant Raman transitions may occur between adjacent Zeeman states  when one irradiates the atom with light beams with well chosen frequencies and polarizations (figure \ref{fig:alcalin}, right):
$|g,F,m_F\rangle \leftrightarrow |g,F,m_F \pm 1\rangle$. Here we will assume for simplicity that the relevant detunings are large compared to the hyperfine splitting, and treat the atomic transition as if the nuclear magnetic moment were zero. 

Taking into account the Clebsch--Gordan coefficients for  $J_g=1/2 \leftrightarrow J_e=1/2$ and $J_g=1/2 \leftrightarrow J_e=3/2$ transitions, the corresponding matrix element is (up to a multiplicative factor of order unity that depends on the choice of polarization and hyperfine structure)
\begin{equation}
\kappa_{\rm eff} \sim \frac{\kappa^2}{6}\left(\frac{1}{\Delta_2}-\frac{1}{\Delta_1}\right).
\label{eq:Raman_alcalin}
\end{equation}
Here $\Delta _1=\omega-\omega_1$ and $\Delta_2=\omega-\omega_2$ are the detunings of the laser with respect to the resonance transitions  $D_1$ et $D_2$. The contributions of these two transitions appear with opposite signs in  (\ref{eq:Raman_alcalin}), because of a `destructive' interference between the two paths going from  $|g,F,m_F\rangle$ to $|g,F,m_F \pm 1\rangle$, and passing via $np_{1/2}$ or $np_{3/2}$. This interference is linked to the electric-dipole character of the atom-light coupling, $q\bs r\cdot \bs{\cal E}$, where 
$\bs r$ is the position of the outer electron and  $\bs{\cal E}$ the electric field of the light. This coupling acts on the orbital degrees of freedom of the electron, but not on its spin. The fact that the electron spin can nevertheless be changed in a Raman transition originates from the spin-orbit coupling, which is responsible for the lift of degeneracy between $np_{1/2}$ and $np_{3/2}$. Now, when the detuning of the light is large compared to the fine structure splitting  ($\Delta_e=\Delta_1 \approx \Delta_2$), the effect of spin-orbit coupling becomes negligible and the Raman coupling amplitude rapidly vanishes  (like $1/\Delta_e^2$),  faster than what one could have naively expected from (\ref{eq:Raman_coupling}).  

The spontaneous emission rate is (again up to a multiplicative factor of order unity):
\begin{equation}
\gamma \sim \frac{\Gamma \kappa^2}{4} \left(\frac{1}{3\Delta_1^2}+ \frac{2}{3\Delta_2^2} \right),
%\abel{}
\end{equation}
where $\Gamma^{-1}$ stands for the radiative lifetime of the excited states $np_{1/2}$ and $np_{3/2}$. 
The Raman matrix element and the spontaneous emission rate are thus both proportional to the light intensity  ($\kappa^2$), so that the merit factor (\ref{eq:merit_factor}) does not depend on this intensity but only on detuning. The optimal merit factor is obtained by taking  $\Delta_1 \approx- \Delta_2\approx \Delta_{\rm f.s.}/2$, where $\Delta_{\rm s.f.}=\Delta_1-\Delta_2$ is the fine structure splitting between $np_{1/2}$ and $np_{3/2}$:
\begin{equation}
{\cal M} =\frac{2}{3} \frac{\Delta_{\rm f.s.}}{\Gamma}.
\label{eq:merit_alcali}
\end{equation} 
 
For rubidium atoms, the merit factor is  ${\cal M}=8\times10^5$. Choosing the effective Rabi frequency $\kappa_{\rm eff}$ equal to the recoil frequency $\omega_{\rm r}=\Er/\hbar=2\pi\times 3.8\,$kHz, we infer the photon scattering rate $\gamma=0.03\,$s$^{-1}$. In average, the random recoil associated with the spontaneous emission of a photon increases the atomic kinetic energy by $\Er$, so that the rate of increase of the energy is $\dot E=\gamma \Er = \kB \times 5\,$nK/s. Consider now the case of $^{40}$K as an example of fermionic species. The fine structure is significantly reduced with respect to rubidium, leading to a lower merit factor, ${\cal M}=2\times10^5$. Choosing again $\kappa_{\rm eff}=\omega_{\rm r}$, the scaling given above leads to a scattering rate that is 10 times larger than for Rb atoms, and the heating rate  reaches $\dot E =   \kB \times120\,$nK/s. This may be too large for a  reliable production of strongly correlated states based on Raman coupling. Note that for a precise comparison between the heating rates for Rb and K, one should also specify the polarisations of the Raman beams, and take into account the difference between the Clebsch--Gordan coefficients originating from the difference in the nuclear angular momenta \cite{Nascimbene:2013c}.

\paragraph{Lanthanides: erbium, dysprosium}

Consider now atomic species like erbium or dysprosium, which have a more complex electronic structure than alkali-metal species. These atoms have two outer electrons  
 and an incomplete inner shell (6s$^2$ and 4f$^{10}$ for Dy). Because of this inner shell,  
 the electronic ground state has a non-zero orbital angular momentum ($L=6$ for Dy). The lower part of the atomic spectrum contains lines corresponding either to the excitation of one of the outer electrons, or of one electron of the inner shell. 

\begin{figure}[t]
\begin{center}
\includegraphics{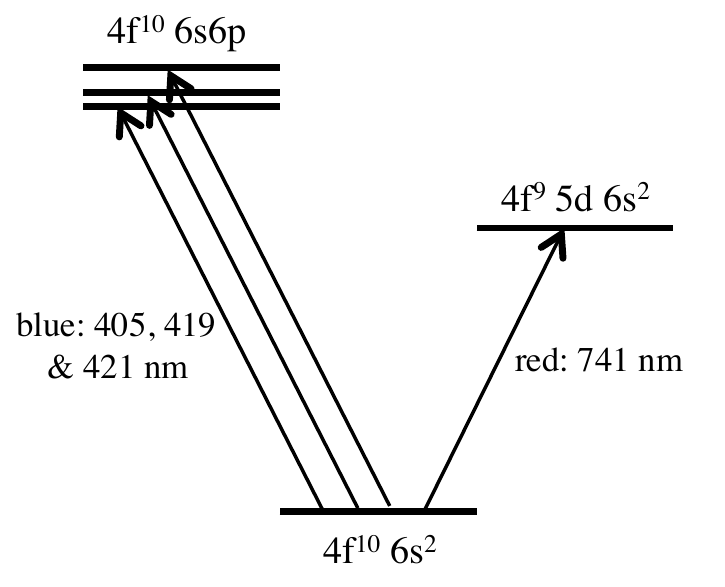}
\end{center}
\caption{A few relevant levels of Dysprosium. The blue resonance line is split  into three components by fine structure coupling. In order to minimize heating due to the random recoils associated with the spontaneous emission of photons, a Raman coupling between the Zeeman sublevels of the electronic ground state can be generated by a laser close to resonance with the narrow red transition at  741 nm.}
\label{fig:Dy}
\end{figure}

We show in fig. \ref{fig:Dy}  a few relevant levels for dysprosium (for erbium, see \cite{Lepers:2014}). The (broad) blue line corresponds to the resonance transition 6s$^2 \to\ $6s\,6p with the natural width $\Gamma_b/2\pi=30$\,MHz; it is split in three components by fine structure coupling. There are also several narrow transitions such as the red line at 741\,nm shown in  fig. \ref{fig:Dy}, corresponding to the excitation of an electron of the inner shell  (4f$^{10} \to\ $4f$^9$5d, term $^5$K$^\circ$, width  $\Gamma_r/2\pi \sim 2$\,kHz) \cite{Cui:2013}. We irradiate the atom with a monochromatic laser in order to induce a Raman transition between ground state sublevels. This laser is characterized by its detunings $\Delta_{b,r}$ from the blue and red transitions, and the corresponding Rabi frequencies $\kappa_{b,r}$. As we show below, one has to choose a small red detuning $|\Delta_r|\ll |\Delta_b|$ in order to maximize the merit factor. 

For the relevant choice of parameters, the dominant term in the Raman coupling comes from the red transition, $\kappa_{\rm eff} \sim \frac{\kappa_r^2}{2\Delta_r}$, and the spontaneous emission rate is obtained by summing the contributions of the blue and red lines\footnote{In order to simplify the algebraic expressions, we give here results obtained within the rotating wave approximation, which is only marginally correct for this large detuning with respect to the blue resonance line. }:
\begin{equation}
\gamma \sim \Gamma_r \frac{\kappa_r^2}{4\Delta_r^2} + \Gamma_b \frac{\kappa_b^2}{4\Delta_b^2}.
%\abel{}
\end{equation}
The Rabi frequencies $\kappa_{b,r}$ associated with the red and blue transitions are proportional to the reduced dipoles $d_{b,r}$ of these transitions, which are themselves proportional to $\sqrt{\Gamma_{b,r}}$, hence:  
\begin{equation}
\frac{\kappa_r^2}{\kappa_b^2}\approx \frac{\Gamma_r}{\Gamma_b}.
%\abel{}
\end{equation}
Using the fact that the ratio $\Gamma_r/\Gamma_b$ is small ($\sim 10^{-4}$), an approximate
calculation leads to the maximal merit factor
\begin{equation}
{\cal M}\approx \frac{\Delta_b}{\Gamma_b},
\label{eq:merit_Dy}
\end{equation}
which is obtained when the detunings are such that 
\begin{equation}
\frac{|\Delta_r|}{ |\Delta_b|} = \frac{\Gamma_r}{\Gamma_b} \ll 1.
%\abel{}
\end{equation}
 The optimal detuning of the Raman coupling laser is only a fraction of angström from the red line, while $\Delta_b$ is of the order of an optical frequency. This leads to a situation that is much more favorable than for alkali-metal species [compare eqs. (\ref{eq:merit_alcali}) and (\ref{eq:merit_Dy})], with a merit factor ${\cal M}\sim 10^7$ and a scattering rate $\gamma \sim 10^{-3}$\,s$^{-1}$ for a Raman coupling equal to the recoil energy. The residual heating is then $\dot E \sim \kB \times 0.1 $\,nK/s for both bosonic and fermionic isotopes.
 
%%%%%%%%%%%%%%%%%%%%%%%%%%%%%%

\section{Non-Abelian potentials and spin-orbit coupling}
\label{sec:non_abelian}

We now generalize the problem of simulating external magnetic fields and ask if it is possible to take advantage of the internal structure of an atom to generate a single-particle Hamiltonian
like
\begin{equation}
\hat H=\frac{\left(
\hp -\bs {\hat {\cal A}}(\hr)
\right)^2}{2M}+\ldots
\label{eq:prototype_non_abelien}
\end{equation}
where $\bs {\hat {\cal A}}(\bs r)$ is not anymore just a vector field, but an ensemble of three matrices $\bs {\hat {\cal A}}=(\hat {\cal A}_x,\hat {\cal A}_y,\hat {\cal A}_z)$ acting in the internal Hilbert space of the atom:
\begin{equation}
\hat {\cal A}_x(\bs r)=[{\cal A}_{x}^{(m,n)}(\bs r)],\quad \hat {\cal A}_y(\bs r)=[{\cal A}_{y}^{(m,n)}(\bs r)], \quad \hat {\cal A}_z(\bs r)=[{\cal A}_{z}^{(m,n)}(\bs r)],
%\abel{}
\end{equation}
where the indices $m,n$ run over a basis of this Hilbert space.

The non-Abelian character is linked to the fact that in a given point $\bs r$, two components generally do not commute:
\begin{equation}
[\hat {\cal A}_x(\bs r),\hat {\cal A}_y(\bs r)]\neq 0.
%\abel{}
\end{equation}
It is clear that this non-Abelian character can enrich the physics of gauge fields. Consider for example a Aharonov--Bohm experiment; in the Abelian case, the presence of a magnetic field entails that the state of the particle acquires a geometric phase when the particle travels around a closed contour ${\cal C}$. In the non-Abelian case, this phase is replaced by a matrix acting in the internal Hilbert space: even if the atom moves arbitrarily slowly, it may not end up in the same internal state after traveling around the contour\footnote{The precise definition of a non-Abelian gauge field considers two contours ${\cal C}_1$ and ${\cal C}_2$ and states that the final internal state of the particle should not be the same when the particle travels over ${\cal C}_1$ and then ${\cal C}_2$, or the reverse \cite{Goldman:2013b}.}. 

\subsection{Non-Abelian potentials in quantum optics}

The idea to generate non-Abelian potentials in a geometric manner was proposed by Wilczek and Zee \cite{Wilczek:1984}. In this work the authors considered a generalization of the adiabatic theorem in the case where the Hamiltonian possesses a group of eigenstates that remain degenerate (or quasi-degenerate) in the course of the evolution, this group being well-separated from the other eigenstates. 
This analysis triggered many studies ranging from molecular physics  to condensed matter physics \cite{Bohm:2003,Xiao:2010}. Here we will focus on atomic physics implementations, initially proposed in 
\cite{Unanyan98OC,Unanyan99PRA,Osterloh:2005,Ruseckas:2005} (for reviews, see \cite{Dalibard:2011,Goldman:2013b}).

In the quantum optics context, we suppose that the internal atomic Hilbert space is a subspace ${\cal E}_q$ of dimension $q$ that is well separated from the rest of the spectrum. We denote $\{ |\psi_{1}(\bs{r})\rangle, \ldots,|\psi_{q}(\bs{r})\rangle \}$ an orthonormal basis of ${\cal E}_q$, composed of eigenvectors of the internal atomic Hamiltonian. The adiabatic approximation assumes that the state of the atom can be written at any time as
\begin{equation}
\Psi(\bs r) \approx \sum_{n=1}^q \phi_{n}(\bs{r})\left|\psi_{n}(\bs{r})\right\rangle.
%\abel{}
\end{equation}
We project the full Schrödinger equation on the subspace ${\cal E}_q$ and obtain $q$ coupled equations:
\begin{equation}
\I \hbar\frac{\partial {\Phi (\bs r,t)}}{\partial t}=\left[\frac{(\hat {\bs p}-\hat {\bs {\cal A}}(\hat{\bs r}))^{2}}{2M}+\hat E(\hat {\bs r})+\hat {\cal V}(\hat {\bs r}) \right]{\Phi}(\bs r,t)\,, 
\qquad 
\Phi (\bs r,t)=\begin{pmatrix}
\phi_1(\bs r,t) \\ \vdots \\ \phi_q(\bs r,t) \end{pmatrix}.
%\abel{}
\end{equation}
Here $\hat E$ stands for the $q\times q$ diagonal matrix formed with the energies $E_n(\bs r)$ and $ \hat {\cal V}(\hat {\bs r})$ is a $q\times q$ matrix representing a geometric scalar potential, generalizing eq.\;(\ref{eq:potentiel_scalaire}). The most important term for our discussion is Berry's connection $\hat {\bs {\cal A}}(\hat{\bs r})$, which is also a $q\times q$ matrix with components:
\begin{equation}
\hat {\bs {\cal A}}^{(n,m)} =  \I \hbar\;\langle\psi_{n}(\bs{r})|\bs \nabla\psi_{m}(\bs{r})\rangle,\label{eq:A-nm}
\end{equation}

\begin{figure}[t]
\begin{center}
\includegraphics[width=13cm]{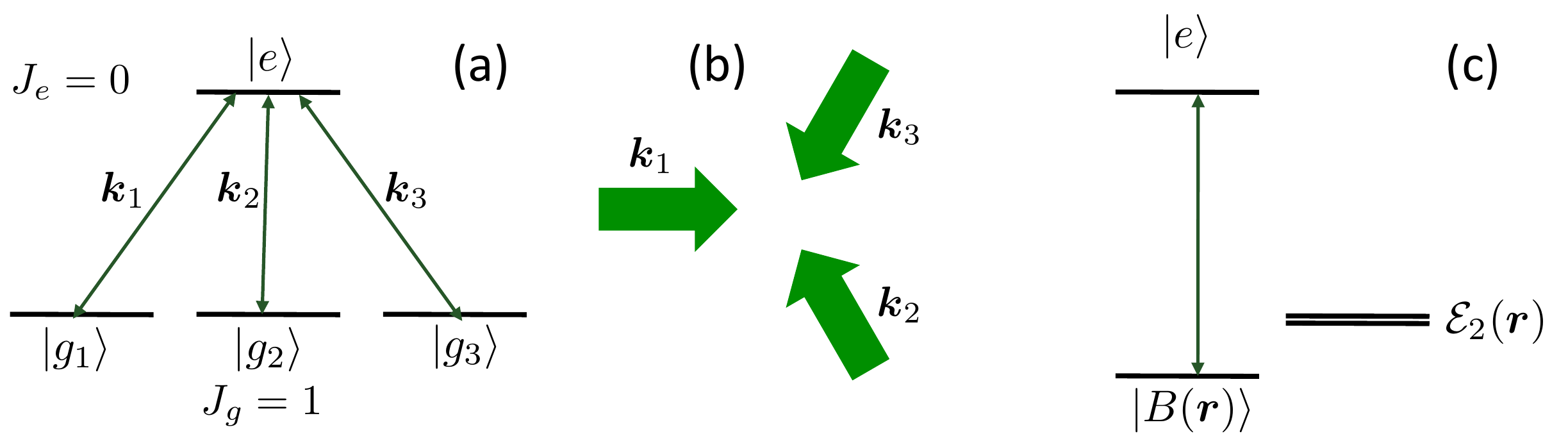}
\end{center}
\caption{(a) Tripod configuration leading to the emergence of a non-Abelian gauge potential. An atom has its electronic ground state with angular momentum $J_g=1$  coupled to an electronic excited state with angular momentum $J_e=0$. (b) Possible laser configuration with three beams (plane waves) at 120 degrees, with orthogonal linear polarizations. Each beam drives one of the three transitions $g_j \leftrightarrow e$. (c) Dressed state picture: one linear combination of the $g_j$ sublevels (bright state) is coupled to $e$. The orthogonal subspace of dimension 2  forms a ``dark" subspace, in which the dynamics is governed by a non-Abelian gauge field.    }
\label{fig:tripod}
\end{figure}

\subsection{Tripod configuration and 2D spin-orbit coupling}

As a concrete example consider the tripod geometry represented in fig.\;\ref{fig:tripod}, where a ground state with angular momentum $J_g=1$ is coupled to an excited state with angular momentum $J_e=0$. The coupling is provided by three laser beams with wave vectors $\bs k_j$ ($j=1,2,3$), with the same frequency and with orthogonal linear polarizations, propagating in the $xy$ plane at 120 degrees from each other. By construction only one linear combination of the ground state manifold is coupled to the excited state. This \emph{bright state} reads:
\begin{equation}
|B(\bs r)\rangle= \frac{1}{\sqrt 3} \left( \E^{-\I \bs k_1\cdot \bs r}|g_1\rangle \;+\;  
\E^{-\I \bs k_2\cdot \bs r}|g_2\rangle \;+\;  
\E^{-\I \bs k_3\cdot \bs r}|g_3\rangle
\right),
%\abel{}
\end{equation}
where $|g_j\rangle$ is an orthonormal basis of the $J_g=1$ manifold, chosen such that the laser beam $j$ drives the transition $g_j \leftrightarrow e$. The energy of the bright state 
is shifted from its initial position by the coupling with the light.
The orthogonal \emph{dark subspace} of the ground state manifold forms the desired space ${\cal E}_q$, with here $q=2$. After some algebra one reaches the expression of the non-Abelian Berry's connection associated to this dark subspace \cite{Juzeliunas:2010}:
\begin{equation}
\bs {\hat {\cal A}}=\frac{\hbar k}{2} \left(  \hat \sigma_x \unitx +\hat \sigma_y \unity \right) =\frac{\hbar k}{2}\;\bs {\hat \sigma}.
%\abel{}
\end{equation} 
where the $\hat \sigma_j$ ($j=x,y,z$) stand for the Pauli matrices and where $\bs u_x,\bs u_y$ are orthonormal vectors in the $xy$ plane containing the three wave vectors $\bs k_j$. We then obtain the following Hamiltonian for an atom moving in the $xy$ plane when its internal state lies in this dark subspace:
 \begin{equation}
\hat H=\frac{\left(
\hp -\bs {\hat {\cal A}}(\hr)
\right)^2}{2M}= \frac{\hat {\bs p^2}}{2M} -\frac{\hbar k}{2M}\left( \hat p_x \hat \sigma_x +\hat p_y \hat \sigma_y \right)+\ldots.
\label{eq:2D_SOC}
\end{equation}

The Hamiltonian (\ref{eq:2D_SOC}) constitutes a prototype of spin-orbit coupling (SOC). The orbital motion of the atom, here its 2D linear momentum $\hat p_{x,y}$, is coupled to its (pseudo) spin degree of freedom described by $\hat \sigma_{x,y}$. We note that this SOC originates from the recoil/Doppler effect; this is very different from the standard SOC in atomic or condensed matter physics, which has a relativistic origin. In standard SOC a charged particle (electron) moves with velocity $\bs v$ in a region with an external electric field $\bs{\cal E}$. In the reference frame of the particle, a motional magnetic field $\bs {\cal B}=\bs v \times \bs {\cal E}$ appears; SOC then results from the interaction between this motional magnetic field and the intrinsic magnetic moment of the particle, proportional to its spin, $\bs \mu=\gamma \bs S$. In atomic physics, the relevant electric field is the Coulomb field generated by the nucleus and it has a radial structure $\bs {\cal E}\; ||\; \bs r$. This leads to a SOC term $\propto (\bs r \times \bs p)\cdot \bs S  = \bs L\cdot \bs S$. In solid state physics, the electric field can be considered as uniform and leads to the so-called Rashba or Dresselhaus Hamiltonians, depending on the origin of $\bs {\cal E}$ (for a review, see \cite{Galitski:2013}). The coupling generated in eq.\;(\ref{eq:2D_SOC}) simulates such a 2D Rashba--Dresselhaus Hamiltonian.

In solid state physics, SOC is at the origin of many applications, in particular in relation with spintronics  \cite{Galitski:2013}. It also plays an important role in fundamental physics since it is at the origin of the notion of topological insulators, which are insulating materials in their bulk but can conduct electricity at their surface \cite{Hasan:2010,Qi:2011}. This surface electricity conduction is topologically robust, as for the Quantum Hall effect. However in contrast with the Quantum Hall effect, there is no external magnetic field that would break time-reversal invariance.  In a Fermi gas with point-like interactions, it can lead to the formation of novel superfluids \cite{Zhang:2008,Nayak:2008b} and possibly the creation of Majorana quasiparticles \cite{Ivanov:2001,Sau:2010,Alicea:2010}.
Finally we note that even at the single-particle level, the SOC Hamiltonian given in eq.\;(\ref{eq:2D_SOC}) can have spectacular consequences, since it changes the nature of the spectrum. In particular the ground state, which is non-degenerate ($\bs p=0$) in the absence of atom-laser interaction, becomes massively degenerate since all states with momentum $\bs p$ with $|\bs p|=\hbar k/2$ can be associated to a spin state such that they have a zero-energy. This massive degeneracy of the single-particle ground state changes the nature of Bose--Einstein condensation for an ideal gas \cite{Zhou:2013}.
 
\subsection{1D version of spin-orbit coupling}

The multi-dimensional implementation of spin-orbit coupling has not yet (to our knowledge) been implemented experimentally. However a 1D version of it, initially identified in \cite{Higbie:2002}, has been realized by several groups, first  with bosons \cite{Lin:2011} and then with fermions \cite{Cheuk:2012,Wang:2012} . 

To describe this implementation, we consider the $\Lambda$ scheme of fig.\;\ref{fig:two_level}b and we suppose that the transitions $g_j \leftrightarrow e$ are driven by two plane waves with wave vectors $\pm \bs k$ along a given direction, say $x$. We suppose that the laser excitation is far detuned from the ground to excited state resonance, and that we can eliminate the excited state in a perturbative manner. We then deal only with the internal subspace associated to the ground states $g_1$ and $g_2$. We introduce the family of states labeled by the momentum $\bs p$:
 \begin{equation}
{\cal F}(\bs p)=\{ |g_1,\bs p -\hbar \bs k\rangle, \ |g_2,\bs p+\hbar \bs k\rangle \},
\label{eq:family}
\end{equation}
which is globally invariant by the action of the atom-laser coupling. This coupling is described in this family by a $2\times 2$ matrix:
\begin{equation}
\hat H(\bs p)=\bpm
(\bs p-\hbar \bs k)^2/2M +\hbar \Delta/2 & \hbar\kappa_0/2 \\
\hbar \kappa_0/2 & (\bs p+\hbar \bs k)^2/2M -\hbar \Delta/2
\epm,
\label{eq:hamiltonien_exact}
\end{equation}
which can also be written in terms of the Pauli matrices:
 \begin{equation}
\hat H(\bs p)=\frac{1}{2M} \left( {\bs p}-\bs{\hat {\cal A}} \right)^2 + \frac{\hbar \Delta }{2} \hat \sigma_z + \frac{\hbar \kappa_0}{2}\hat \sigma_x\qquad \mbox{with}\quad \bs{\hat {\cal A}}=\hbar \bs k\, \hat \sigma_z.
%\abel{}
\end{equation}
This corresponds to the desired form, since the ``vector potential" is an operator with respect to the internal degrees of freedom. 

\begin{figure}[t]
\begin{center}
\includegraphics{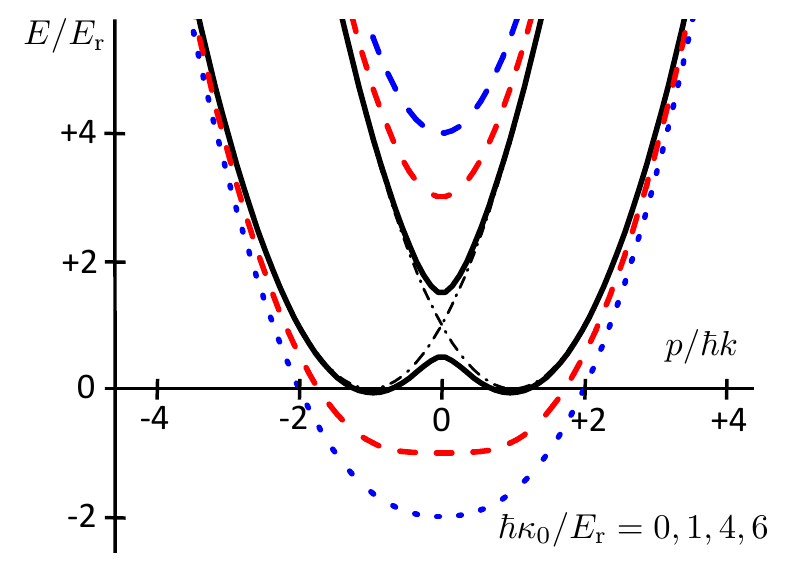}
\end{center}
\caption{Energy levels of the 1D SOC Hamiltonian given in eq.\;(\ref{eq:hamiltonien_exact}), labeled by the momentum index $p$ of the family (\ref{eq:family}). The detuning $\Delta$ is chosen equal to zero. The Rabi frequency is given by $\hbar \kappa_0/\Er=0$ (dash-dotted), 1 (continuous), 4 (dashed) and 6 (dotted).
}
\label{fig:soc_calcul}
\end{figure}

It is clear that the three components of $\bs{\cal A}$ commute with each other in this particular case, so that it does not correspond to a non-Abelian coupling. However in spite of this restriction, some characteristics of SOC remain relevant. For example, for $\Delta=0$ and a relatively small  amplitude of the Raman coupling, one still obtains a non-unique single-particle ground sate. We plotted in Figure \ref{fig:soc_calcul} the energy levels obtained from the diagonalization of the Hamiltonian (\ref{eq:hamiltonien_exact}):
\begin{equation}
E_\pm(p)=\frac{p^2}{2M}+E_{\rm r} \pm\frac{\hbar}{2} \left[\kappa_0^2+4\left( \frac{kp}{M}  \right)^2\right]^{1/2} .
%\abel{}
\end{equation}
The lowest level, $E_-(p)$, has two symmetric minima in $p$ and $-p$ if the Rabi frequency $\kappa_0$  is such that
\begin{equation}
\hbar \kappa_0 < 4\Er.
%\abel{}
\end{equation}
Above this value, the minimum of $E_-(p)$ is located in $p=0$. In the limit $\kappa_0 \gg \Er$, we recover the situation where the adiabatic approximation is valid [cf. eq. (\ref{eq:valid_adiab})]. The interesting case for SOC, \emph{i.e.}, a degenerate ground state, thus corresponds to the regime opposite to the one studied in subsect.~\ref{subsec:adiab_following}. 

At the single-particle level, the achievement and the description of 1D SOC is thus quite simple. It is essentially a reinterpretation of the usual dressed atom diagram, in which the coupling   between internal and external degrees of freedom is due to the Doppler shift in  a plane running wave. At the many-body level, SOC is at the origin of several interesting phenomena even this conceptually simple 1D configuration (for a review, see \cite{Li:2014}). Several questions are still open concerning the phases that may appear for bosons (see for example \cite{Li:2012}) as well as for fermions, with the possibility to generate Majorana particles at the edges of a chain of atoms in a topological superconductor \cite{Kitaev:2000,Wilczek:2009,Oreg:2010,Lutchyn:2010,Sau:2010b,Gangadharaiah:2011,Nascimbene:2013,Kraus:2013,Ruhman:2014}.

%%%%%%%%%%%%%%%%%%%%%%%%%%%%%%%%%%%%%%%%%%%%%%%%%%%

\section{Gauge fields on a lattice}
\label{sec:lattice}

In the two previous sections we considered the generation of a gauge field for an atom moving freely in space. Another important class of problems deals with the magnetic phenomena that can appear in the presence of a spatially periodic potential. This question emerges in particular when one studies the effect of a magnetic field on the electron fluid of a crystal. 

The richness of the problem is linked to the existence of two length scales, which can ``compete" which each other and lead to \emph{frustration phenomena}. The first length scale is the period $a$ of the potential (in the following we consider a 2D square lattice). The second length scale is the magnetic length that we already introduced $\lm=\sqrt{{\hbar}\,/\,{qB}}$. The ratio of these two length scales can be written in terms of the flux $\Phi=Ba^2$ of the magnetic field through the unit cell of the lattice and of the flux quantum $\Phi_0=h/q$:
\begin{equation}
\frac{a^2}{\lm^2}=\frac{qBa^2}{\hbar}=2\pi \frac{\Phi}{\Phi_0}.
\label{eq:flux}
\end{equation}
The ratio $a^2/\lm^2$ is thus equal to the Aharonov--Bohm phase accumulated by the particle when it travels along the sides of the unit cell.  

As long as the flux $\Phi$ remains small compared to $\Phi_0$, one does not expect spectacular modifications compared to the free particle case: the lattice step is much smaller than the typical size of a cyclotron orbit, which is thus only marginally affected by the discretization of space. This regime $\Phi\ll \Phi_0$ corresponds to the case of ``ordinary solids" (lattice period of a few Angströms) and of magnetic fields that can be achieved in the laboratory (a few tens of Teslas). On the opposite, for large magnetic fields such that $\Phi\sim\Phi_0$, the competition between these two length scales is at the origin of new phenomena, such as a fractal structure for the single-particle spectrum. This regime can be explored with synthetic materials (see \emph{e.g.} \cite{Pannetier:1984,Albrecht:2001,Dean:2013}) or with artificial magnetic fields, in particular with cold atoms in an optical lattice. 

A key notion that will appear in the following is the topology of an energy band. It is characterized by an integer number, the Chern index, which indicates if a given filled band can contribute to a particle current along a given direction, say $y$, when a force is applied along the orthogonal direction $x$. This notion is directly inspired from the study of Hall conductivity \cite{Thouless:1982a}; we will introduce it here by considering the case of a square lattice placed in a uniform magnetic field, the so-called \emph{Hofstadter model} \cite{Hofstadter:1976}, that was recently implemented with cold atoms in optical lattices \cite{Aidelsburger:2013,Miyake:2013}. This topological characterization of bands via their Chern index can be generalized to other classes of lattice configurations, such as the honeycomb model introduced by Haldane \cite{Haldane:1988} that has also been realized recently with cold atomic gases \cite{Jotzu:2014}.

\subsection{Tight-binding model}
In most of this section, we will consider for simplicity a single-band, tight-binding model in a two-dimensional square lattice (Hubbard model). We denote by $|{j,l}\rangle$ the state of the particle localized on the site $\bs r=a(j\unitx +l \unity)$ ($j,l\in \ZZ$) and we assume that the particle moves in the lattice via tunneling from a given site $(j,l)$ to the four neighboring sites $j\pm 1,l$ and $j, l\pm 1$. Denoting the tunnel amplitude by $J$, the Hamiltonian in the absence of magnetic field  is:
\begin{equation}
\hat H=-J \sum_{j,l} \left(  
|{j+1,l}\rangle \langle {j,l}| + 
|{j,l+1}\rangle \langle {j,l}| 
\right)\ +\ \mbox{h.c.}
\label{eq:H_reseau_2D}
\end{equation}
The eigenstates of this Hamiltonian are the Bloch states $|\psi({\bs q})\rangle$ labelled by the Bloch vector
 $\bs q=(q_x,q_y)$ with energy $E(\bs q)$
\begin{equation}
|\psi({\bs q})\rangle=\sum_{j,l} \E^{\I a\left( jq_x +lq_y)\right)}|{j,l}\rangle,\qquad
E(\bs q)=-2J\left[ \cos (aq_x) + \cos(aq_y) \right].
\label{eq:spectre_Hubbard_2D}
\end{equation}
The energy spectrum is thus a band centered on $E=0$ with a full width $8J$. Since two Bloch vectors differing by a vector of the reciprocal lattice correspond to the same eigenstate, we can restrict the domain accessible to $\bs q$ to 
the first Brillouin zone $-\pi/a < q_j \leq \pi/a$, $j=x,y$, in which case all $|\psi({\bs q})\rangle$ are independent and form a basis of the Hilbert space.

We take into account the presence of the magnetic field by assigning a complex value to the tunneling matrix elements: 
\begin{equation}
-J\,|{j+1,l}\rangle \langle {j,l}| \quad \longrightarrow \quad -J\,\E^{\I \phi(j,l \to j+1,l)}\;|{j+1,l}\rangle ,\langle {j,l}| 
\label{eq:Peierls}
\end{equation}
and similarly for the links in the other direction. We also set $\phi(j+1,l \to j,l)=-\phi(j,l \to j+1,l)$ to ensure that the Hamiltonian remains Hermitian. This technique of using complex tunnel coefficients is well known in condensed-matter physics and is called the Peierls substitution. In the case of a real magnetic field $\bs B(\bs r)$, the Peierls substitution states that the phase $\phi(j,l \to j+1,l)$ is calculated (once a gauge choice has been made) using the vector potential:
 \begin{equation}
\phi(\bs r\to \bs r')=\frac{q}{\hbar} \int_{\bs r}^{\bs r'} \bs A\cdot \D \bs r .
\label{eq:phases_accumulees}
\end{equation} 
The justification of the Peierls justification is far from trivial and we refer the reader to \cite{Luttinger:1951,Nenciu:1991} for a discussion. However in the context of cold atoms, the possible difficulties associated to this justification are not relevant; indeed one looks for a direct implementation of the complex tunnel matrix elements of eq.\;(\ref{eq:Peierls}), without deriving them from a continuous model with external $\bs B(\bs r)$ and $\bs A(\bs r)$.

As for the continuous case, there is a gauge freedom for the spatially discretized problem. An infinite number of choices for the phases $\{ \phi(j,l \to j+1,l), \phi(j,l \to j,l+1)\}$ lead to the same physical situation, and one can switch from one set to the other by changing the phases of the $|{j,l}\rangle$'s. One can show quite generally that the gauge invariant quantities for the discrete problem are the sums:
\begin{eqnarray}
\frac{q}{\hbar}\Phi_{j,l}&=&\phi(j,l \to j+1,l) + \phi(j+1,l \to j+1,l+1) \nonumber \\
&+& \phi(j+1,l+1 \to j,l+1) + \phi(j,l+1 \to j,l).  
\label{def_Phi}
\end{eqnarray}
Each sum represents the accumulated phase when the particle travels counterclockwise around the cell having the site $(j,l)$ in its lower left corner. Using eq.\;(\ref{eq:phases_accumulees}), one sees that $\Phi_{j,l}$ is just the flux of the magnetic field $\bs B(\bs r)$ through the considered plaquette.

We will be interested in the following in the case of a uniform field on the lattice, \emph{i.e.}, the same magnetic flux $\Phi$ across each lattice cell. Using eq.\;(\ref{eq:phases_accumulees}) with the Landau gauge $\bs A=-By\,\unitx$, we find that this amounts to take
\begin{equation}
\phi(j,l \to j,l+1)=0,\qquad \phi(j,l \to j+ 1,l)=- 2\pi \alpha\, l.
%\abel{}
\end{equation}
One can immediately check that eq.\;(\ref{def_Phi}) yields in this case  the same flux $\Phi_{j,l}\equiv \Phi=\alpha\Phi_0 $ for all cells.
With this gauge choice, the target Hamiltonian reads 
\begin{equation}
\hat H=-J \sum_{j,l} \left(  \E^{-\I\,2\pi \alpha\,l}
|{j+1,l}\rangle \langle {j,l}| \ +\  
|{j,l+1}\rangle \langle {j,l}| 
\right)\ +\ \mbox{h.c.}
\label{eq:H_reseau_2D_B}
\end{equation}

\subsection{Hofstadter butterfly}
\label{subsec:Hofstadter}

Before looking at the possible implementations of the Hamiltonian (\ref{eq:H_reseau_2D_B}) with a cold atom setup, we briefly discuss the main properties of its spectrum and of its eigenstates. First we note that this Hamiltonian is invariant in the change $\alpha \to \alpha+1$. Therefore   it is enough\footnote{One can even restrict the domain of study to the interval $\alpha\in [0,1/2]$, since one switches between $\alpha$ and $-\alpha$ by changing the orientation of the $x$ axis.} for this spatially discretized problem to study the spectrum of the Hamiltonian for $\alpha$ between 0 and $1$ . 

For the gauge choice leading to the Hamiltonian  (\ref{eq:H_reseau_2D_B}), the presence of the magnetic field breaks the translational invariance along $y$, but preserves the invariance along $x$. Hence one can still look for the eigenstates of $\hat H$ as Bloch functions along $x$:
\begin{equation}
|\Psi\rangle=\sum_{j,l} C_l\; \E^{\I jaq_x }|{j,l}\rangle,
%\abel{}
\end{equation}
where the coefficients $C_l$ are unknown at this stage. Inserting $|\Psi\rangle$ in the eigenvalue equation for $\hat H$, we find the recursion equation called the \emph{Harper equation}
\begin{equation}
C_{l-1} +2 C_l \cos(2\pi \alpha\,l+aq_x) + C_{l+1}= -(E/J)\, C_l.
%\abel{}
\end{equation}
One is then left with the diagonalization of a tri-diagonal matrix.

An important simplification occurs when $\alpha$ is a rational number,  $\alpha=p'/p$, where $p,p'$ are coprime positive integers. Then one recovers a periodic problem also along $y$, but with an increased spatial period $pa$ instead of $a$. Indeed one gets in this case
\begin{eqnarray}
\phi(j,l+p \to j+1,l+p)&=& -2\pi \alpha (l+p)=-2\pi\alpha l - 2\pi p' \nonumber \\
&=& \phi(j,l \to j+1,l) \quad \mbox{modulo}\ 2\pi.
%\abel{}
\end{eqnarray}
We are then back to the problem of the motion of a particle on a periodic lattice, but with a unit cell of size $a \times (pa)$ and $p$  non-equivalent sites per unit cell. In this case the general result is that the initial energy band (of width $8J$) gets fragmented in $p$ subbands, generally separated by forbidden gaps\footnote{These gaps may be replaced by a contact with a Dirac point, like around $E=0$ for $\alpha$ even.}.

\begin{figure}[t]
\begin{center}
\includegraphics[height=5cm]{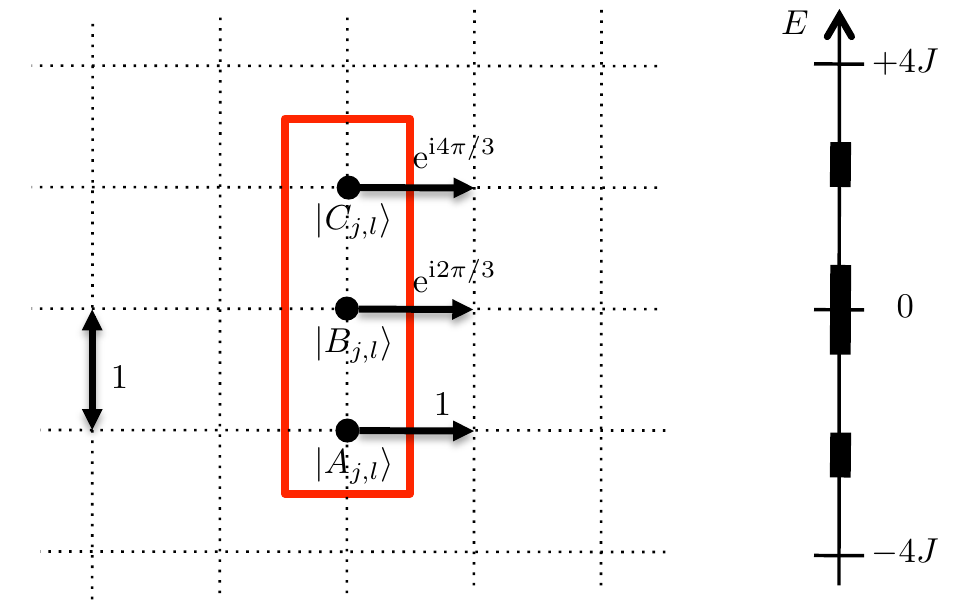}
\end{center}
\caption{Left: Choice of a unit cell of size $a\times 3a$ for a magnetic flux $\alpha=1/3$. Right: the three subbands resulting from the fragmentation of the band of width $8J$.}
\label{fig:cellule_3}
\end{figure}

\begin{figure}[t]
\begin{center}
\includegraphics[width=9cm]{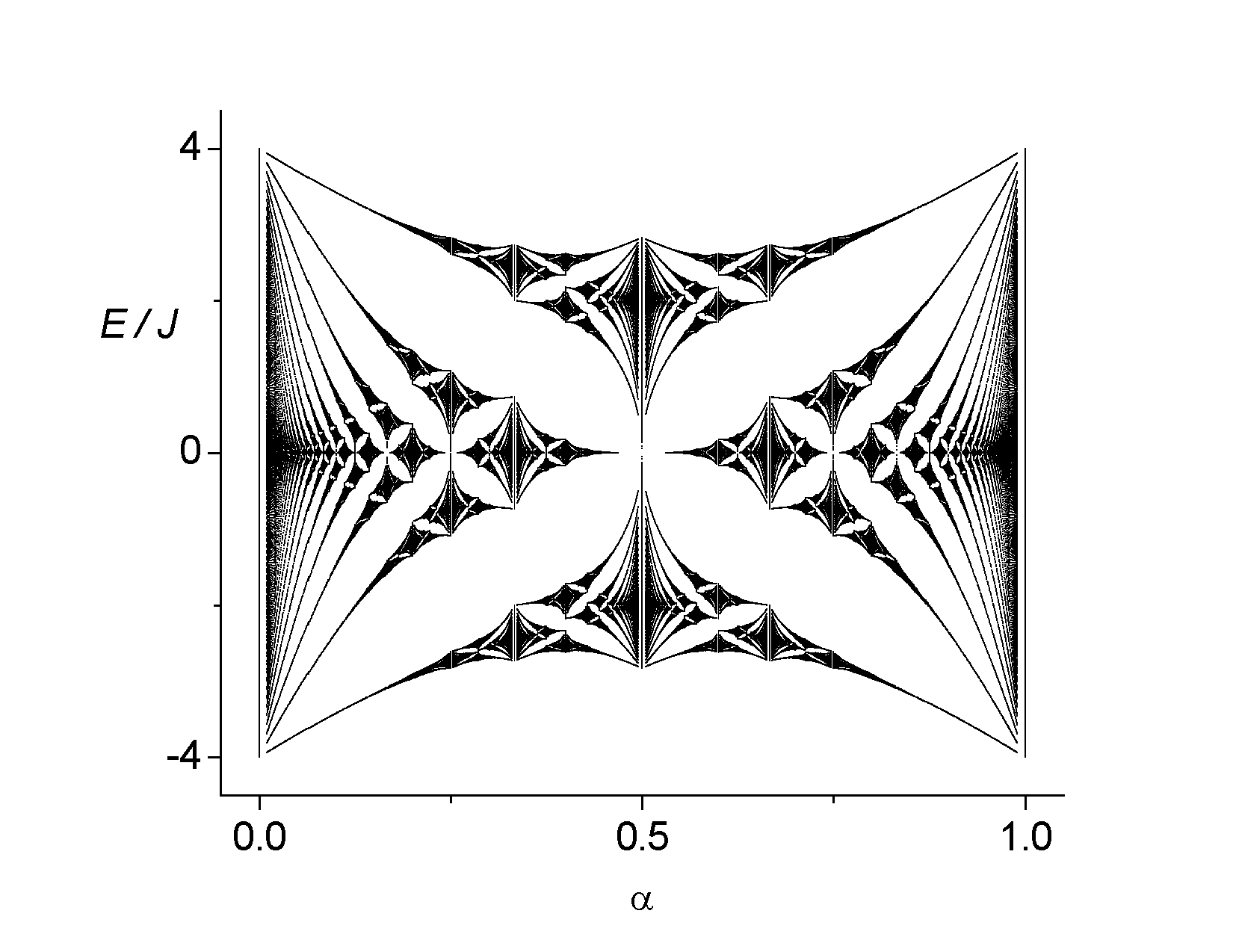}
\end{center}
\caption{Hofstadter butterfly: energy spectrum for a particle moving on a square lattice in the tight-binding approximation in the presence of a uniform magnetic field. The flux $\Phi$ of the magnetic field across a unit cell is such that  $\Phi=\alpha \,\Phi_0$, where $\Phi_0=h/q$ is the flux quantum \cite{Hofstadter:1976}. The calculation has been done for rational values of $\alpha$, $\alpha=p'/p$, with denominators $p$ up to 100. }
\label{fig:papillon}
\end{figure}

We choose the case $\alpha=1/3$ as a concrete example and refer the reader to Appendix 2 for a general discussion. The unit cell has a length $a$ along $x$ and $3a$ along $y$. It contains three sites denoted $|A\rangle, |B\rangle,|C\rangle$ in fig.\;\ref{fig:cellule_3}. Each cell is labelled by the indices $j,l'$, where we take by convention the $A$ site at point $a\left[j\unitx+ 3l'\unity\right]$. We look for the eigenstates of the Hamiltonian as Bloch functions labelled by the Bloch vector $\bs q=(q_x,q_y)$:
\begin{equation}
|\Psi({\bs q})\rangle=\sum_{j,l'}\E^{\I a(j q_x+3l' q_y)}\left(  
 \beta_1 |A_{j,l'}\rangle+\E^{ \I a q_y}\beta_2 |B_{j,l'}\rangle +\E^{2 \I a q_y}\beta_3 |C_{j,l'}\rangle
\right)
%\abel{}
\end{equation}
with $|A_{j,l'}\rangle=|j,3l'\rangle$, $|B_{j,l'}\rangle=|j,3l'+1\rangle$,
$|C_{j,l'}\rangle=|j,3l'+2\rangle$ and
\begin{equation}
q_x\in]-\pi/a,\pi/a],\qquad q_y\in ]-\pi/(3a),\pi/(3a)].
\label{eq:intervalle_q}
\end{equation}
The coefficients $\beta_r$ are obtained from the eigenvalue equation $\hat H |\Psi\rangle=E|\Psi\rangle$, which amounts to look for the eigenstates of the Hamiltonian in reciprocal space:
\begin{equation}
\hat {\cal H}(\bs q)=-J \bpm
2 \cos (aq_x ) & \E^{\I aq_y} & \E^{-\I aq_y} \\
\E^{-\I aq_y} & 2 \cos(aq_x+2\pi/3) & \E^{\I aq_y} \\
\E^{\I aq_y} & \E^{-\I aq_y} & 2\cos(aq_x+4\pi/3)
\epm
%\abel{}
\end{equation}
written here in the basis $\{|A\rangle, |B\rangle,|C\rangle\}$. The diagonalization of this $3\times 3$ matrix gives three eigenvalues, which are functions of $\bs q$. When $q_x$ and $q_y$ vary in the intervals (\ref{eq:intervalle_q}), one finds the three energy bands indicated in 
fig.\;\ref{fig:cellule_3}. The energy spectrum is symmetric with respect to $E=0$ and its total width is reduced with respect to the value $8J$ found for $\alpha=0$. The lowest subband has a width $\sim0.7\, J$ and it is separated from the middle band by a gap of $\sim1.3 \,J$.

The spectrum of $\hat H$ for arbitrary $\alpha$ is represented in Figure \ref{fig:papillon}. It has a very specific fractal structure called \emph{Hofstadter butterfly} \cite{Hofstadter:1976}. One can understand the origin of this structure by comparing two close values of the flux like $1/3$ and $10/31$, and noting that they lead to very different results, the first one with three subbands 
(fig.\;\ref{fig:cellule_3}), the second one with 31 subbands. 

\begin{figure}[t]
\begin{center}
\includegraphics[width=9cm]{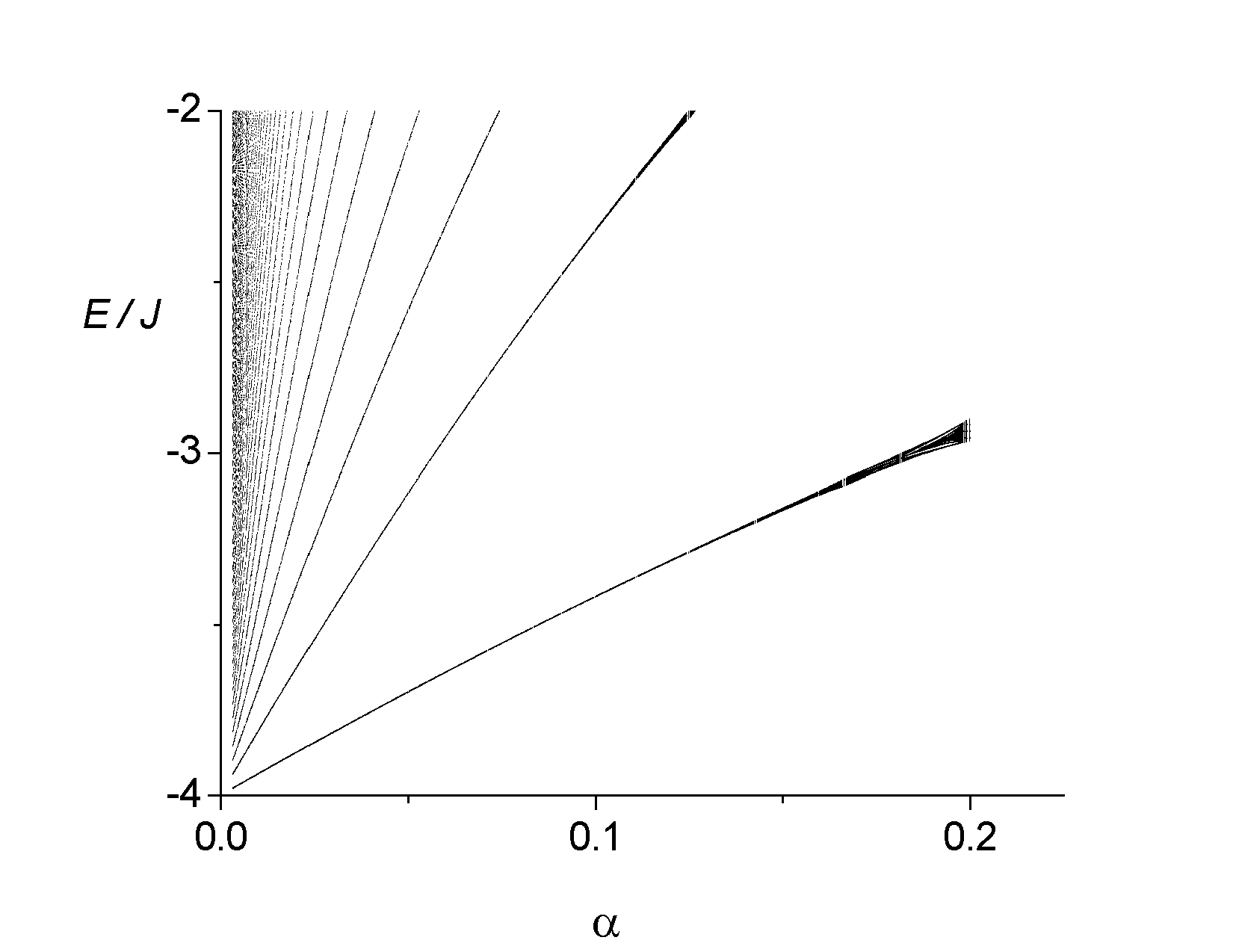}
\end{center}
\caption{The Landau levels recovered for the Hofstadter butterfly (zoom on the lower left corner of fig.\;\ref{fig:papillon}). The low energy levels form a quasi-discrete spectrum with $E_n/J\approx -4+ 4\pi (n+1/2)\alpha$, $n$ integer, corresponding to the Landau spectrum in $(n+1/2) \hbar \omega_{\rm c}$, where the cyclotron frequency $\omega_{\rm c}$ is determined using the effective mass $M_{\rm eff}=\hbar^2/(2Ja^2)$.}
\label{fig:zoom}
\end{figure}

It is interesting to connect the result obtained in the lattice case for low magnetic fluxes ($\alpha \ll 1$) to the Landau level structure of the continuum case. A zoom on the lower left corner of the Hofstadter butterfly is shown in fig.\;\ref{fig:zoom}. One sees that each bandwidth gets very small compared to the band spacing, so that we recover in the limit $\alpha \to 0$ the notion of well defined energy levels. The levels are approximately equidistant with an energy that can be written $E_n\approx -4J+(n+1/2) \hbar \omega_{\rm c}$, with $\omega_{\rm c}=qB/M_{\rm eff}$. Here the effective mass $M_{\rm eff}$ is obtained from the dispersion relation at the bottom of the band [see eq.\;(\ref{eq:spectre_Hubbard_2D})]: $M_{\rm eff}=\hbar^2/(2Ja^2)$.

\subsection{Chern number for an energy band}

The subbands that result from the fragmentation of the initial  band in the presence of a magnetic field  have specific, non-trivial topological properties. These properties are at the origin of a quantized Hall conductance; if one applies a constant force $F$ along one axis of the lattice, say $x$, then a current of particles flows in the orthogonal direction. Suppose for example that the lowest subband is filled with fermionic particles, while the others subbands are empty. In the absence of topological properties, this would correspond to a band insulator and no current would be expected.  However due to the topology of the lattice, a quantized non-zero current along $y$ appears, which is characterized by the so-called \emph{Chern index} ${\cal C}$, an integer number associated to the considered subband. The Chern index is defined such that ${\cal C}$ particles cross a given $x$-oriented link of the lattice during the time interval $t_{\rm B}=h/aF$. This time $t_{\rm B}$ is the period of the Bloch oscillation phenomenon that occurs in the presence of the force $F$ for the lattice of spatial period $a$. The proof of this remarkably simple result, as well as relevant references, is given in Appendix 2. The important relation between the topological bulk property described by the Chern index and the edge currents that appear in a finite-size sample is addressed in \cite{Hatsugai:1993,Qi:2006}.

A Chern index can also be associated to the Landau levels of a free particle in the presence of a uniform field (subsect.~\ref{subsec:cycloton_LL}). In particular the Chern index of the lowest Landau level is ${\cal C}=1$. Consequently the lowest subband in a lattice configuration is said to be topologically equivalent to the LLL if it also has ${\cal C}=1$. This is the case of the lowest subband of the Hofstadter butterfly when one chooses  $\alpha=1/p$ with $p$ integer.  

As explained in Appendix 2, the Chern number is equal to the integral over the magnetic Brillouin zone of the Berry curvature ${\cal B}(\bs q)$ associated to the variations of the Bloch states $|\Psi_{\bs q}\rangle$ with the Bloch vector $\bs q$. The Berry curvature can itself be determined from the dynamics of an atomic wave packet in the lattice, as proposed in \cite{Price:2012,Dauphin:2013,Cominotti:2013}. This technique was recently implemented by the Munich group (see the contribution of Immanuel Bloch to this volume and \cite{Aidelsburger:2014}) using a topological band produced by lattice modulation, which we describe in the next section. 

%%%%%%%%%%%%%%%%%%%%%%%%%%%%%%%%%%%%%%%%%%%%%%%%%

\section{Generation of lattice gauge fields via shaking or modulation}
\label{sec:lattice_shaking}

We now turn to the description of some possible ways to generate non-real tunneling matrix elements for an atom moving in an optical lattice. Many possible implementations have been proposed in the literature and our  presentation will not be exhaustive. In particular we will not describe methods based on lattice rotation, for which the basic principles are similar to the bulk case described in subsect.~\ref{subsec:rotations} (see \emph{e.g.} \cite{Sorensen:2005,Tung:2006b,Hemmerich:2007,Kitagawa:2010} and refs. in).

\subsection{Rapid shaking of a lattice}

We consider in this section a Hamiltonian which depends explicitly on time and corresponds to a shaken lattice. The potential reads in the one-dimensional case
\begin{equation}
{\cal V}(x,t)=V[x-x_0(t)],
%%\abel{}
\end{equation}
where $x_0(t)$ is a time-periodic function with period $T$. We will suppose in this subsection  that $T$ is very small compared to the other time scales of the problem, such as $\hbar/J$. A rigorous treatment of the problem is based on a Floquet-type approach. More precisely, one can develop a method that separates in a non-ambiguous way the dynamics related to an effective, time-independent Hamiltonian and the micro-motion caused by the fast modulation $x_0(t)$ \cite{Rahav:2003,Goldman:2014c,Bukov:2014}. Here for simplicity we will not follow such a rigorous treatment, but we will take a time-average of the Hamiltonian $\hat H(t)$, once it has been written in a suitable form. The link between this simple-minded approach and rigorous treatments is made in \cite{Goldman:2014c}, for example.

It is not immediate to transcribe the Hamiltonian of a particle in a shaken lattice
\begin{equation}
\hat H_1(t)=\frac{\hat p^2}{2M}+V[x-x_0(t)]
\label{eq:shaken_lattice}
\end{equation}
in the tight-binding approach. The simplest strategy is to use the unitary transformation generated by
\begin{equation}
\hat U(t)=\exp[\I x_0(t)\hat p/\hbar],
\label{eq:unitary_for_shaking}
\end{equation}
so that the Hamiltonian after transformation
\begin{equation}
\hat H(t)= U(t)\;\hat H_1(t)\;\hat U^\dagger(t)\ +\ \I \hbar \frac{\D \hat U(t)}{\D t}\hat U^\dagger(t)
%%\abel{}
\end{equation}
reads
\begin{equation}
\hat H(t)=\frac{\left[ \hat p-A(t) \right]^2}{2M}+V(x), \qquad A(t)=M\dot x_0(t).
%%\abel{}
\end{equation}
This amounts to applying a time-dependent vector potential in the direction of the shaking\footnote{The term $A^2(t)/2M$ is a mere time-dependent global shift of the energies. It can be eliminated by a unitary transform and has no physical consequence.}. Then the tight-binding version of the Hamiltonian is obtained in a straightforward way:
\begin{equation}
\hat H (t)= -J \;\E^{\I  Ma \dot x_0(t)/\hbar} \sum_{j}  |{j+1}\rangle \langle {j}| + \mbox{h.c.}
%\abel{}
\end{equation}
Now using the Magnus expansion at the lowest order (see \emph{e.g.} \cite{Goldman:2014c} and refs. in), we simply time-average this Hamiltonian over one period of the fast oscillation. We obtain the renormalized tunnel matrix element:
\begin{equation}
\bar J= J \;\langle \E^{\I \,M a \dot x_0(t)/\hbar}  \rangle . 
%%\abel{}
\end{equation}

In the context of cold atom physics, this result has first been considered as a way to change the magnitude of the tunnel matrix element by using a sinusoidal variation of the lattice displacement $x_0(t)$ \cite{Eckardt:2005}:
\begin{equation}
\frac{Ma}{\hbar}\dot x_0(t)=\xi_0\,\sin(\Omega t+\phi) \quad \rightarrow \quad \bar J=J\; {\cal J}_0(\xi_0),
\label{eq:sinusoidal}
\end{equation}
where ${\cal J}_n$ designates the $n$-th Bessel function of the first kind. This modification of $J$ in a 1D vibrating lattice has been accurately checked by the Pisa group \cite{Lignier:2007}. 

Here we are interested in obtaining complex tunnel matrix elements, which is not the case for the sinusoidal modulation of eq.\;(\ref{eq:sinusoidal}). A non-symmetric temporal modulation of $x_0(t)$ provides the desired result \cite{Struck:2012}.  For example one can divide the time-period $T$ in two unequal time $T_1$ and $T_2$ ($T_1+T_2=T$) and have the lattice moving at uniform velocity $v_i$ during $T_i$ ($i=1,2$) with $v_1T_1+v_2T_2=0$ so the displacement of the lattice is zero in one full period $T$. Then the renormalized tunnel coefficient obtained from this saw-tooth modulation reads
\begin{equation}
\frac{\bar J}{J}= \langle \E^{\I\,Ma \dot x_0(t)/\hbar} \rangle =\frac{T_1}{T}\E^{\I Mav_1/\hbar}+\frac{T_2}{T}\E^{\I Mav_2/\hbar},
\end{equation}
which is generally non-real. This technique was implemented by the Hamburg group in a 1D lattice, and the displacement in momentum space of the bottom of the lowest band, which constitutes a signature of a non-real effective tunnel coupling, was observed  \cite{Struck:2012}. A different implementation of a 1D tight-binding Hamiltonian with complex tunneling coefficients, based on a combination of Raman coupling and radio-frequency fields, was realized at NIST \cite{Jimenez:2012}.

The technique of a fast global shaking of the lattice can be generalized in two dimensions to generate a non-zero flux through  the cells of the lattice. A close look at the effect of the modulation shows that it does not lead to the desired flux if the sides of the unit cell are parallel to each other. In this case, the phases on two parallel sides indeed cancel each other, when one looks at the total phase accumulated over the border of  a unit cell as in (\ref{def_Phi}). On the contrary, this method does provide a non-zero flux for a triangular lattice, as demonstrated by the Hamburg group  \cite{Struck:2013}. Note however that the obtained flux is not uniform but staggered, with equal magnitude and opposite sign for two adjacent triangles. Using also a fast modulation and starting from a honey-comb type lattice, the ETH group could recently implement experimentally the Haldane model \cite{Haldane:1988,Jotzu:2014}.

\subsection{Resonant shaking/modulation}

Another possibility for generating complex tunneling matrix elements is to shake the lattice system at a resonant frequency. To reach this goal, one first adds to the lattice an external, static potential that lifts the degeneracy between adjacent sites by a energy offset $\hbar \Omega$. Then one shakes the lattice at frequency $\Omega$, and the phase of the function describing the shaking gets ``printed" on the corresponding tunnel matrix element. To prove this result we follow again a simple approach based on the lowest order of the Magnus expansion. A more rigorous, systematic expansion in powers of $1/\Omega$ can be found in \cite{Goldman:2014d,Bukov:2014}.

Let us consider for example the case of a 1D lattice, where the energy offset $\hbar \Omega$ between adjacent sites is obtain by applying a linear potential $-\hbar \Omega\, \hat x/a$, where $\hat x$ is the position operator and $a$ the lattice constant. The global shaking of the lattice is described by the function $x_0(t)$ as in eq.\;(\ref{eq:shaken_lattice}) and we proceed with two successive unitary transformations. The first one uses the same operator as in eq.\;(\ref{eq:unitary_for_shaking}) and leads  (up to additive constants) to
\begin{equation}
\hat H_1 =-J \sum_{j} \left( \E^{\I M a \dot x_0(t)/\hbar }\;|{j+1}\rangle \langle {j}| \ +\    \mbox{h.c.}\right) - 
\hbar \Omega\sum_j  j\; |{j}\rangle \langle {j}|.
\end{equation}
The second unitary transform eliminates the linear potential and replaces it by an extra time-dependence of the tunnel coefficients [see eqs.\;(\ref{eq:HtotF}) and (\ref{eq:HBloch}) in Appendix 2 for details]:
\begin{equation}
\hat H(t) =-J \;\E^{\I \left[M a \dot x_0(t)/\hbar -\Omega t\right] }\; 
\sum_{j} |{j+1}\rangle \langle {j}| \ +\  \mbox{h.c.}
\label{eq:time-modulated-hamiltonian}
\end{equation}
We assume the same sinusoidal modulation of the lattice as in eq.\;(\ref{eq:sinusoidal}) and  take the time average of the Hamiltonian (\ref{eq:time-modulated-hamiltonian}) over one period of modulation $2\pi/\Omega$. We obtain the renormalized tunneling coefficient
\begin{equation}
\bar J = J\; \langle \E^{\I \left[ \xi_0\, \sin(\Omega t+\phi) -\Omega t \right]}\rangle=J\,{\cal J}_1(\xi_0)\;\E^{\I \phi}.
\end{equation}
This result differs from the one obtained for rapid shaking [eq. (\ref{eq:sinusoidal})] in two ways. First the zeroth order Bessel function ${\cal J}_0$ is replaced the first order function ${\cal J}_1$; in particular no tunneling occurs in the absence of modulation (${\cal J}_1(0)=0$), because of the energy offset between adjacent sites. Second and most importantly, the phase $\phi$ of the modulation is now ``printed" on the renormalized tunnel coefficient, which thus becomes complex. 

For a 1D lattice, this printed phase can be viewed as a mere gauge transform in which one redefines the relative phases of adjacent Wannier functions. The procedure has wider applications in 2D since it can lead to a non-zero, uniform phase on each lattice cell. A first generalization of this method to the two-dimensional case  was proposed in \cite{Kolovsky:2011}, but it presented some drawbacks discussed in \cite{Creffield:2013}. An alternative method that provides the desired uniform flux over the lattice was proposed  \cite{Aidelsburger:2013b} and implemented by the Munich and MIT groups  \cite{Aidelsburger:2013,Miyake:2013}. They relaxed the condition of a uniform shaking of the lattice sites and implemented a modulation that varies from site to site. This implementation was made possible using the dipole potential created by an extra pair of laser beams, which created the necessary modulation of the lattice sites\footnote{Since the modulation is induced by an auxiliary laser field, this scheme is sometimes called \emph{laser-induced tunneling}. Although this terminology is certainly correct in the present case, we prefer to reserve it for schemes where a laser beam is used to simultaneously (i) induce a transition between internal atomic states, and (ii) induce a jump of the atomic center-of-mass between adjacent lattice sites, as described in subsect.~\ref{subsec:ladder}.}. 
%%%%%%%%%%%%%%%%%%%%%%%%%%%%%%%%%%%%%%%%%%%%%%%%%%%%%%%

\section{Generation of lattice gauge fields via internal atomic transitions}
\label{sec:lattice_internal}

The possibility to use internal states in an optical lattice brings some new flexibility for the generation of complex tunnel coefficients in a lattice. The process in play is \emph{laser assisted tunneling}, which we will first present on a simple, one-dimensional system, before switching to the case of an infinite two-dimensional lattice.

\begin{figure}[t]
\begin{center}
\includegraphics[height=55mm]{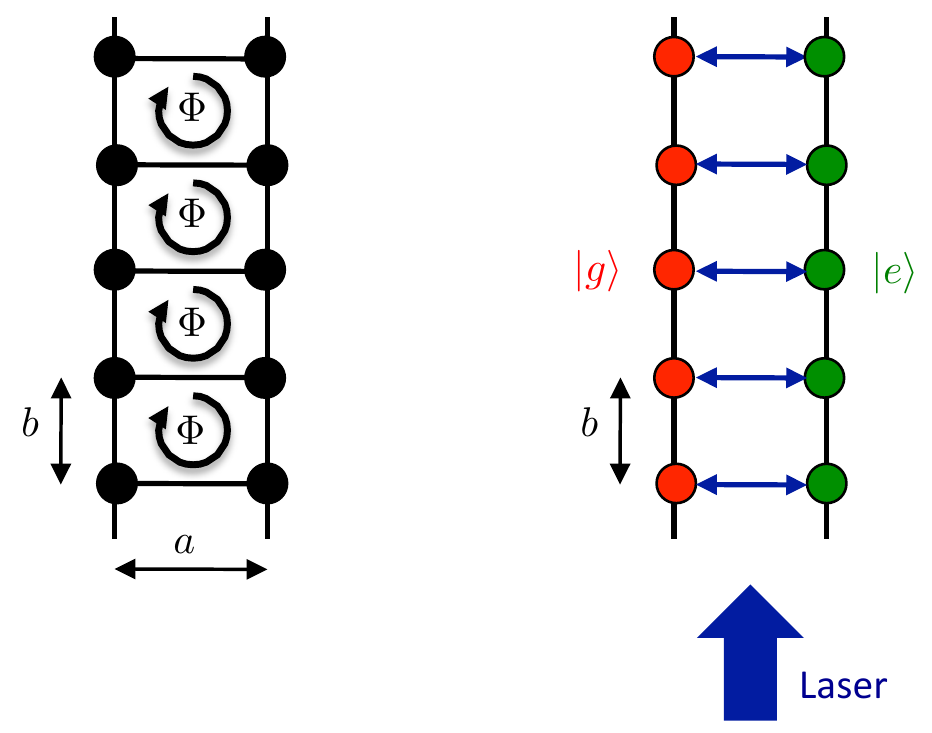}
\end{center}
\caption{Left: Infinite ladder with two sides separated by the distance $a$. The ladder rungs are equidistant, separated by a distance $b$. The particle can jump from one site to the next by tunneling. A uniform magnetic field creates a flux $\Phi$ on each cell. Right: Simulation of this ladder with a double optical lattice. A laser beam propagating along the lattice direction induces transitions  $\ketg \leftrightarrow \kete$, similar to the tunneling along the ladder rungs in the left picture. The laser phase $\phi=ky$ varies linearly with the rung index $j$ and gets ``printed" on the corresponding matrix element.}
\label{fig:echelle}
\end{figure}

\subsection{Laser assisted tunneling in a 1D ladder} 
\label{subsec:ladder}

We are interested here in a ladder with two parallel sides as in fig.\;\ref{fig:echelle}. This ladder is submitted to a uniform magnetic field and we suppose that it is well described by the same tight-binding approximation as above. The magnetic field is characterized by the flux $\Phi$ across a unit cell.

To simulate this ladder we use an atom with two internal states $\ketg$ and $\kete$. A trapping potential created by an optical lattice along $y$ localizes the atoms along the rungs of the ladder. We assume that this potential is different for $\ketg$ and $\kete$, so that each internal state is trapped on a given side of the ladder ($\ketg$ on the left, $\kete$ on the right). We suppose that these two internal states are stable (no spontaneous emission) and we consider a laser excitation that allows one to induce a resonant transition between them. As explained in subsect.~\ref{subsec:two_level}, this atomic level scheme can be obtained with species with two outer electrons, as well as with alkali metals, if one uses a Raman transition between sublevels of the atomic ground state.  

Using a single-band model, we denote $|j\rangle_g$ and $|j\rangle_e$, $j\in \ZZ$, the spatial states associated to the two internal states $\ketg$ and $\kete$. To simplify the algebra we suppose that the corresponding wavefunctions are identical, except for the translation over $a$:
\begin{equation}
w_j^{g}(\bs r)=w_0(x,y- jb),\qquad  w_j^{e}(\bs r)=w_0(x-a,y-jb).
\end{equation}
In the absence of laser excitation, we then have two independent 1D lattices with the Hamiltonian:
\begin{equation}
\hat H= \hat H_g +\hat H_e, \qquad \hat H_\alpha =-J_y \sum_{j\in \ZZ} |{j+1}\rangle_\alpha \;_\alpha\langle j| +\mbox{h.c.}, \quad \alpha=g,e.
\end{equation}
In order to couple these lattices, we use a  laser beam that is resonant with the transition $\ketg \leftrightarrow \kete$, with a wave vector $\bs k$ parallel to the $y$ axis. The atom--light coupling reads:
\begin{equation}
\hat V= \frac{\hbar \kappa}{2} \,\E^{\I ky}\,|e\rangle \langle g| + \mbox{h.c.}
\end{equation}
Let us rewrite this coupling in terms of the spatial states $|j\rangle_\alpha$. Using the closure relation for internal and external states in the single band approximation:
\begin{equation}
\hat 1= |g\rangle \langle g| \otimes \left( \sum_{j\in \ZZ} |j\rangle_g \;_g\langle j|\right) \ +\ |e\rangle \langle e|\otimes\left( \sum_{j\in \ZZ} |j\rangle_e\;_e \langle j|\right).
\end{equation}
we obtain
\begin{equation}
\hat V=  \sum_{j,j'} {\cal V}_{j,j'}|j\rangle_e\; _g\langle {j'}| + \mbox{h.c.}
\end{equation}
with
\begin{equation}
{\cal V}_{j,j'}=
\frac{\hbar \kappa}{2} \int w_j^{e}(\bs r)\;\E^{\I ky} \;w_{j'}^{g}(\bs r)\;\D^2 r\ .
\end{equation}
In the tight-binding approximation the wave functions $w_j^{e}(\bs r)$ and $w_{j'}^{g}(\bs r)$ are well localized and on can neglect their overlap as soon as $j\neq j'$. This allows us to simplify the expression of ${\cal V}_{j,j'}$ 
\begin{eqnarray}
{\cal V}_{j,j'} &\approx& \delta_{j,j'} \frac{\hbar \kappa}{2}  \int w_0(x-a,y-jb)\;\E^{\I ky} \;w_0(x,y-jb)\;\D^2r \nonumber \\
&\approx& \delta_{j,j'} \,J_x\, \E^{\I jkb},
\label{eq:coef_tunnel_assiste}
\end{eqnarray}
where the tunnel coefficient $J_x$ is proportional to the Rabi frequency and to the overlap of the Wannier functions for the internal states $\ketg$ and $\kete$:
\begin{equation}
J_x=\frac{\hbar \kappa}{2}  \int w_0(x-a,y)\;\E^{\I ky} \;w_0(x,y)\;\D^2r.
\end{equation}
For a function $w_0(x,y)$ that is even in the change $y \to -y$, this coefficient $J_x$ is real:
\begin{equation}
J_x=\frac{\hbar \kappa}{2}  \int w_0(x-a,y)\;\cos(ky) \;w_0(x,y)\;\D^2r.
\end{equation} 

The expression  (\ref{eq:coef_tunnel_assiste}) has the desired structure. The tunnel matrix element is complex, with a phase that increases linearly with the index $j$. The total phase accumulated along the contour of a cell 
\begin{equation}
|g,j\rangle \overset{J_y}{\longrightarrow} 
|g,j+1\rangle \overset{J_x\E^{\I(j+1)kb}}{\longrightarrow} 
|e,j+1\rangle \overset{J_y}{\longrightarrow} 
|e,j\rangle \overset{J_x\E^{-\I j kb}}{\longrightarrow} 
|g,j\rangle.
\end{equation}
is non-zero and it takes the same value $(j+1)kb-jkb=kb$ for all cells. One simulates in this way a uniform magnetic field over the whole ladder.

\subsection{Lattice with artificial dimension}
\begin{figure}[t]
\begin{center}
\includegraphics[height=55mm]{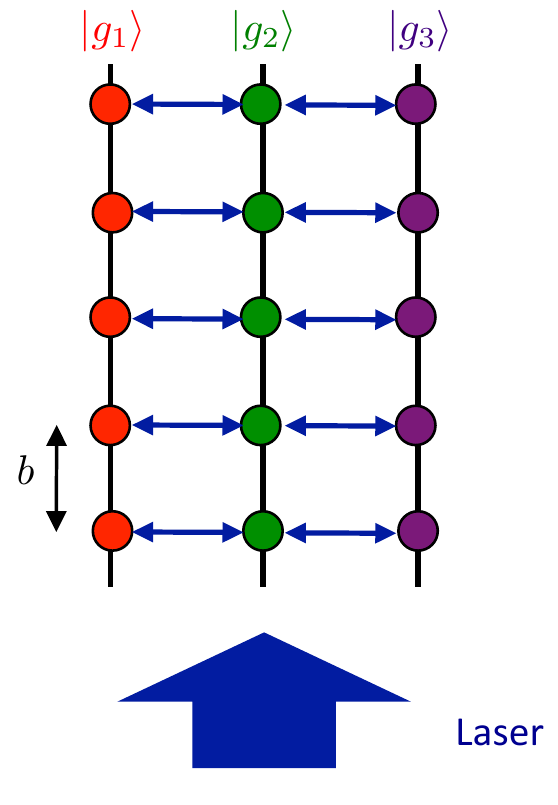}
\end{center}
\caption{Simulation of a double ladder for an atom with three internal states
 \cite{Celi:2014}. This scheme can give rise to a single-particle spectrum similar to the Hofstadter butterfly and to edge states with opposite currents along the ``edges" associated to the states $|g_1\rangle$ and $|g_3\rangle$.  }
\label{fig:echelle2}
\end{figure}

One can notice that in the preceding description of a two-leg ladder, nothing requires the length $a$ to be non-zero. The ladder can have a dual nature: real along the $y$ direction, and fictitious along $x$. The requirement for a non-zero $a$ will come only later, when we try to increase the number of sites along the $x$ direction. Starting from this remark, it was proposed in \cite{Celi:2014} to extend the preceding reasoning to the case of an atom with $N$ internal states, 
$|g_n \rangle$, $n=1,\ldots, N$, by choosing an atom--light coupling that induces the transitions
\begin{equation}
|g_n,j \rangle   \overset{J_x\E^{\I jkb}}{\longrightarrow} |g_{n+1},j\rangle.
\end{equation}
This allows one to simulate a ladder with $N$ legs, each leg being associated to one internal state.  We illustrate this in fig.\;\ref{fig:echelle2} for the case $N=3$, taking for example the three Zeeman states $m=0,\pm 1$ of an atomic ground state with angular momentum 1.

\subsection{Laser-induced tunneling in a 2D lattice}

We now extend the scheme outlined above to the two dimensional case. Now the length $a$ is not fictitious and we consider a series of parallel 1D lattices, corresponding alternatively to the internal states   $\ketg$ and $\kete$. One forms in this case a 2D lattice in which a site $(j,l)$ is occupied by an atom in state $\ketg$ (resp. $\kete$) if $j$ is even (odd).

\begin{figure}[t]
\begin{center}
\includegraphics[height=55mm]{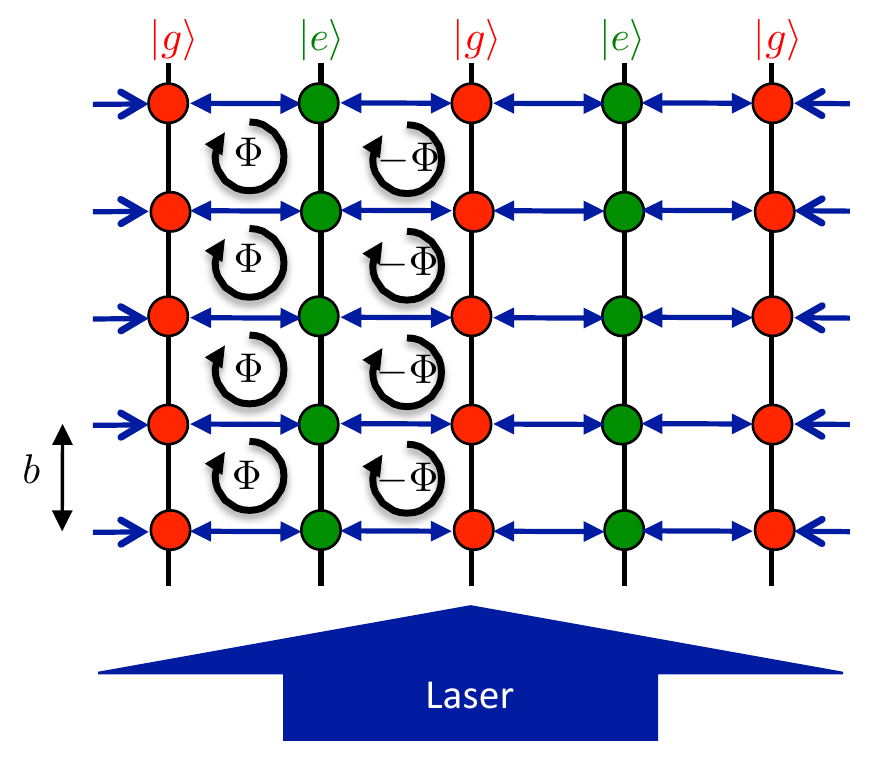}
\end{center}
\caption{Realization of a square lattice with a staggered flux, using a single light beam to induce laser-assisted tunneling.}
\label{fig:flux_alterne}
\end{figure}

The analysis of the phases associated to laser induced tunneling shows that the lattice then corresponds to a staggered flux. We recall that  a uniform flux is obtained by taking
\begin{equation}
|j,l\rangle \overset{\E^{\I\,lkb}}{\longrightarrow} |j+1,l\rangle 
\end{equation}
whereas we have here
\begin{equation}
|j,l\rangle \overset{\E^{\I\,lkb}}{\longrightarrow} |j+1,l\rangle \quad \mbox{if \emph{j} even}
\end{equation}
corresponding to the transition $\ketg \longrightarrow \kete$ and 
\begin{equation}
|j,l\rangle \overset{\E^{-\I\,lkb}}{\longrightarrow} |j+1,l\rangle \quad \mbox{if \emph{j} odd}
\end{equation}
corresponding to the transition $\kete \longrightarrow \ketg$. The sign of the accumulated phase on the contour of a cell alternates from one column to the next.

This lattice with a staggered flux has interesting properties when interactions are taken into account \cite{Moller:2010}, but it does not correspond to the desired simulation of uniform magnetism with non-trivial topological properties. One thus needs to find a way to ``rectify" the magnetic field, in order to obtain a flux with the same sign over each cell. Several techniques have been proposed to reach this goal. The first one \cite{Jaksch:2003} consists in adding an extra linear potential in order to lift the degeneracy between the transitions $\ketg \longrightarrow \kete$ going to the right ($2j\to 2j+1$) and those going to the left ($2j\to 2j-1$). Another version, well adapted to alkaline-earth species, uses a super-lattice along the $x$ direction with a spatial period $2a$ \cite{Gerbier:2010}. In both versions, the resulting different transitions $\ketg \longrightarrow \kete$  are provided by laser beams propagating along $+\unity$ and $-\unity$, ensuring that the phases accumulated on all cells of the lattice always have the same sign.

\subsection{Optical flux lattices}

We now present a different way to simulate magnetism on a lattice, which consists in imposing that the light configuration is fully periodic and stationary in time. We consider a situation where all laser beams are monochromatic and form standing waves, and we ask if some interesting properties, \emph{i.e.} energy bands with non-trivial topological features, can be found in this case. We will see that this is indeed the case, as it was  first proposed by N.R. Cooper in 2011 \cite{Cooper:2011}.

We note first that due to the periodicity of the lattice, the phase accumulated by a particle when it travels along the sides of a unit cell must be a multiple of $2\pi$. In the tight-binding limit, we know from our study of subsect.~\ref{subsec:Hofstadter} that for such a phase, the physics is the same as for a zero magnetic flux, hence it is not relevant for our present goal. From now on, we will therefore assume that the potential created by the laser light is relatively weak, so that the lattice operates outside the tight-binding regime. In practice this means that the Rabi frequency characterizing the atom laser coupling should be on the order of (or smaller than) the recoil frequency $E_{\rm r}/\hbar$.
 
To proceed, we consider a situation where the internal dynamics is well described by  the two-level approach of subsect.~\ref{subsec:two_level}, and we suppose that the parameters $\Omega$, $\theta$ and $\E^{\I \phi}$ entering in the $2\times 2$ matrix for the atom--laser coupling are periodic functions of space. Assuming for the moment that the adiabatic approximation holds, the vector potential $\bs {\cal A}(\bs r)$ --Berry's connection--  given in eq.\;(\ref{eq:connexion_Berry}) is also a periodic function of space. The contour integral of $\bs {\cal A}(\bs r)$ over the unit cell must then be zero. If the Stokes theorem holds, the flux of Berry's curvature $\bs {\cal B}(\bs r)$ through the unit cell is also zero, meaning that the lattice does not create any average magnetism. However there exist situations where the Stokes theorem is not valid, leading to a non-zero magnetic flux in spite of the periodicity of $\bs {\cal A}$: this is precisely the definition of an optical flux lattice\footnote{There is a duality between the concept of an optical flux lattice in real space, and the concept of a Chern band in momentum space. In the case of a Chern band, the physical problem is periodic over the Brillouin zone but one can still have a non-zero flux of Berry's curvature $\bs {\cal B}(\bs q)$, corresponding to a non-zero Chern index ${\cal C}$ [cf. eqs.\;(\ref{eq:Berry_curvature}) and (\ref{eq:def_Chen}) in Appendix 2].}.  

To obtain a ``violation" of Stokes theorem, one needs to have singularities of ${\cal A}(\bs r)$, ensuring that
\begin{equation}
\oint \bs {\cal A}(\bs r)\cdot \D \bs r \neq \intint {\cal B}_z \;\D x \,\D y.
\label{eq:stokes}
\end{equation}
Let us take an example extracted from \cite{Cooper:2011}, in which the coefficients of the $2\times 2$ matrix giving the atom--laser coupling vary as
\begin{equation}
\Omega \cos \theta \propto \sin(kx)\sin(ky),\qquad \Omega \sin\theta\; \E^{\I \phi }\propto
\cos(kx)+\I \cos(ky), \qquad \Omega>0 .
\label{eq:coupling_for_OFL}
\end{equation}
The unit cell of the lattice is $-\pi < kx,ky \leq \pi$, and the phase $\phi$ is ill-defined at the four points in the cell where $\sin\theta$ vanishes: $(kx,ky)=(\pm \pi/2,\pm \pi/2)$. This entails that Berry's connection $ \bs {\cal A}$, whose expression contains the term $\pm (\hbar/2) \cos \theta\, \bs \nabla \phi$ [see eqs.\;(\ref{eq:connexion_Berry}) and (\ref{connexion_Berry_2})], may also be singular in these points. 

There are various ways to handle these singularities and determine the magnetic properties of this lattice configuration. One option is to evaluate directly $\bs {\cal B}$ from eq.\;(\ref{eq:courbure_Berry}), and check that $\bs {\cal B}$ is actually non singular at those points. A second possibility is to make different gauge choices $\bs {\cal A}^{I}$, $\bs {\cal A}^{II}$,\ldots  over different regions of the unit cell, such that no singularity occurs over any region. The gauge transformations at the boundaries between these regions provide the desired flux\footnote{For the coupling (\ref{eq:coupling_for_OFL}), the gauge choice (\ref{eq:connexion_Berry}) leads to no singularity for $\bs {\cal A}$ at the two points $(kx,ky)=(\pi/2,\pi/2)$ and $(-\pi/2,-\pi/2)$ because $(\cos \theta-1)$ vanishes in these two points. Similarly the gauge choice (\ref{connexion_Berry_2}) leads to no singularity for $\bs {\cal A}$ in the two other points $(\pi/2,-\pi/2)$ and $(-\pi/2,\pi/2)$ because $(\cos \theta+1)$ vanishes in those points.} \cite{Kohmoto:1985}. A third approach consists in taking a contour for $\bs {\cal A}$ that avoids the (discrete) singularities of $\phi$, so that the equality in eq.\;(\ref{eq:stokes}) holds.  The flux of $\bs {\cal B}$ is then given by the contribution of these singularities, each being a multiple of $\pi \hbar$. Still another method is to use the concept of Dirac strings to account for these singularities \cite{Juzeliunas:2012}. All these approaches of course lead to the same result, with a flux of  $\bs {\cal B}$ that can be non-zero and that is always a multiple of $2\pi \hbar$.

\begin{figure}[t]
\begin{center}
\includegraphics{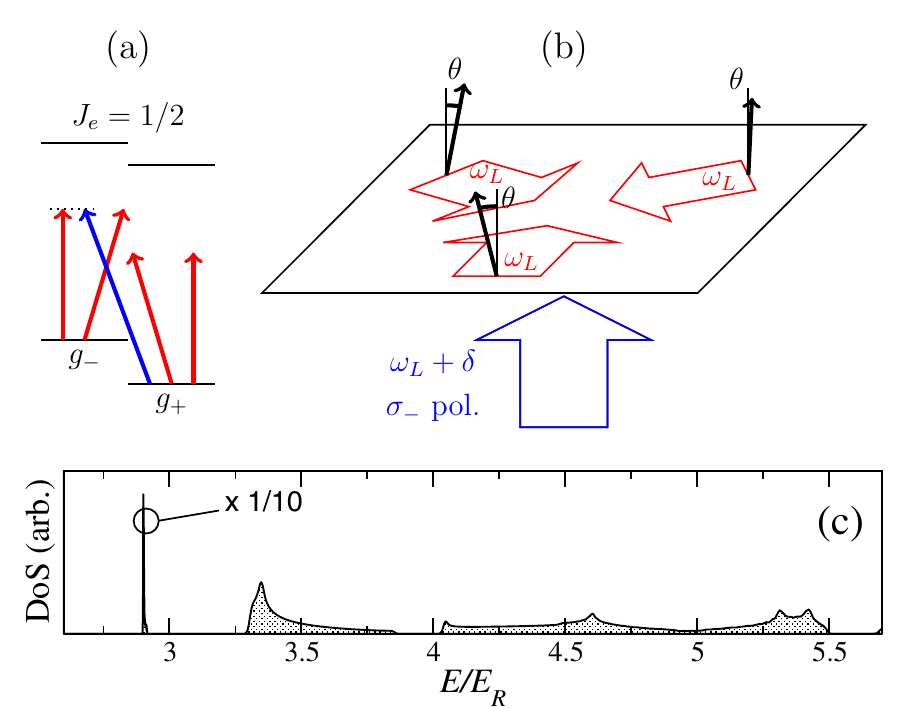}
\end{center}
\caption{Optical flux lattice configuration for an atom with a spin $1/2$ ground state. Top left: A magnetic field lifts the degeneracy between the two Zeeman sublevels $g_\pm$. Top right: Three identical laser beams propagate in the $xy$ plane at 120 degrees from each other, with a linear polarization at an angle $\theta$ with respect to the $z$ direction. A fourth laser beam propagates along the $z$ axis and induces, together with the three horizontal beams, a resonant Raman coupling between $g_+$ and $g_-$. Bottom: resulting density of states (DoS) for a proper choice of the Rabi frequencies of the laser beams and their polarization orientation (for details, see \cite{Cooper:2011b}).}
\label{fig:OFL}
\end{figure}

The quantitative optimization of an optical flux lattice configuration amounts to make it as similar as possible to the case of a particle moving freely in a uniform magnetic field, in which case the energy eigenstates group together to form massively degenerate Landau levels (sect.~\ref{sec:free_atom}). For this optimization, it is preferable not to rely on the adiabatic approximation since the relatively small laser coupling makes the validity of this approximation marginal [see eq.\;(\ref{eq:valid_adiab})]. Consequently, one solves the band problem for the Hamiltonian (\ref{eq:Htot}) and tries to fulfill simultaneously three conditions, focusing for example on the lowest band: (i) The band should have a non-trivial topology, characterized by a non-zero Chern index (we remind that the Chern index for the Lowest Landau Level is 1). (ii) This lowest band should be very narrow (in units of the recoil energy) to mimic the flatness of a Landau level. (iii) It should be separated from the first excited band by a large gap, so that the dynamics can safely be restricted to this lowest band even in the presence of interactions between particles. We show in fig.\;\ref{fig:OFL} the result of an optimization for a two-level atom, with an in-plane triangular laser configuration that fulfills these three criteria: the lowest band has a Chern index of 1, a width of $0.01\,\Er$ only, and it is separated from the next band by a gap 40 times larger than its width \cite{Cooper:2011b}. 

The notion of an optical flux lattice can be generalized in various ways. First it is possible to design optical flux lattice configurations with Chern indices larger than 1. Also one can identify  situations where: (i) the lowest band has a non-zero Chern number (ii) Berry's curvature (\ref{eq:courbure_Berry}) calculated for the adiabatic dressed state has a zero flux across the unit cell in real space \cite{Cooper:2012}. One obtains in this case a situation that is reminiscent of Haldane's model \cite{Haldane:1988}. Another generalization consists in using an atomic structure with more internal states, and designing a time-reversal symmetric configuration corresponding to a $\ZZ_2$ topological insulator \cite{Beri:2011}. 

\section{Conclusion}

\begin{figure}[t]
\begin{tabular*}{134.7mm}{p{22mm}|p{50mm}|p{50mm}|}
 & {Time-independent \linebreak Hamiltonian} & {Time-dependent\linebreak Hamiltonian (frequency $\Omega$)} \\
 \hline
 No use of  \hfill \hfill  \linebreak internal states & $\bs p\cdot \bs A$ as a Lagrange  \hfill \hfill \linebreak parameter in rotation & 
 $\Omega \sim \omega_c$: rotation  \hfill \hfill \linebreak
 $\Omega\gg \oc$: lattice shaking 
 \\
\hline
Using internal states & 
Berry's phase  \hfill \hfill \linebreak 
Spin-orbit coupling \hfill \hfill \linebreak
Laser assisted tunnelling  \hfill \hfill \linebreak
Optical flux lattices  %\hfill \hfill  \linebreak
  & Spin-orbit coupling \\
\hline
\end{tabular*}
\caption{Classification of the various procedures used in cold-atom setups to simulate magnetism (cyclotron frequency $\oc$) or spin-orbit coupling. Schemes in the lower right box have not been described in this lecture but are addressed in \cite{Xu:2013,Anderson:2013,Goldman:2014c}.}
\label{fig:bilan}
\end{figure}

We have presented in these notes a series of methods that allow one to simulate, at the single-particle level and with neutral atoms, the physics of a charged particle in a static external magnetic field. Several of these schemes have been successfully implemented in the laboratory and others are currently being investigated. A tentative classification of these schemes is presented in fig.\;\ref{fig:bilan}, where we sort the various procedures along two criteria: does it take advantage of the internal atomic structure? Is it based on an explicitly time-dependent Hamiltonian? 

It is essential to recall that we described here only a fraction of the activity in this field of research. First, as mentioned above, the space limitation did not allow us to discuss all the schemes that have been proposed so far at the single-atom level. For example we focused here onto the Hamiltonian approach to the problem. Another strategy to the general search for `topological quantum matter' is instead to reach these states by dissipation:  One has to engineer a master equation for which the desired topological states are `dark', \emph{i.e.} they lie in the kernel of this master equation and are reached after some relaxation time period (see \cite{Bardyn:2013} and refs. in).
Second, many other systems such as photonics devices are also currently investigated to simulate these magnetic-like topological effects. We refer the reader to the recent review articles \cite{Carusotto:2013,Hafezi:2014} for discussion and references on these photonic implementations of artificial gauge fields. 

We did not address in these notes the role of interactions between particles. These interactions are of course crucial if one wants to produce many-body states similar to those appearing in the fractional quantum Hall regime. For homogeneous systems, we refer the reader to the review \cite{Cooper:2008} where the similarities and differences with respect to quantum Hall states are discussed in detail, both in terms of the statistical nature of the particles -- fermions or bosons --, and of the type and range of the interaction potential. For optical flux lattices, which have properties analogous to the Landau states of the free-particle case, one can identify many-body states that are similar to those of bulk systems \cite{Cooper:2013}. For optical lattices in the tight-binding regime the situation is much more open \cite{Moller:2009,Hormozi:2012}. Finally let us emphasize that in all our discussions, we considered the gauge fields as static and imposed from the outside. An important challenge for the cold atom community is to make these gauge fields dynamical, by coupling their states/values to the dynamics of the particles, see e.g. \cite{Tagliacozzo:2013,Glaetzle:2014} and refs. in. 

\acknowledgments

This work is supported by IFRAF, ANR (ANR-12- 247 BLANAGAFON) and ERC (Synergy grant UQUAM). The author thanks Jérôme Beugnon, Tom Bienaimé, Lauriane Chomaz, Nigel Cooper, Laura Corman, Fabrice Gerbier, Nathan Goldman and Sylvain Nascimbene for many helpful discussions and remarks on these lecture notes. 

 \newcommand{\kket}[1]{|#1\; \rangle\hskip-2mm\rangle}
 \newcommand{\bbra}[1]{\langle\hskip-2mm\langle\;  #1 |}
 \newcommand{\bbrab}[1]{\langle\hskip-2mm\langle\;  #1 }
 \newcommand{\tB}{t_{\rm B}} 

\section*{Appendix 1: Landau levels}

\subsection*{Eigenstates with the Landau gauge}

\begin{figure}[t]
\begin{center}
\includegraphics[height=6cm]{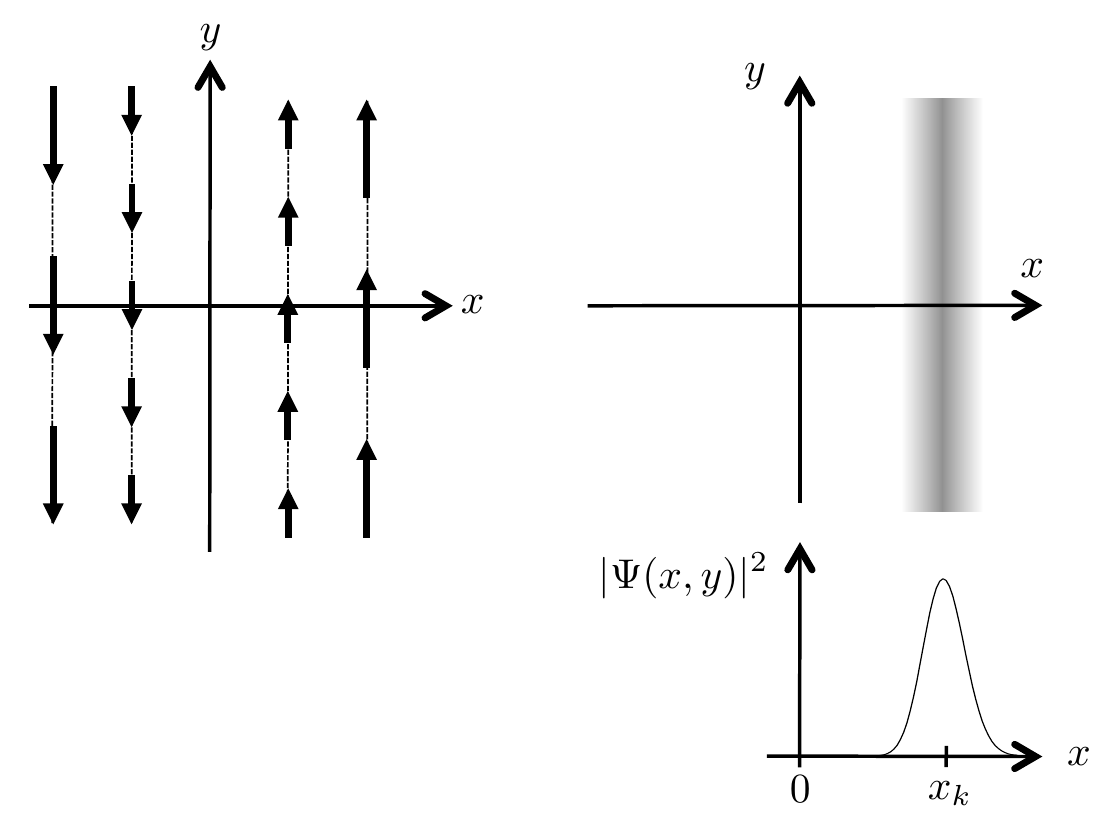}
\end{center}
\caption{Left: vector potential for the Landau gauge. Right: probability density of a LLL eigenstate of the Hamiltonian,  $\Psi(\bs r)= \psi_k(x)\,\E^{\I ky}$, where
$\psi_k(x)$ is a gaussian of width $\lm$ centered in $x_k=\hbar k/(qB)=k\lm^2$.}
\label{fig:jauge_Landau}
\end{figure}

Using the Landau gauge $\bs A(\bs r)= Bx\,\bs u_y$ (cf. fig.\;\ref{fig:jauge_Landau}), the Hamiltonian can be written
\begin{equation}
\hat H=\frac{\hat p_x^2}{2M}+\frac{\left( \hat p_y -qB\hat x\right)^2}{2M}.
\label{eq:H_jauge_Landau}
\end{equation}
Since it does not depend on the position operator $\hat y$ but only on  $\hat p_y$, we can look for an eigenstate basis in the form of plane waves along $y$, $\Psi_k(\bs r)=\psi_k(x) \;\E^{\I k y}$. We suppose that the sample has a finite size $L_y$ along $y$ and we take periodic boundary conditions along this axis. The quantum number $k$ is thus quantized
$k=\left({2\pi}/{L_y}\right)\,n_y$ {with} $ n_y \in \ZZ$.

The function $\psi_k(x)$ is a solution of:
\begin{equation}
-\frac{\hbar^2}{2M} \psi_k''(x)+ \frac{\left( \hbar k -qB x\right)^2}{2M} \psi_k(x) = E\;\psi_k(x)
\label{eq:hamiltonien_reduit_selon_x}
\end{equation}
which can be conveniently written
\begin{equation}
-\frac{\hbar^2}{2M} \psi_k''(x)+ \frac{1}{2}M\omega_c^2\left( x-x_k\right)^2 \psi_k(x) = E\;\psi_k(x) 
\end{equation}
with $x_k={\hbar k}\,/\,{qB}=k\lm^2$. For each $k$, this corresponds to the eigenvalue equation for a harmonic oscillator centered in $x_k$, with frequency  $\oc$. We thus recover the Landau level spectrum of eq.\;(\ref{eq:Landau_level_spectrum}). 

Several remarks are in order at this stage:
\begin{itemize}
 \item
 Although the wave number $k$ along $y$ is generally non-zero, there is no kinetic energy 
$\hbar^2 k^2/2M$ associated to it. The presence of the magnetic field entails that the only influence of this wave number is to shift the center of the fictitious oscillator along $x$. We shall deepen this point later, when we calculate the average velocity for a particle in state $\Psi_k$.

 \item
 States belonging to the ground level, also called the lowest Landau level (LLL), are obtained by taking linear combinations of the ground-state wavefunctions of the various oscillators.
Each ground-sate wavefunction is a gaussian centered  in $x_k$ with a width $\lm$:
\begin{equation}
\Psi_k(\bs r) \propto \E^{-(x-x_k)^2/2\lm^2} \;\E^{\I ky}.
\label{eq:etat_propre_jauge_Landau}
\end{equation}

\item
The various $\Psi_k$'s are orthogonal because of their different variations with $y$:
\begin{equation}
\int \Psi_k^*(\bs r)\,\Psi_{k'}(\bs r)\;\D^2r
\propto \int \E^{-\I ky}\E^{\I k' y}\;\D y =L_y \delta_{n_y,n'_y},
\end{equation}
where  $n_y$ and  $n'_y$ are the two integers characterizing the quantization of the wave numbers $k$ and $k'$.

\item
Two consecutive values of $k$ are separated by $2\pi/L_y$, which leads to very close values for the centers of the corresponding oscillators:
\begin{equation}
x_k-x_{k'}=2\pi \frac{\lm^2}{L_y} \qquad \mbox{if}\quad\ k-k'=\frac{2\pi}{L_y},
\end{equation}
which is very small compared to $\lm$ if the size $L_y$ is large compared to $\lm$.
\end{itemize}

Finally we can estimate the degeneracy ${\cal N}$ of a given Landau level, for example the LLL, for a rectangular sample of size $L_x\times L_y$. First, a given eigenstate $\Psi_k$ will play a significant role in the expression of a physical wave packet only if the center $x_k$ of the oscillator is located inside the rectangle. This relates the wave number $k$ and the size $L_x$:
\begin{equation}
0 \leq x_k \leq L_x \quad \Rightarrow \quad 0\leq k \leq \frac{L_x}{\lm^2}. 
\label{eq:condition_k_Lx}
\end{equation}
Second, the finite size of $L_y$ imposes that $k$ is discrete and a multiple of  $2\pi/L_y$. The number of relevant independent states $\Psi_k$ is then:
\begin{equation}
{\cal N} \approx \frac{L_x}{\lm^2}\;\frac{L_y}{2\pi}= \frac{{\cal A}}{2\pi \lm^2}
\label{eq:LLL_degeneracy}
\end{equation}
where ${\cal A}=L_xL_y$ is the sample area. 

\subsection*{Probability current in a Landau state}
In order to understand why there is no kinetic energy term $\hbar^2k^2/(2M)$ in the expression of the total energy of the state $\Psi_k$, we  evaluate the average kinetic momentum $\langle \hat {\bs \Pi}\rangle$ in this state. Using the expression (\ref{eq:kinetic_momentum}) for $\hat {\bs \Pi}=M\hat {\bs v}$, we find:
\begin{equation}
\langle \hat {\bs \Pi}\rangle= M\langle \hv \rangle =\int \Psi_k^*(\bs r)\;\left(-\I \hbar \bs \nabla -q\bs A(\bs r) \right)\; \Psi_k(\bs r)\;\D^2r,
\end{equation}
where the expression of $\Psi_k$ for a state of the LLL is given in eq.\;(\ref{eq:etat_propre_jauge_Landau}). 

Clearly the component along $x$ of $\langle \hat {\bs \Pi}\rangle$ is zero; indeed  $A_x=0$ and the average of the momentum $p_x$ in the ground state of the oscillator Hamiltonian is also zero.  The treatment of the $y$-component is less straightforward. The action of  $\hat p_y$ on $\Psi_k(\bs r)$ gives a factor $\hbar k$ so that 
\begin{equation}
M  \langle v_y(k)\rangle= \int \Psi_k^*(\bs r)\;\left( \hbar k -q A_y \right)\; \Psi_k(\bs r)\;\D^2r.
\end{equation}
The integration along $y$ can be done explicitly to give 
\begin{equation}
 \langle v_y(k)\rangle=\oc \int |\psi_k(x)|^2 \;(x_k-x)\;\D x.
\end{equation}
For the ground state of the oscillator centered in  $x_k$, this integral is zero because of the symmetry of this state with respect to $x_k$. Therefore the average velocity and the average kinetic momentum are zero in spite of the presence of 
 $\E^{\I ky}$ in the expression of the eigenstate: the contribution of the canonical momentum operator is exactly compensated by that of the vector potential.
 
\begin{figure}[t]
\begin{center}
\includegraphics[height=6cm]{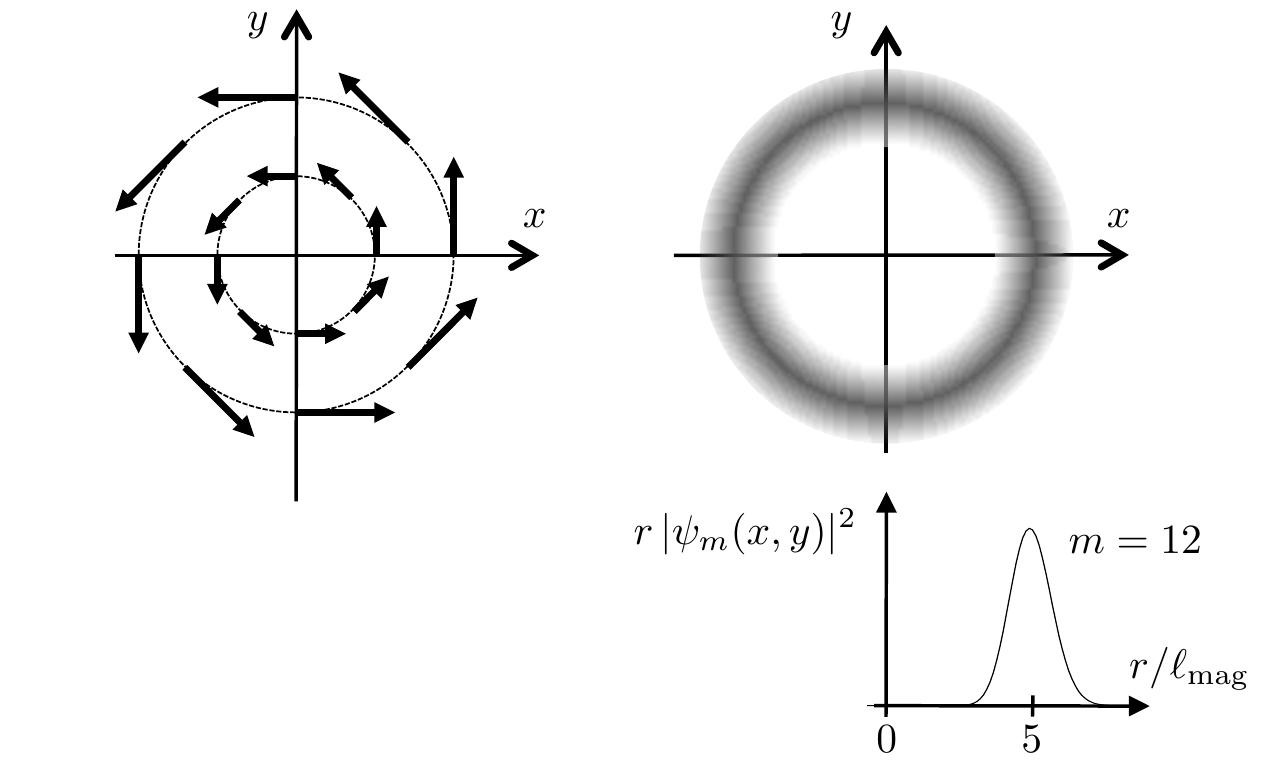}
\end{center}
\caption{Left: Vector potential for the symmetric gauge. Right: probability density for an eigenstate $\psi_m$ in the LLL [cf. eq.\,(\ref{eq:etats_extremes})], plotted here for $m=12$. }
\label{fig:jauge_sym}
\end{figure} 
 
\subsection*{Eigenstates with the symmetric gauge}
We now choose the symmetric gauge $\bs A(\bs r)=B (-y\,\bs u_x+x\,\bs u_y)/2$ leading to the Hamiltonian
\begin{equation}
\hat H=\hat H_0 -\frac{\oc}{2}\hat L_z, \qquad \mbox{with}\quad \hat H_0=\frac{\hat p^2}{2M}+\frac{1}{8}M\oc^2 r^2, \quad \hat L_z=\hat x\hat p_y-\hat y\hat p_x.
\label{eq:hamiltonien_jauge_symetrique}
\end{equation}
This Hamiltonian is the sum of two operators which commute, so that we can find a common eigenbasis to $\hat H$, $\hat H_0$ and $\hat L_z$. Let us discuss first separately the two components of $\hat H$.  The operator $\hat L_z$ is the $z$-component of the orbital angular momentum and its expression is quite simple in polar coordinates, $\hat L_z=-\I \hbar \partial_\varphi$,
so that its eigenstates are functions of the type $F(r )\,\E^{\I m\varphi}$, with $m\in \ZZ$ 
and $F$ an arbitrary function of the radial coordinate. The Hamiltonian $\hat H_0$ describes a 2D isotropic harmonic oscillator with frequency $\oc/2$, and its eigenvalues are $(n_0+1) \hbar \oc/2$, $n_0\in \NN$. 

To proceed further we have to inject the generic form $F(r )\,\E^{\I m\varphi}$ into the eigenvalue equation for $\hat H_0$ and find the functions $F(r)$. The solution is relatively involved in the general case (see \emph{e.g.}  \cite{Cohen_MQ_en}), but it becomes quite simple if we restrict to the eigenfunctions of the LLL. Considering the set of functions
\begin{equation}
\psi_m(x,y)= (x+ \I y)^m\, \E^{-r^2/4\lm^2}=r^m\,\E^{\I m \varphi}\, \E^{-r^2/4\lm^2}
\label{eq:etats_extremes}
\end{equation}
for $m\in \NN$, one can check relatively easily that these functions are both eigenstates of $\hat H_0$ with  energy $(m+1)\hbar \oc/2$ and of $\hat L_z$ with eigenvalue $m\hbar$. Therefore they are all eigenstates of $\hat H$ with the same energy $\hbar \oc/2$, corresponding to the lowest Landau level in eq.\;(\ref{eq:Landau_level_spectrum}). We plotted in fig.\;\ref{fig:jauge_sym} the probability density  $r|\psi_m|^2$. It is rotationally invariant around the $z$ axis,  it is maximal on a circle of radius $r_m = \sqrt{2m+1}\; \lm$ and  its width $\Delta r \sim \lm$ does not depend on $m$. Note that for a disk shape sample centered at the origin, we can recover the LLL degeneracy (\ref{eq:LLL_degeneracy}) by counting the number of $\psi_m$'s whose maximum $r_m$ is located inside the disk. 

The general form of a LLL state using the symmetric gauge is a linear combination of the $\psi_m$'s with arbitrary amplitudes $C_m$:
\begin{equation}
\psi(x,y)=\sum_{m}C_{m} \psi_{m} (x,y)=F(u)\,\E^{-r^2/4\lm^2},
\label{eq:forme_LLL}
\end{equation}
where $u=x+\I y$ and $F(u)= \sum_{m}C_{m} u^{m}$ is a polynomial or an analytic function of the complex variable $u$. With this writing, one sees that the restriction to the LLL corresponds to an effective passage from the 2D quantum motion in the $xy$ plane to the 1D motion described by functions of the $u$ variable only.

\section*{Appendix 2: Topology in the square lattice}

In this appendix we intend to give an ``elementary" derivation of the expression of the Hall conductivity for a square lattice of period $a$ pierced by a uniform flux. More precisely we want to show that for an insulator-type filling of some of the subbands of the lattice, the conductivity is characterized by an integer number, the \emph{Chern index}, with the following physical meaning: when one applies a force $F$ along one direction of the lattice, say $x$, the Hamiltonian -- in the proper frame of reference -- is time-periodic with period $\tB=h/(aF)$. For a zero flux, this is at the origin of the well-known Bloch oscillation phenomenon. For a non-zero flux, a current of particles appears along the direction $y$, such that the number of particles crossing a $x$-oriented link of the lattice $(j,l) \to (j+1,l)$ during the time duration $\tB$ is equal to the Chern index. The results derived here can of course be found in many instances in the literature (see \emph{e.g.} \cite{Thouless:1982a,Avron:1983,Kohmoto:1985,Kohmoto:1989}, and \cite{Xiao:2010} for a review), but we thought it might be useful for the reader to have a self-consistent derivation of this important result within the present set of lecture notes.

\subsection*{Band structure and periodicity in reciprocal space}
\begin{figure}[t]
\begin{center}
\includegraphics{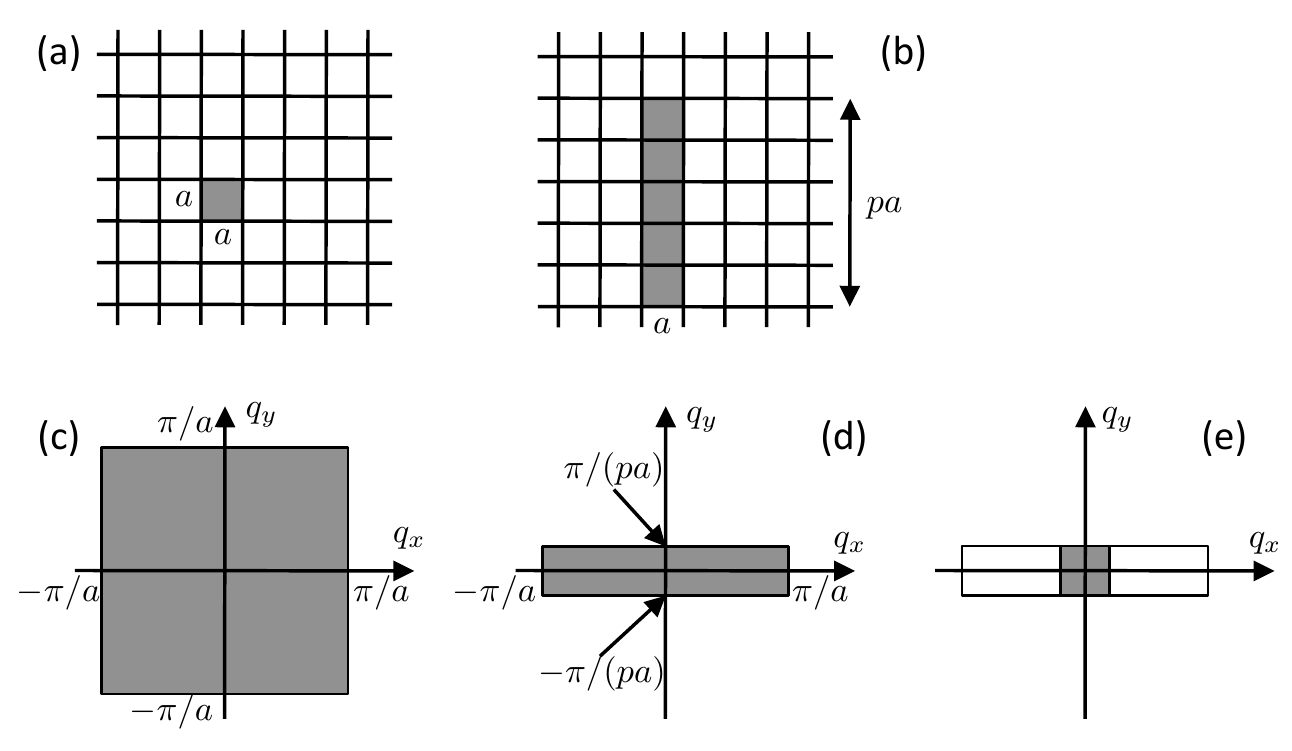}
\end{center}
\caption{Relevant regions in real (a-b) and reciprocal (c-d-e) spaces for a square lattice in the tight-binding, single mode approximation. (a) Square lattice with a unit cell of size $a\times a$. (b) Square lattice in a uniform magnetic field. The flux $\Phi$ through a initial unit cell $a\times a$ is $\Phi=\alpha\phi_0$, where $\Phi_0$ is the flux quantum. The coefficient $\alpha$ is supposed to be a rational number, $\alpha=p'/p$, so that one recovers the spatial periodicity with an increased unit cell, the so-called \emph{magnetic cell}, with a size $a\times (pa)$ for the gauge choice of the text. Here $p=5$. (c) First Brillouin zone (FBZ) for the square lattice in the absence of magnetic flux. (d) First Brillouin zone in the presence of magnetic flux (\emph{magnetic Brillouin zone}). (e) \emph{Reduced magnetic Brillouin zone}: choosing  Bloch vectors in this reduced zone is sufficient to span the whole spectrum of the Hamiltonian. }
\label{fig:zones}
\end{figure}

In this section we consider non-interacting particles moving on a square lattice in the tight-binding approximation (fig.\;\ref{fig:zones}a). In the presence of a uniform flux, we write the single-particle Hamiltonian using the Landau gauge $\bs A=-By\,\unitx$:
\begin{equation}
\hat H[\alpha,\nu]=-J \sum_{j,l} \left(  \E^{-\I\,2\pi (\alpha\,l+\nu)}
|{j+1,l}\rangle \langle {j,l}| \ +\  
|{j,l+1}\rangle \langle {j,l}| 
\right)\ +\ \mbox{h.c.}
\label{eq:H_reseau_2D_B_app}
\end{equation}
where we have added the parameter $\nu$ with respect to the Hamiltonian considered in eq.\;(\ref{eq:H_reseau_2D_B}). In principle $\nu$ can be eliminated by redefining the phases of the basis states $|{j,l}\rangle$, but it is convenient to keep it explicitly for the calculations below.

We restrict ourselves to rational values of $\alpha$, $\alpha=p'/p$ with $p,p'$ coprime positive integers, so that the Hamiltonian is spatially periodic with period $a$ along $x$ and $pa$ along $y$. Let us take the rectangle $a\times pa$ as the unit cell in the presence of the magnetic flux.  This \emph{magnetic unit cell}  is represented in fig.\;\ref{fig:zones}b in the case $p=5$. A given lattice site $(j,l)$ is assigned to the cell labelled by the integers $(j,l')$, with $l=pl'+r-1$, $r=1,\ldots,p$. We suppose that the sample is formed by $N$ such unit cells, with $N$ integer $\gg1$. We look for the eigenstates of $\hat H[\alpha,\nu]$ in the form of Bloch states:
\begin{equation}
|\psi[\alpha,\nu,\bs q]\rangle =\frac{1}{\sqrt N}\sum_{j,l'}\E^{\I a (j q_x+pl' q_y)} |u_{j,l'}[\alpha,\nu,\bs q]\rangle
\label{eq:Bloch_form}
\end{equation}
 where the normalized state $|u_{j,l'}[\alpha,\nu,\bs q]\rangle$ inside the unit cell $(j,l')$ repeats itself in a periodic manner over the whole lattice, \emph{i.e.,} it can be written
 \begin{equation}
|u_{j,l'}[\alpha,\nu,\bs q]\rangle =\sum_{r=1}^p \beta_r[\alpha,\nu,\bs q] \ \E^{\I (r-1) a q_y}\ |j,l=pl'+r-1\rangle,
\label{eq:periodic}
\end{equation}
where the coefficients $\beta_r$ are independent of the cell index $(j,l')$. Inserting the expression (\ref{eq:Bloch_form}) into the equation $\hat H|\psi\rangle=E\;|\psi\rangle$, we arrive for each set $\alpha,\nu,\bs q$ at the $p\times p$ eigenvalue problem for the Hamiltonian in reciprocal space $\hat{\cal H}[\alpha,\nu,\bs q]$:
\begin{equation}
\hat{\cal H} \;  \kket{\bs \beta[\alpha,\nu,\bs q] } = E[\alpha,\nu,\bs q]\; \kket{\bs \beta [\alpha,\nu,\bs q] }.
\label{eq:p_times_p_EV}
\end{equation}
Setting 
\begin{equation}
\Delta_r=2 \cos[aq_x+2\pi((r-1)\alpha+\nu)],
\label{eq:def_Delta}
\end{equation}
 this can be written explicitly
\begin{equation}
\hat{\cal H} =-J
\begin{pmatrix} 
\Delta_1 & \E^{\I aq_y}  & 0 & \ldots & 0 &\E^{-\I aq_y} \\
\E^{-\I aq_y}  & \Delta_2& \E^{\I aq_y} & \ldots & 0 & 0 \\
0 & \E^{-\I aq_y}  &  \Delta_3 &  \ldots & 0 & 0 \\
\vdots & \vdots & \vdots & &\vdots & \vdots\\
0 & 0 & 0 & \ldots & \Delta_{p-1} & \E^{\I aq_y} \\
\E^{\I aq_y} & 0 &0 & \ldots & \E^{-\I aq_y}  & \Delta_p \\
\end{pmatrix},
\qquad 
\kket{\bs \beta } =\begin{pmatrix} \beta_1 \\ \beta _2 \\ \vdots \\ \beta_p \end{pmatrix}.
\label{eq:H_matrix}
\end{equation}
Here we introduced the notation of curly symbols and double brackets for denoting $p$-component operators and vectors in the reciprocal space (\emph{i.e.,} at a fixed $\bs q$), in order to distinguish them from operators and vectors in the full Hilbert space.

\begin{figure}[t]
\begin{center}
\includegraphics{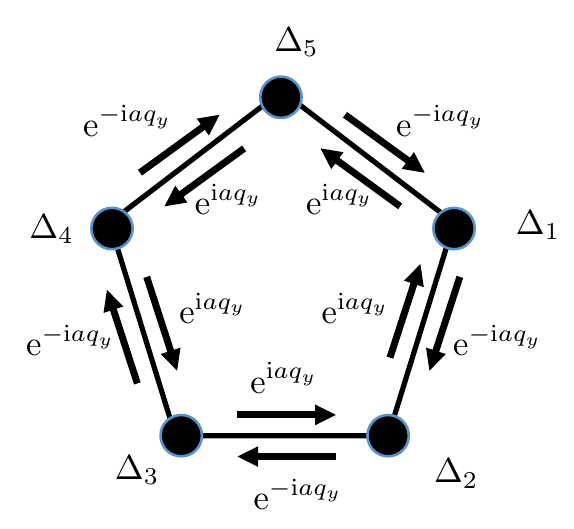}
\end{center}
\caption{The 1D cyclic chain described by the Hamiltonian $\hat {\cal H}$ of eq.\; (\ref{eq:H_matrix}) in the case $p=5$. The on-site energies are $-J\Delta_r$, $r=1,\ldots,p$, and the nearest neighbor couplings are $-J\E^{\pm \I aq_y}$.}
\label{fig:pentagone}
\end{figure}

The eigenvalue system (\ref{eq:p_times_p_EV}--\ref{eq:H_matrix}) corresponds to the problem of a particle moving on a 1D cyclic chain with $p$ sites, with on site energies $-J\Delta_r$, $r=1,\ldots,p$, and nearest neighbor couplings $-J\E^{\pm \I aq_y}$ (see 
fig.\;\ref{fig:pentagone}). It gives rises to $p$ eigenvalues $E^{(s)}[\alpha,\nu,\bs q]$, $s=1,\ldots,p$, setting by convention $E^{(1)} \leq E^{(2)}\leq \ldots \leq E^{(p)}$. For each energy $E^{(s)}$, we have one eigenvector $\kket{\bs \beta^{(s)}\!}$ for the Hamiltonian in reciprocal space $\hat{\cal H}$, hence one Bloch vector $|\psi^{(s)}\rangle $ eigenstate of $\hat H$.

When $\bs q$ is varied continuously \footnote{For a finite-size sample, the values of $q_x,q_y$ are discrete, but they become more and more closely spaced as the number of cells $N$ increases.} in the first Brillouin zone (FBZ)  associated to the magnetic unit cell (the so-called \emph{magnetic Brillouin zone}, see fig.\;\ref{fig:zones}d):
\begin{equation}
  -\frac{\pi}{a} < q_x \leq \frac{\pi}{a} \qquad  -\frac{\pi}{pa} <q_y \leq \frac{\pi}{pa}, 
%\abel{}
\end{equation}
one finds that these $p$ energies give rise to $p$ (non-overlapping) subbands -- called \emph{magnetic Bloch bands} -- for each couple $[\alpha,\nu]$. Actually due to the structure of $\hat {\cal H}$ [see eqs.\;(\ref{eq:def_Delta}) and (\ref{eq:H_matrix})], it is clear that the $E^{(s)}$'s are not independent functions of $q_x$ and $\nu$, but only depend on $aq_x+2\pi \nu$. Hence the band structure obtained when $\bs q$ varies in the FBZ is independent of $\nu$. 

With the representation in Figure \ref{fig:pentagone}, it is clear that the eigenvalue problem for $\hat {\cal H}$ depends only on the phase $\exp(\pm \I paq_y)$. Indeed the only relevant (gauge-invariant) parameter for this cyclic chain problem is the total phase accumulated in a round trip around the chain. Hence this eigenvalue problem is unchanged in the shift $q_y \to q_y + 2\pi/pa$, ensuring that the lower and upper sides of the FBZ of fig.\;\ref{fig:zones}d are equivalent. Also, since $\hat {\cal H}$ is invariant in the change $q_x \to q_x+2\pi/a$, it is clear that the left and right sides of the FBZ are equivalent. This equivalence between opposite sides of the FBZ will be essential later, when we show that the Chern number must be an integer.

A remarkable property of the eigenvalue problem of eqs.\;(\ref{eq:p_times_p_EV}) and (\ref{eq:H_matrix}) is that it is also invariant in the change $q_x \to q_x+2\pi p'/(pa)$ provided one shifts simultaneously all indices $r\to r-1$ modulo $p$ (we recall that $\alpha=p'/p$). This means that a given eigenvalue of $\hat{\cal H}$ is (at least) $p$ times degenerate, since it can be obtained from $p$ independent Bloch vectors $[q_x,q_y]$, $[q_x+2\pi/pa,q_y]$, $[q_x+4\pi/pa,q_y]$, \ldots. Therefore if one is interested in the spectrum of the Hamiltonian, one can restrict the search by choosing Bloch vectors in the \emph{reduced magnetic Brillouin zone}, where both $q_x$ and $q_y$ span an interval of width $2\pi/pa$ (see fig.\;\ref{fig:zones}e).

\subsection*{Constant force and unitary transformation}
We suppose now that a constant force $F$ is added along $x$, resulting for example from a electric field in a Hall-type experiment. The single-particle Hamiltonian then reads
\begin{equation}
\hat H_{\rm tot}^{(0)}[\alpha,\nu,F]=\hat H[\alpha,\nu] -F \hat X
\label{eq:HtotF}
\end{equation}
where the position operator $\hat X$ in the tight-binding model is
\begin{equation}
\hat X=a \sum_{j,l} j \;|j,l\rangle \langle j,l|. 
%\abel{}
\end{equation}
With the form (\ref{eq:HtotF}) we lose the spatial periodicity of the problem along $x$. However we can recover it, thanks to a unitary transformation generated by the time-dependent operator $\hat U(t)= \exp(-\I \hat X F t/\hbar)$. The total Hamiltonian after the transformation is
\begin{eqnarray}
\hat H_{\rm tot}(t)&=& U(t)\;\hat H_{\rm tot}^{(0)}\;\hat U^\dagger(t)\ +\ \I \hbar \frac{\D \hat U(t)}{\D t}\hat U^\dagger(t)
\nonumber \\
&=&
-J \sum_{j,l} \left(  \E^{-\I\,2\pi (\alpha\,l+\nu +t/\tB)}
|{j+1,l}\rangle \langle {j,l}| \ +\  
|{j,l+1}\rangle \langle {j,l}| 
\right)\ +\ \mbox{h.c.}
\label{eq:HBloch}
\end{eqnarray}
where we introduced the Bloch time
\begin{equation}
\tB= \frac{h}{aF}.
%\abel{}
\end{equation}
As announced we recover a spatially periodic problem, with the same unit cell $a\times (pa)$ as before. The presence of the uniform force $F$ along the $x$ axis is described by the additional, time-dependent  complex phase  $2\pi t/\tB$ for the couplings along the $x$ direction. 

It appears clearly in eq.\;(\ref{eq:HBloch}) that the total Hamiltonian is not an independent function of $\nu$, $F$ and $t$; the presence of the force $F$ along $x$ is indeed summarized in the modification:
\begin{equation}
\hat H_{\rm tot}[t,\alpha,\nu_0] \longrightarrow \hat H_{\rm tot}[t=0,\alpha,\nu(t)],\qquad \mbox{with}\quad \nu(t)=\nu_0+\frac{t}{\tB}.
%\abel{}
\end{equation}
Since the spatial periodicity of the Hamiltonian is conserved, it is still meaningfull to consider the Hamiltonian in reciprocal space $\hat {\cal H}$. Its structure is identical to eq.\;(\ref{eq:H_matrix}), except that the diagonal coefficients $\Delta_r$ are now time-dependent:
\begin{equation}
\Delta_r= 2 \cos \left\{aq_x +2\pi [(r-1)\alpha + \nu(t)]\right\},
\label{eq:Cr_coefficient}
\end{equation}
hence a time-periodicity $\tB$. The study of the response of the particles to the force $F$  therefore amounts to studying the dynamics associated to the time-periodic $p\times p$ Hamiltonian $\hat {\cal H}$, with (in particular) the possibility for the state of the particle to follow adiabatically one of the $p$ eigenstates of this Hamiltonian, when the phase $2\pi\nu$ entering in its diagonal coefficients varies linearly in time \cite{Thouless:1983}. 

\subsection*{Bloch oscillations and adiabatic following}

In the absence of magnetic flux ($\alpha=0$), the Hamiltonian (\ref{eq:HBloch}) is separable as a sum of two terms describing the motions along $x$ and $y$. In this case it is well known that the force $F$ causes the phenomenon of Bloch oscillations, which takes a particularly simple form in the single-band, tight-binding approximation. Suppose that one starts from the Bloch state\footnote{For zero flux, the FBZ is $(-\pi/a,\pi/a]\times (-\pi/a,\pi/a]$ (fig.\;\ref{fig:zones}c) and there is only one Bloch state, $|\psi(\bs q)\rangle=\sum_{j,l} \E^{\I a (jq_x+lq_y)}|j,l\rangle$, associated to a given quasi momentum $\bs q$.}  $|\psi(\bs q)\rangle$ at time $t=0$. In the transformed frame, the Hamiltonian $\hat H_{\rm tot}(t)$ is translationally invariant so $\bs q$ remains a good quantum number: the state is at any time proportional to $|\psi(\bs q)\rangle$ and the only dynamics is contained in the time-dependent phase acquired by this state. Alternatively, in the initial frame\footnote{In this case the unitary transformation associated to $\hat U$ is a mere shift of the quasi-momentum $\bs q \to \bs q+\unitx\,Ft/\hbar$.} where the Hamiltonian is $\hat H_{\rm tot}^{(0)}$, one finds that the state remains at any time a Bloch state with a time-dependent quasi-momentum $\bs q(t)$ moving linearly in time across the Brillouin zone at speed $\dot q_x=F/\hbar$; a duration $\tB$ is then necessary for the particle to travel across the full FBZ and come back to its initial value. First observations of Bloch oscillations with atoms in optical lattices were reported in  \cite{Dahan:1996,Wilkinson:1996}  and they have found many applications over the last decade, from matter-wave interferometry and metrology (see e.g. \cite{Denschlag:2002,Roati:2004,Clade:2006a}) to the identification of Dirac points in band structures \cite{Tarruell:2012,Duca:2015}.

The presence of a magnetic flux complicates the matter, but some essential features of the Bloch oscillation phenomenon remain valid \cite{Price:2012,Dauphin:2013,Cominotti:2013}. Suppose again that at the initial time $t=0$, the particle is prepared in a state $|\Psi(0)\rangle$ equal to a Bloch state $|\psi^{(s)}\rangle$ with quasi-momentum $\bs q$, \emph{i.e.}, one of the $p$ eigenstates of $\hat H[\alpha,\nu_0]$ associated to the energy $E^{(s)}[\alpha,\nu_0,\bs q]$. Since the Hamiltonian $\hat H_{\rm tot}[\alpha,\nu(t)]$ preserves the spatial periodicity $a \times (pa)$, the state at time $t$ remains a Bloch state with the same quasi-momentum $\bs q$. 

In the general case, this Bloch state is a linear combination of the various eigenstates of $\hat H$ for this Bloch vector $\bs q$:
\begin{equation}
|\Psi(t)\rangle= \sum_{s'=1}^p \gamma_{s'}(t)\  |\psi^{(s')}[\alpha,\nu(t),\bs q]\rangle , \qquad \nu(t)=\nu_0+\frac{t}{\tB}.
\label{eq:Psi_t}
\end{equation}
If the force $F$ is weak enough (or equivalently $\tB$ large enough), the state $|\Psi(t)\rangle$ follows quasi-adiabatically the subband $s$ when the parameter $\nu$ varies in time, so that we will have at any time $|\gamma_s| \approx 1$ and $|\gamma_{s'}|\ll 1$ for $s'\neq s$. 
%However it would be too naive to set a strictly zero value to the coefficients $\gamma_{s'}$. 
For the following discussion it will be sufficient to give an approximate value for $\gamma_{s'}(t)$ at the lowest non-vanishing order. We will see that the non-zero Hall current in the $y$ direction originates from the ``weak contamination" of the state $|\Psi(t)\rangle$ by the Bloch vectors of the subbands $s'\neq s$.

An approximate value for $\gamma_{s'}$ can be obtained by a perturbative expansion of the time-dependent Schrödinger equation (see \emph{e.g.} the appendix of  \cite{Xiao:2010}):
\begin{equation}
\gamma_{s'}(t) \approx \I\, \hbar \dot \nu \ \gamma_s(t)\;\frac{\langle \psi^{(s')}|\partial_\nu \psi^{(s)}\rangle}{E^{(s')}-E^{(s)}} \qquad \mbox{for}\quad s'\neq s.
\label{eq:corr_adiab}
\end{equation}
We can express the scalar product in the numerator of eq.\;(\ref{eq:corr_adiab})  in terms of the coefficients $\bs \beta$ of the periodic part of the Bloch vector, and take advantage of the fact that they depend only on the quantity $aq_x-2\pi \nu$:
\begin{equation}
\gamma_{s'}(t) \approx \I \, F \ \gamma_s(t)\;\frac{\bbrab{\bs \beta^{(s')}}  \kket{ \partial_{q_x} \bs \beta^{(s)} }  }  
{E^{(s')}-E^{(s)} } \qquad \mbox{for}\quad s'\neq s.
\label{eq:corr_adiab2}
\end{equation}

\subsection*{The velocity operator and its matrix elements} Ultimately we want to evaluate the average Hall current that appears along the direction $y$ when the force $F$  is applied along $x$. We define the velocity operator $\hat V_y$ as the time derivative (in the Heisenberg picture) of the position operator $\hat Y$:
\begin{equation}
\hat V_y=\frac{\D \hat Y}{\D t}=\frac{\I}{\hbar} \left[ \hat H, \hat Y \right], \qquad \hat Y=a \sum_{j,l} l\; |j,l\rangle \langle j,l|.
%\abel{}
\end{equation}
The expression of $\hat V_y$ does not depend on the parameters $\alpha$, $\nu$ and $F$ entering in the definition of the Hamiltonian and it reads in the tight-binding approximation:
\begin{equation}
\hat V_y =\I\,\frac{ Ja}{\hbar} \sum_{j,l} \left( |j,l+1\rangle \langle j,l|- |j,l\rangle \langle j,l+1| \right).
%\abel{}
\end{equation}
The velocity operator is invariant in a translation over the square lattice. Hence  the subspace associated to a given Bloch vector $\bs q$ is globally invariant under the action of $\hat V_y$. In other words, for the problem under consideration here, the velocity operator is fully characterized by the $p\times p$ matrix elements  between Bloch states
\begin{equation}
\langle \psi^{(s')}[\alpha,\nu,\bs q]| \ \hat V_y\  | \psi^{(s)}[\alpha,\nu,\bs q]\rangle, \quad s,s'=1,\ldots,p.
%\abel{}
\end{equation}
It is convenient to rewrite this matrix element in terms of the coefficients $\bs \beta^{(s)},\bs \beta^{(s')}$ giving the periodic part of the Bloch vector:
\begin{equation}
\langle \psi^{(s')}| \ \hat V_y\  | \psi^{(s)}\rangle=\bbra {\bs \beta^{(s')}} \ \hat {\cal V}_y\  \kket{ \bs \beta^{(s)} }
\label{eq:matrix_velocity}
\end{equation}
where we have introduced the $p\times p$ matrix $\hat {\cal V}_y$ giving the velocity operator in reciprocal space:
\begin{equation}
\hat{\cal V}_y =\I \frac{Ja}{\hbar}
\begin{pmatrix} 
0 & -\E^{\I aq_y}  & 0 & \ldots & 0 &\E^{-\I aq_y} \\
\E^{-\I aq_y}  & 0 & -\E^{\I aq_y} & \ldots & 0 & 0 \\
0 & \E^{-\I aq_y}  &  0 &  \ldots & 0 & 0 \\
\vdots & \vdots & \vdots & &\vdots & \vdots\\
0 & 0 & 0 & \ldots & 0 & -\E^{\I aq_y} \\
-\E^{\I aq_y} & 0 &0 & \ldots & \E^{-\I aq_y}  & 0 \\
\end{pmatrix}.
%\abel{}
\end{equation}
One can immediately check that this velocity operator is connected to the Hamiltonian $\hat {\cal H}$ by the simple relation
\begin{equation}
\hat{\cal V}_y = \frac{1}{\hbar}\;\frac{\partial \hat {\cal H}}{\partial q_y} .
%\abel{}
\end{equation}
This expression allows us to give an explicit expression for the matrix elements (\ref{eq:matrix_velocity}) that will be useful in the following. Starting from the eigenvalue equation $\hat {\cal H}\kket{\bs \beta^{(s)}}=E^{(s)} \kket{\bs \beta^{(s)}}$\,, taking its derivate with respect to $q_y$ and multiplying with $\bbra{\bs \beta^{(s')}}$, we obtain 
\begin{eqnarray}
\bbra {\bs \beta^{(s')}} \ \hat {\cal V}_y\  \kket{ \bs \beta^{(s)} } &=& \frac{1}{\hbar}\;(E^{(s)}-E^{(s')}) \ 
\bbrab {\bs \beta^{(s')}} \kket{ \partial_{q_y}\bs \beta^{(s)} } \qquad \mbox{for}\quad   s\neq s' ,\\
\bbra {\bs \beta^{(s)}} \ \hat {\cal V}_y\  \kket{ \bs \beta^{(s)} } &=& \frac{1}{\hbar} \partial_{q_y}E^{(s)}.
%\abel{}
\end{eqnarray}

\subsection*{The Berry curvature}
Since we  now have at our disposal the velocity operator along the $y$ direction, we can calculate the average flux along this direction when the force $F$ is applied along $x$ and the system is prepared 
in a given Bloch vector $|\psi^{(s)}\rangle$. We suppose that the force $F$ is small enough for the adiabatic approximation to hold, so that we can use the perturbative expansion of eqs.\;(\ref{eq:Psi_t}) and (\ref{eq:corr_adiab2}) for the state of the system $|\Psi(t)\rangle$. Using $|\gamma_s|^2\approx 1$, we obtain
\begin{equation}
\langle \Psi(t) |\hat V_y |\Psi(t)\rangle \approx \frac{1}{\hbar}\partial_{q_y}E^{(s)}+ \frac{\I F}{\hbar}
\sum_{s'\neq s}\left\{ \bbrab {\partial_{q_x} \bs \beta^{(s)} } \kket {\bs \beta^{(s')}} \ \bbrab{ \bs \beta^{(s')} } 
\kket{\partial_{q_y} \bs \beta^{(s)} } - \mbox{c.c.} \right\}.
%\abel{}
\end{equation}
We can formally add to the sum over $s'$  the contribution of the term $s'=s$, since this term is actually zero\footnote{The two quantities $\bbrab {\partial_{q_x} \bs \beta^{(s)} } \kket {\bs \beta^{(s)}}$ and $\bbrab {\partial_{q_y} \bs \beta^{(s)} } \kket {\bs \beta^{(s)}}$ are purely imaginary  
since $\kket {\bs \beta^{(s)}}$ is normalized.}. Using a closure relation we then obtain
\begin{equation}
\bar V_y=\langle \Psi(t) |\hat V_y |\Psi(t)\rangle \approx \frac{1}{\hbar}\partial_{q_y}E^{(s)}+ \frac{ F}{\hbar} {\cal B}^{(s)}(\bs q),
\label{eq:barVy}
\end{equation}
where we have introduced the Berry curvature for a given $\alpha$ and for the subband $s$
\begin{equation}
{\cal B}^{(s)}(\bs q)=\I \left( \bbrab {\partial_{q_x} \bs \beta^{(s)} } 
\kket{\partial_{q_y} \bs \beta^{(s)} } - \bbrab {\partial_{q_y} \bs \beta^{(s)} } 
\kket{\partial_{q_x} \bs \beta^{(s)} } \right).
\label{eq:Berry_curvature}
\end{equation}
Eq.\;(\ref{eq:barVy}) is the starting point of the semi-classical study of the dynamics of an electron in a magnetic field \cite{Xiao:2010}. It shows that the velocity of the particle includes, in addition to the usual group velocity $\bs \nabla_{\bs q} E^{(s)}/\hbar$,  a second term proportional to the Berry curvature $ {\cal B}^{(s)}$  of the band \cite{Karplus:1954}, which is sometimes called the  \emph{anomalous velocity} \cite{Xiao:2010}.

From the average velocity $\bar V_y$ in eq.\;(\ref{eq:barVy}), we can deduce the average flux through a given horizontal link $j,l\to j+1,l$ of the lattice. Since a Bloch state corresponds to one particle that is delocalized over $N$ lattice cells, with each cell having an area $pa^2$, the number of particles crossing the considered link of length $a$ in a time interval $\delta t$ is 
\begin{equation}
\delta n=\frac{a \bar V_y }{Npa^2}\;\delta t .
\label{eq:flux_one_particle}
\end{equation}

\subsection*{Conduction from a filled band and Chern number}
The last step in our reasoning is to consider a situation where a macroscopic number ($\sim N$) of non-interacting fermionic particles are simultaneously present and to evaluate the flux in this case. For simplicity we consider a situation with exactly $N$ particles and assume that the gas is initially in its ground state, with only the lowest subband $s=1$ populated and all the other subbands empty. We also assume that the subband $s=1$ is separated from the next subband $s=2$ by a gap\footnote{This would not hold for $\alpha=1/2$ where the two subbands touch at a Dirac point.}. Then, for a small enough force $F$, the population essentially remains in the lowest subband at any time.

Starting from a quantity $\Phi(\bs q)$ calculated for one particle in a given Bloch state $\psi(\bs q)$, we obtain the contribution of the $N$ particles with their Bloch vectors spanning the FBZ by the following integral:
\begin{equation}
\Phi(\bs q) \longrightarrow N \intint_{\rm FBZ} \frac{\D q_x}{2\pi/a}\;\frac{\D q_y}{2\pi/pa}\ \Phi(\bs q).
%\abel{}
\end{equation}
Here, using the expression of the flux (\ref{eq:flux_one_particle}) for a single Bloch state, we obtain the flux of particles crossing a given horizontal link of length $a$ when the subband $s=1$ is filled:
\begin{equation}
\delta N=\frac{{\cal C}^{(s=1)}}{\tB} \;\delta t
\label{eq:flux_total}
\end{equation}
where we have defined the Chern number associated to the subband $s$
\begin{equation}
{\cal C}^{(s)}=\frac{1}{2\pi} \intint_{\rm FBZ} {\cal B}^{(s)}(\bs q)\; \D q_x\,\D q_y.
\label{eq:def_Chen}
\end{equation} 
Note that only the second term in the right-hand side expression (\ref{eq:barVy}) of $\bar V_y$ contributes to $\delta N$, since the first term proportional to $\partial_{q_y}E^{(s)}(\bs q)$ has a zero-average over the FBZ. 

The expression (\ref{eq:flux_total}) leads to the interpretation of the Chern number announced in the introduction: when one applies a force $F$ along $x$, the Bloch oscillation phenomenon occurs with the time period $\tB=h/aF$. In the presence of a flux through the lattice, the Hall current in the $y$ direction is such that ${\cal C}^{(s=1)}$ particles cross a given horizontal link of length $a$ during the time interval $\tB$.

As it is defined, the Chern number clearly depends on the flux $\alpha$. One could think that it is also a function of $\nu_0$ and $F$; however  ${\cal C}^{(s)}$ is actually independent of these quantities. Indeed they enter in the Hamiltonian $\hat {\cal H}$ only via the coefficients $\Delta_r$ defined in  eq.\;(\ref{eq:Cr_coefficient}), through the linear combination $aq_x -2\pi(\nu_0-t/\tB)$. Since one performs an integration over $q_x$ on the FBZ, the values of $\nu_0$ and $\tB$ (hence $F$) are irrelevant for the value of the integral, hence for the determination of the Chern number.

\subsection*{The Chern number is an integer}
In order to show this general property, we first introduce \emph{Berry's connection} in reciprocal space
\begin{equation}
\bs {\cal A}^{(s)} (\bs q) = \I \;\bbrab{ \bs \beta^{(s)} } \kket{ \bs \nabla_{\bs q} \bs \beta^{(s)}}\ ,
\label{eq:Berry_connection}
\end{equation}
which is a vector in the $q_x,q_y$ plane such that
\begin{equation}
{\cal B}^{(s)}(\bs q) = {\bs u_z} \cdot \left(  \bs \nabla_{\bs q} \times \bs {\cal A}^{(s)} (\bs q) \right).
%\abel{}
\end{equation}
It is then ``tempting" to replace the surface integral  (\ref{eq:def_Chen}) of ${\cal B}^{(s)}$ on the FBZ by the contour integral of $\bs {\cal A}^{(s)} $ around the edge of the FBZ:
\begin{equation}
{\cal C}^{(s)}=\frac{1}{2\pi}  \oint_{\partial{\rm FBZ}} \bs {\cal A}^{(s)} (\bs q)\cdot \D \bs q.
\label{eq:contour}
\end{equation}
However this requires some care since, as emphasized by Kohmoto in \cite{Kohmoto:1989},  eq.\;(\ref{eq:def_Chen}) has a subtle topological nature. First we note that Berry's curvature (\ref{eq:Berry_curvature}) is gauge-independent, \emph{i.e.,} it is not modified in the change $\kket{ \bs \beta } \to \E^{\I \theta} \kket{ \bs \beta }$, where $\theta$ is a smooth function of $\bs q$. On the contrary, Berry's connection is gauge-dependent so that eq.\;(\ref{eq:contour}) could \emph{a priori} depend on the gauge choice. Second we notice that if there exists a gauge choice which defines a global, single-valued phase of the $\kket{\bs \beta}$\;'s over the whole FBZ, then the contour integral (\ref{eq:contour}) must be zero; indeed the FBZ has the structure of a torus (opposite sides correspond to the same physical situation) and its ``edge" has thus a zero length. 

In the presence of a magnetic field, the situation is made subtle by the fact that it is generally not possible to define a smooth global and single-valued phase for the $\kket{\bs \beta}$\;'s over the FBZ. When looking for such a definition, a possible strategy could consist in setting one of the coefficients of $\kket{\bs \beta}$ in eq.\;(\ref{eq:H_matrix}), say the first one $\beta_1$, to be real and positive everywhere in the FBZ. But this strategy fails if there exists points in the FBZ, where this first component vanishes: the phase of 
$\kket{\bs \beta}$ is ill-defined at these points, which introduces singularities in $\bs {\cal A}^{(s)}$. 

Two options for circumventing this difficulty and using eq.\;(\ref{eq:contour}) have been developed. The first one consists in dividing the FBZ in various zones over which the phase of  $\kket{\bs \beta}$ is separately well-defined, the remaining task being to properly account (via a gauge change) for the discontinuities of $\bs {\cal A}^{(s)} $ at the boundaries between these zones \cite{Kohmoto:1985}. The other option is to relax the condition that $\kket{\bs \beta}$ should be periodic over the Brillouin zone, \emph{i.e.,} single-valued over the torus. In this case $\bs {\cal A}^{(s)} $ can be chosen as a smooth function over the FBZ  \cite{Thouless:1990},  and the contour in the integral (\ref{eq:contour}) becomes a `true' rectangle (and not a zero-length line on a torus); the challenge in this case is to properly take into account the connexion between the choices for the $\kket{\bs \beta}$\;'s on opposite sides of the FBZ. This is the strategy that we adopt now.

\begin{figure}[t]
\begin{center}
\includegraphics{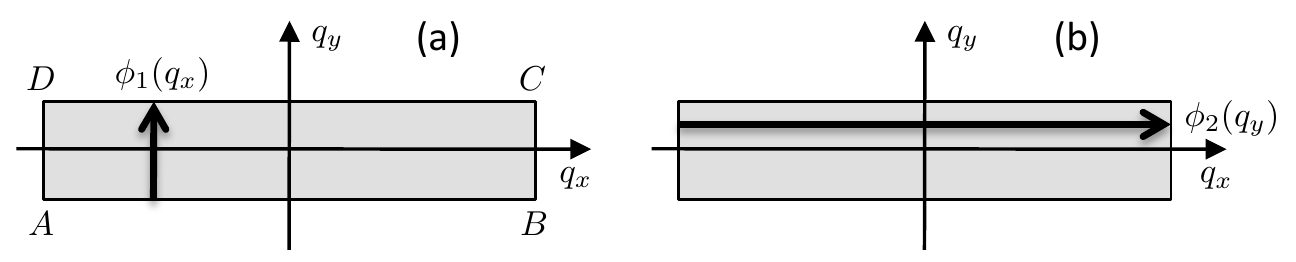}
\end{center}
\caption{Connecting opposite sides of the FBZ. Due to the periodicity of the problem, the states on opposite sides of the FBZ can differ only by a phase factor, which we denote $\E^{\I \phi_1(q_x)}$ for the upper-lower sides, and $\E^{\I \phi_2(q_y)}$ for the left-right sides.}
\label{fig:connecting_sides}
\end{figure}

The periodicity of the Hamiltonian in reciprocal space entails that its eigenstates on two opposite sides of the FBZ differ only by a phase factor. More precisely, there must exist a function $\phi_1(q_x)$ that relates the lower and upper sides of the FBZ (fig.\;\ref{fig:connecting_sides}a):
\begin{equation}
\kket{ \bs \beta^{(s)} (q_x, \frac{\pi}{pa})} = \E^{\I \phi_1(q_x)} \kket{ \bs \beta^{(s)} (q_x, -\frac{\pi}{pa})} 
%\abel{}
\end{equation} 
and a function $\phi_2(q_y)$ that relates the left-hand and the right-hand sides of the FBZ 
(fig.\;\ref{fig:connecting_sides}b):
\begin{equation}
\kket{ \bs \beta^{(s)} (\frac{\pi}{a}, q_y)} = \E^{\I \phi_2(q_y)} \kket{ \bs \beta^{(s)} ( -\frac{\pi}{a},q_y)}. 
%\abel{}
\end{equation} 
When one evaluates the contour integral (\ref{eq:contour}) following the path $ABCD$, one can regroup the contribution of the segments $AB$ and $CD$:
\begin{equation}
\left( \int_{AB}+\int_{CD }\right) \bs {\cal A}^{(s)} (\bs q)\cdot \D \bs q=\int_{-\pi/a}^{\pi/a} \phi_1'(q_x)\;\D q_x=
\phi_1(\pi/a)-\phi_1(-\pi/a),
\label{eq:line1}
\end{equation}
and the contributions of the segments $BC$ and $DA$:
\begin{equation}
\left( \int_{BC}+\int_{DA }\right) \bs {\cal A}^{(s)} (\bs q)\cdot \D \bs q=-\int_{-\pi/pa}^{\pi/pa} \phi_2'(q_y)\;\D q_y=
\phi_2(-\pi/pa)-\phi_2(\pi/pa).
\label{eq:line2}
\end{equation}

The last step in the reasoning is to notice that the states at each corner of the FBZ are all equal up to a phase factor. In particular there are two ways to relate the state in the lower left corner $A$ and the one in the upper right corner $C$:
\begin{eqnarray*}
\kket{ \bs \beta^{(s)} (\frac{\pi}{a}, \frac{\pi}{pa})} &=&\E^{\I \phi_1(\pi/a) } \ \kket{ \bs \beta^{(s)} (\frac{\pi}{a}, -\frac{\pi}{pa})} = \E^{\I \left[\phi_1(\pi/a)+\phi_2(-\pi/pa)\right] } \ \kket{ \bs \beta^{(s)} (-\frac{\pi}{a}, -\frac{\pi}{pa})}\ , \\
 &=&\E^{\I \phi_2(\pi/a) } \ \kket{ \bs \beta^{(s)} (-\frac{\pi}{a}, \frac{\pi}{pa})} = \E^{\I \left[\phi_1(-\pi/a)+\phi_2(\pi/pa)\right] }\ \kket{ \bs \beta^{(s)} (-\frac{\pi}{a}, -\frac{\pi}{pa})}\ .
\end{eqnarray*}
These two ways are equivalent if and only if
\begin{equation}
\phi_1(\pi/a)+\phi_2(-\pi/pa) = \phi_1(-\pi/a)+\phi_2(\pi/pa)\quad \mbox{modulo}\ 2\pi.
%\abel{}
\end{equation}
This entails that the sum of the line integrals of $\bs {\cal A}^{(s)}$ on the four segments $AB$, $BC$, $CD$, and $DA$ obtained by adding eqs.\;(\ref{eq:line1}) and (\ref{eq:line2}) is a multiple of $2\pi$, hence the Chern number (\ref{eq:contour}) is an integer. We have proven this result in the specific case of a square lattice, but it can be generalized to more complex geometries (see \cite{Xiao:2010} and refs. in).

One can go one step further and determine the value of the Chern number for a given flux $\alpha=p'/p$ and a subband $s$ \cite{Thouless:1982a}. We give here the result without proof for the lowest subband; the Chern number appears in the solution of the Diophantine equation 
\begin{equation}
1=p' {\cal C}^{s=1}+ p {\cal D}
%\abel{}
\end{equation}
where ${\cal D}$ is an integer such that $|{\cal D}|\leq p'/2$. For the particular case $\alpha=1/p$, {\emph i.e.,} $p'=1$, this gives the Chern number ${\cal C}^{s=1}=1$ (and ${\cal D}=0$).

\bibliographystyle{varenna}
%\bibliography{//users/jean/Documents/biblioJD}

\end{document}